\documentclass[10pt,twocolumn,american,prx,simglecolumn,amsmath,amssymb,superscriptaddress]{revtex4-1}
\usepackage[T1]{fontenc}
\usepackage[latin9]{inputenc}
\usepackage{xcolor}
\usepackage{babel}
\usepackage{prettyref}
\usepackage{mathtools}
\usepackage{amsmath}
\usepackage{stackrel}
\usepackage{graphicx}
\usepackage{wasysym}
\usepackage[unicode=true,pdfusetitle,
 bookmarks=true,bookmarksnumbered=false,bookmarksopen=false,
 breaklinks=True,pdfborder={0 0 1},backref=false,colorlinks=false]
 {hyperref}
\hypersetup{
 colorlinks=true,linkcolor=black,citecolor=black,urlcolor=black,filecolor=black}
\usepackage{breakurl}

\makeatletter

\newcommand{\lyxdot}{.}

\usepackage{MnSymbol}
\usepackage{notoccite}

\newrefformat{cap}{\hyperref[#1]{Figure~\ref{#1}}}
\newrefformat{fig}{\hyperref[#1]{Figure~\ref{#1}}}
\newrefformat{tab}{\hyperref[#1]{Table ~\ref{#1}}}
\newrefformat{sec}{\hyperref[#1]{Section~\ref{#1}}}
\newrefformat{sub}{\hyperref[#1]{Section~\ref{#1}}}
\newrefformat{cha}{\hyperref[#1]{Chapter~\ref{#1}}}
\newrefformat{app}{\hyperref[#1]{Appendix~\ref{#1}}}

\renewcommand{\textendash}{--}

\makeatother

\begin{document}
\global\long\def\D{\mathcal{D}}
\global\long\def\bx{\boldsymbol{x}}
\global\long\def\bl{\boldsymbol{l}}
\global\long\def\bh{\boldsymbol{h}}
\global\long\def\bJ{\boldsymbol{J}}
\global\long\def\N{\mathcal{N}}
\global\long\def\hh{\hat{h}}
\global\long\def\bhh{\hat{\boldsymbol{h}}}
\global\long\def\T{\mathrm{T}}
\global\long\def\Tbi{\mathrm{T}_{\mathrm{p}}}
\global\long\def\TI{\mathrm{T}_{\mathrm{I}}}
\global\long\def\by{\mathrm{\mathbf{y}}}
\global\long\def\diag{\mathrm{diag}}
\global\long\def\th{\tilde{h}}
\global\long\def\tj{\tilde{j}}

\global\long\def\Gammafl{\Gamma_{\mathrm{fl}}}
\global\long\def\gammafl{\gamma_{\mathrm{fl}}}
\global\long\def\R{\mathbb{R}}
\global\long\def\Z{\mathcal{Z}}
\global\long\def\jt{j^{\mathrm{T}}}
\global\long\def\Tb{T}
\global\long\def\MR{\boldsymbol{R}}
\global\long\def\MQ{\boldsymbol{Q}}
\global\long\def\Qi{{\cal Q}}
\global\long\def\Qs{Q}
\global\long\def\Qr{Q_{12}}
\global\long\def\Qri{Q_{12}^{\mathrm{I}}}
\global\long\def\Qrb{Q_{12}^{\mathrm{b}}}
\global\long\def\Qzb{Q_{0}^{\mathrm{b}}}
\global\long\def\Qzi{Q_{0}^{\mathrm{I}}}
\global\long\def\qr{q_{12}}
\global\long\def\qrb{q_{12}^{\mathrm{b}}}
\global\long\def\qri{q_{12}^{\mathrm{I}}}
\global\long\def\xe{x^{1}}
\global\long\def\xz{x^{2}}
\global\long\def\he{h^{1}}
\global\long\def\hz{h^{2}}
\global\long\def\Qrd{\epsilon\left(s\right)}
\global\long\def\Qrid{\epsilon^{\mathrm{I}}\left(s\right)}
\global\long\def\erf{\mathrm{erf}}

\global\long\def\b#1{\boldsymbol{#1}}
 \global\long\def\av#1#2{\left\langle #1\right\rangle _{#2}}
\global\long\def\cum#1#2{{\left\llangle #1\right\rrangle }_{#2}}
 \global\long\def\D{\mathcal{D}}
 \global\long\def\d{\mathrm{d}}
 \global\long\def\at#1#2{\left.#1\right|_{#2}}
 \global\long\def\abs#1{\left|#1\right|}

\global\long\def\bR{\mathbb{R}}

\title{Transient Chaotic Dimensionality Expansion by Recurrent Networks}

\author{Christian Keup}

\thanks{C.K. and T.K. contributed equally to this work.}

\affiliation{Institute of Neuroscience and Medicine (INM-6) and Institute for
Advanced Simulation (IAS-6) and JARA Institut Brain Structure-Function
Relationships (INM-10), J\"ulich Research Centre, J\"ulich, Germany}

\affiliation{RWTH Aachen University, Aachen, Germany}

\author{Tobias K\"uhn}

\thanks{C.K. and T.K. contributed equally to this work.}

\affiliation{Institute of Neuroscience and Medicine (INM-6) and Institute for
Advanced Simulation (IAS-6) and JARA Institut Brain Structure-Function
Relationships (INM-10), J\"ulich Research Centre, J\"ulich, Germany}

\affiliation{RWTH Aachen University, Aachen, Germany}

\affiliation{Laboratoire de Physique de l'ENS, Laboratoire MSC de l'Université
de Paris, CNRS, Paris, France}

\author{David Dahmen}

\affiliation{Institute of Neuroscience and Medicine (INM-6) and Institute for
Advanced Simulation (IAS-6) and JARA Institut Brain Structure-Function
Relationships (INM-10), J\"ulich Research Centre, J\"ulich, Germany}

\author{Moritz Helias}

\affiliation{Institute of Neuroscience and Medicine (INM-6) and Institute for
Advanced Simulation (IAS-6) and JARA Institut Brain Structure-Function
Relationships (INM-10), J\"ulich Research Centre, J\"ulich, Germany}

\affiliation{Department of Physics, Faculty 1, RWTH Aachen University, Aachen,
Germany}

\date{\today}

\maketitle

\section*{Abstract}

Neurons in the brain communicate with spikes, which are discrete
events in time and value. Functional network models often employ rate
units that are continuously coupled by analog signals. Is there a
qualitative difference implied by these two forms of signaling? We
develop a unified mean-field theory for large random networks to show
that first- and second-order statistics in rate and binary networks
are in fact identical if rate neurons receive the right amount of
noise. Their response to presented stimuli, however, can be radically
different. We quantify these differences by studying how nearby state
trajectories evolve over time, asking to what extent the dynamics
is chaotic. Chaos in the two models is found to be qualitatively different.
In binary networks we find a network-size-dependent transition to
chaos and a chaotic submanifold whose dimensionality expands stereotypically
with time, while rate networks with matched statistics are nonchaotic.
Dimensionality expansion in chaotic binary networks aids classification
in reservoir computing and optimal performance is reached within about
a single activation per neuron; a fast mechanism for computation that
we demonstrate also in spiking networks. A generalization of this
mechanism extends to rate networks in their respective chaotic regimes.

\section{Introduction}

While biological neurons communicate by spikes, which are discrete
all-or-nothing events, artificial neural networks overwhelmingly use
continuous-valued units, commonly referred to as ``rate neurons''.
The ramifications of this fundamental distinction between discrete
and continuous signaling have been debated concerning learning algorithms
\citep{Kempter98_nips,Pfeiffer18}, energy efficiency \citep{Laughlin01},
and information coding \citep{Corticonics,Softky93,Koenig96,Shadlen98,Shadlen99,rolls10_noisybook,Brette15_151,Deneve16}.
\begin{figure}
\begin{centering}
\includegraphics{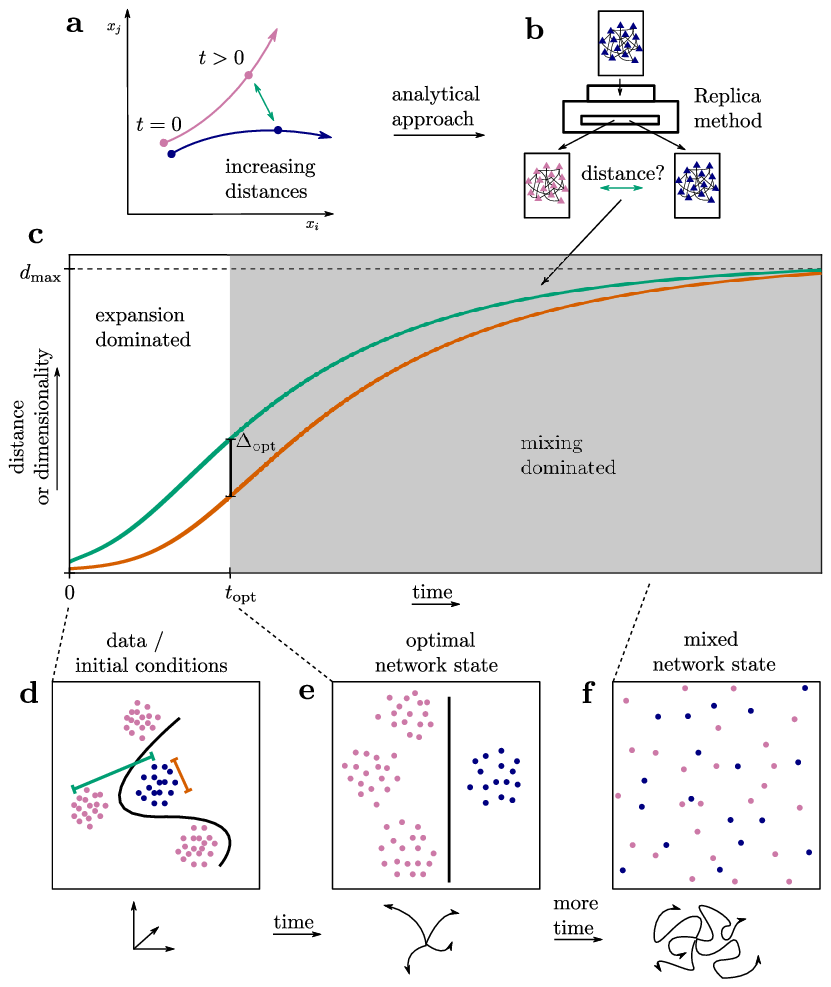}
\par\end{centering}
\caption{\textbf{Transient chaotic dimensionality expansion. }The divergence
of trajectories with different initial conditions in a chaotic network
(\textbf{a}) can be computed by a replica calculation (\textbf{b}),
leading to a curve describing the average temporal evolution of the
distance (\textbf{c}). Sketch of linearly nonseparable data (\textbf{d})
supplied as initial condition to the dynamics (\textbf{a}). Chaotic
dynamics expands the representation into spaces with increasing dimension
(\textbf{c}). Interclass separation (green) initially grows quicker
than variability within class (dark orange) up to point $t_{\mathrm{opt}}$
of optimal separability (\textbf{e}). Subsequently, chaotic mixing
causes separability to decline, ultimately leading to a completely
mixed state (\textbf{f}).\label{fig:Transient-chaotic-dimensionality-summary}}
\end{figure}
Here we study how differences in signaling impact network dynamics
underlying classification performance in a reservoir setting \citep{Buonomano95_1028,Jaeger01_echo,Maass02_2531}:
Input stimuli influence the dynamical state of a randomly connected
network which then acts as the representation, from which the desired
output is extracted by a linear readout. For a classification task,
the representation thus needs to allow a linear separation of classes.
Dynamics promotes this separability by nonlinearly embedding the input
into its high-dimensional state space. This embedding is analogous
to the kernel trick used in support vector machines \citep{Vapnik98}:
A generic mapping into a high-dimensional nonlinear feature space
tends to improve separability, because in $N$ dimensions dichotomies
of $2N$ random points can be linearly separated with high probability
\citep{Cover1965_326}. Presenting input stimuli as initial conditions
to the dynamics of a network, the nonlinear transformation of the
representation is determined by the subsequent temporal evolution
(\prettyref{fig:Transient-chaotic-dimensionality-summary}\textbf{a,d}).\textcolor{black}{{}
For example, consider stimuli belonging to different classes, each
given by a centroid and local noise (\prettyref{fig:Transient-chaotic-dimensionality-summary}}\textbf{\textcolor{black}{d}}\textcolor{black}{).
Two properties are needed for classificat}ion: Differences between
stimulus classes must be maintained or amplified to foster discrimination
(\prettyref{fig:Transient-chaotic-dimensionality-summary}\textbf{c},
green). Similar stimuli, however, should lead to similar representations
to support generalization; the distance between trajectories of data
points belonging to the same class should have limited growth (\prettyref{fig:Transient-chaotic-dimensionality-summary}\textbf{c},
dark orange). This view exposes the tight link to chaos, the sensitivity
of the dynamics to initial conditions. For rate networks close to
the edge of chaos, separation and generalization are well balanced,
leading generally to optimal performance \citep{Toyoizumi11_051908,Bertschinger04_1413,Legenstein07_127,Legenstein07_323}.
While the theory of deterministic \citep{Sompolinsky88_259,Crisanti18_062120}
and stochastic rate networks \citep{Schuecker18_041029} is well understood
and predicts a clearly defined transition to chaos, its link to chaos
in binary networks \citep{Vreeswijk96,Vreeswijk98} remains elusive.
Binary networks are the simplest class of models with discrete signaling
between neurons.

Here we develop a systematic and model-independent approach to derive
mean-field theories for large random networks (\prettyref{sec:Field-theoretic-description-of-binary-neurons}).
The formalism finds the same set of mean-field equations simultaneously
describing binary and rate networks. It shows that a stochastic rate
network with properly chosen noise has the same first- and second-order
activity statistics as a binary network. The approach allows for replica
calculations, the study of ensembles of pairs of networks with identical
connectivity in each realization, but different stimuli, as required
to assess chaos and computation (\prettyref{fig:Transient-chaotic-dimensionality-summary}\textbf{b}).
For stochastic dynamics one compares two systems with slightly different
initial conditions but identical realization of stochasticity \citep{Baxendale92_3}.
The replica theory exposes that chaos and signal processing in statistically
matched rate and binary networks are qualitatively different: Binary
networks show a transition to chaos that depends on network size (\prettyref{sec:Binary-networks-always-chaotic-in-thermlim}
and \prettyref{sec:Transition-to-chaos-binary-finite-size}). In the
chaotic regime, distances between states in binary networks increase
transiently in a stereotypical manner, confined to a chaotic submanifold
whose dimension depends on the coupling strength and is a fraction
of the entire state space ($d_{\text{max}}$ in \prettyref{fig:Transient-chaotic-dimensionality-summary}\textbf{c},
\prettyref{sec:dynamics-submanifold}). Rate networks with statistically
matched activity, in contrast, are nonchaotic (\prettyref{sec:Same_stat_Different-mechanisms-of-chaos}).
Giving up on the statistical match, rate networks with weak noise
in their corresponding chaotic regime show a qualitatively different
divergence of state trajectories that sensitively depends on the coupling
strength (\prettyref{sec:Same_stat_Different-mechanisms-of-chaos}).
Given a distribution of input data whose within-class variability
is smaller than the average between-class distances (\prettyref{fig:Transient-chaotic-dimensionality-summary}\textbf{d},
dark orange and green), the dimensionality expansion of presented
stimuli by chaotic binary networks leads to a separation that is optimal
for classification after $t_{\mathrm{opt}}/\tau=2\ln2\simeq1.4$ activations
per neuron (\prettyref{fig:Transient-chaotic-dimensionality-summary}\textbf{c},\textbf{e},
\prettyref{sec:Computation-by-transient-dim-expansion}). Subsequently,
the chaotic mixing leads to a gradual decline of separability (\prettyref{fig:Transient-chaotic-dimensionality-summary}\textbf{f}).
Despite the qualitative differences between rate and binary networks,
both mechanisms of chaos can be employed to increase classification
performance deep in the chaotic regime in a wide range of networks
models, including long-short-term-memory (LSTM) and spiking networks
(\prettyref{sec:Generalization-SNR-amplificaiton}).

\section{Results\label{sec:Results}}

\subsection{Model-independent field theory of neuronal networks\label{sec:Field-theoretic-description-of-binary-neurons}}

Here we derive a framework to compute the statistics of neuronal networks
in a manner that is largely independent of the employed neuron model.
Such a framework is needed to systematically compare different model
classes and to assess the generality of results. It must be flexible
enough to enable the use of methods such as disorder averages and
replica calculations; techniques that are required to systematically
derive mean-field equations that allow us to compare networks on a
statistical level and to assess how distances between different dynamical
states evolve over time and classification of input signals can be
achieved (\prettyref{fig:Transient-chaotic-dimensionality-summary}).

We consider a network of $N$ neurons with connectivity matrix $\b J$,
where individual entries are independently and identically distributed
as $J_{ij}\stackrel{\text{i.i.d.}}{\sim}\mathcal{N}\left(\frac{\bar{g}}{N},\frac{g^{2}}{N}\right)$;
assumptions on the statistics can easily be relaxed as long as higher-order
cumulants are suppressed by the large network size. The $N$ neurons
have inputs $\b h=\left(h_{1}(t),...,h_{N}(t)\right)$ and outputs
$\b x=\left(x_{1}(t),...,x_{N}(t)\right)$. The input-to-output relation
of a neuron is often stochastic, so that a conditional probability
$\rho[\boldsymbol{x}|\boldsymbol{h}]$ of the output given the input
is the most general description of the neural dynamics. The joint
statistics of input and output is then
\begin{equation}
\rho[\boldsymbol{x},\boldsymbol{h}]=\rho[\boldsymbol{x}|\boldsymbol{h}]\,\rho[\boldsymbol{h}],\label{eq:decomp_input_output}
\end{equation}
amounting to a separation of the neurons' input-output functional
$\rho[\boldsymbol{x}|\boldsymbol{h}]:=\prod_{i}\rho[x_{i}|h_{i}]$
and the input statistics $\rho[\b h]$. Here we denote functionals
by angular brackets and vectors of neuron indices by bold-font symbols.

\begin{figure}
\begin{centering}
\includegraphics[width=3.4in]{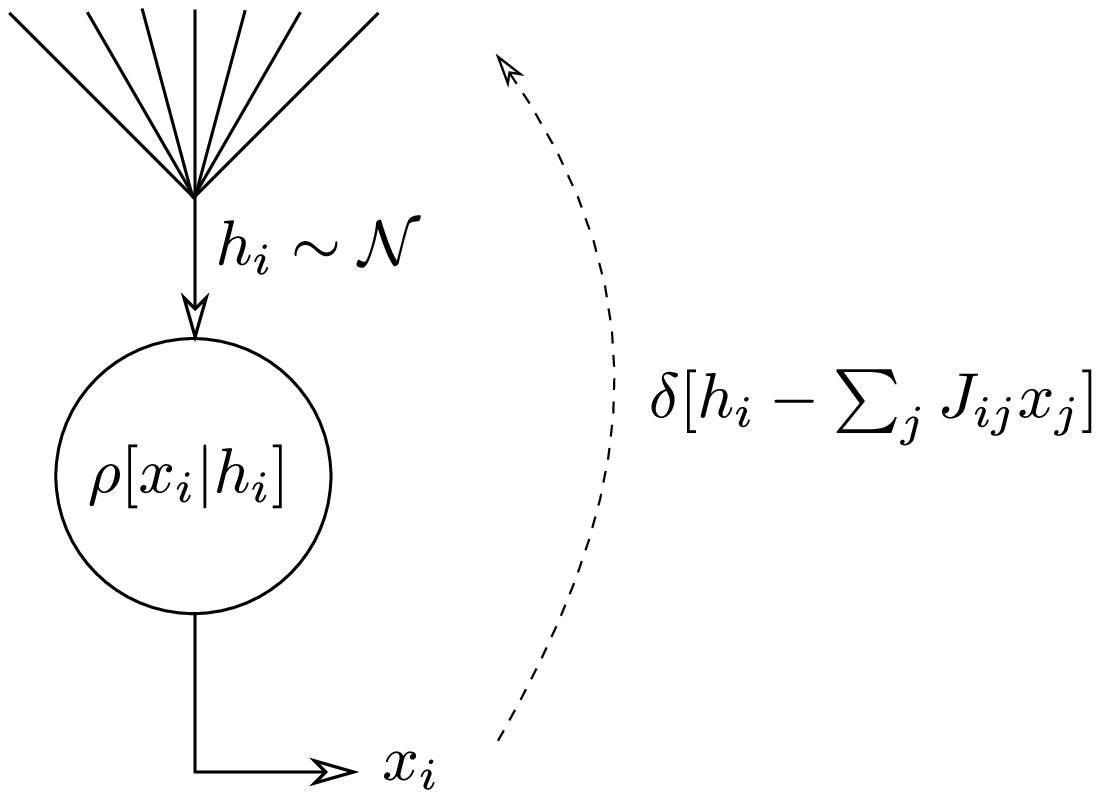}
\par\end{centering}
\centering{}\caption{\textbf{Summary of model-independent field theory.} Conceptual idea
to split network into neuronal dynamics, described by conditional
probability $\rho[x_{i}|h_{i}]$ of neuronal output $x_{i}$ given
its input $h_{i}$, and the mapping of output to input by connectivity
$J_{ij}$. The connectivity average affects only the output-to-input
mapping and can thus be performed without specifying the neuron model.
The formal saddle-point approximation in auxiliary fields $\mathcal{R}(t)\simeq R(t)$
and $\mathcal{Q}(t,s)\simeq Q(t,s)$, eqs. \prettyref{eq:saddle_point_R-1}
and \prettyref{eq:saddle_point_Q-1}, amounts to a Gaussian approximation
of the input $h_{i}\sim\protect\N(R,Q)$.\label{fig:concepts_MIFT_and_bin-rate_mapping}}
\end{figure}

Any observable $O$ of a neuronal network can be expressed as a functional
of the inputs $\boldsymbol{h}$, which have the advantage of being
closer to a Gaussian distribution than $\b x$, due to the convergence
of many outputs on one input. Because we do not know the disorder
realization (e.g. of the connectivity) in detail, but at most its
statistics, we can access only quenched disorder-averaged quantities
like
\[
\left\langle O\left[\bh\right]\right\rangle _{\b J,\bh}\coloneqq\int{\cal D}\bh\,\left\langle \rho\left[\bh\right]\left(\bJ\right)\right\rangle _{\bJ}\,O\left[\bh\right].
\]
The description of the network dynamics is self-consistently closed
by using a delta distribution $\rho[\bh]=\delta[\b h-\b J\b x]$ to
enforce that the input to each neuron is composed of a sum of outputs
weighted by the synaptic connectivity $\bJ$. The idea of splitting
the system into a neuron and a coupling model is illustrated in \prettyref{fig:concepts_MIFT_and_bin-rate_mapping}.
Note that \prettyref{eq:decomp_input_output} is not a circular definition
because $\rho[\boldsymbol{x}|\boldsymbol{h}]$ is a causal functional
and $\rho[\bh]=\delta[\b h-\b J\b x]$ couples only equal time points
of $\boldsymbol{h}$ and $\boldsymbol{x}$, so that the concatenation
in \prettyref{eq:decomp_input_output} can be understood as a spiral
moving forward in time (see also \prettyref{app:Model-independent-mean-field-the_appendix}).

Using the Fourier representation of $\rho[\bh]$ we obtain, at the
expense of introducing the response fields $\hat{\b h}$, the disorder-averaged
input statistics
\begin{align*}
\av{\rho[\bh](J)}{\b J}= & \left\langle \int\D\bx\,\rho[\bx,\bh]\right\rangle _{\bJ}\\
= & \int\D\bhh\,\exp\left(\bhh^{\T}\bh\right)\\
 & \times\int\D\bx\,\left\langle \exp\left(-\bhh^{\T}\bJ\bx\right)\right\rangle _{\bJ}\rho[\bx|\bh],
\end{align*}
obtained by marginalizing over $\bx$. The connectivity average acts
only on the interaction term, which now has the form of a moment-generating
function of $\b J$. In its cumulant expansion, intensive parameters
of the system are the first and second cumulant $\bar{g}/N$ and $g^{2}/N$,
respectively. Higher cumulants would also be suppressed if one assumes
the commonly chosen scaling $\propto N^{-\frac{1}{2}}$ of synaptic
weights \citep{Vreeswijk96,VanVreeswijk98_1321}.

The cumulant expansion suggests to do a Hubbard-Stratonovich transformation
with the auxiliary fields $\mathcal{R}(t):=\frac{\bar{g}}{N}\sum_{i}x_{i}(t)$
and $\mathcal{Q}(t,s):=\frac{g^{2}}{N}\sum_{i}x_{i}(t)x_{i}(s)$,
as outlined in \prettyref{app:Model-independent-mean-field-the_appendix},
so that a saddle-point approximation gives self-consistency relations
for the mean inputs and mean time-lagged autocorrelations, a dynamical
mean-field theory (DMFT)
\begin{align}
R(t) & =\bar{g}\av{x(t)}{\Omega(R,Q)},\label{eq:saddle_point_R-1}\\
\Qs\left(t,s\right) & =g^{2}\left\langle x(t)x(s)\right\rangle _{\Omega(R,Q)},\label{eq:saddle_point_Q-1}
\end{align}
where the average $\langle\ldots\rangle_{\Omega(R,Q)}$ is defined
in \prettyref{eq:def_av_Omega} of \prettyref{app:Model-independent-mean-field-the_appendix}
as an average over $x\sim\langle\rho[x|h]\rangle_{h}$ and $h$ is
a Gaussian process $h\sim\mathcal{N}\left(R,Q\right)$. We may think
of $x(t)$ as the representative neuron of a homogeneous population,
because all neurons with statistically identical connectivity and
properties are identical after the disorder average. On the intuitive
level, DMFT corresponds to modeling the inputs of all neurons as independent
Gaussian processes $h\sim\mathcal{N}\left(R,\Qs\right)$. 

Thus, one obtains the DMFT using only the output-to-input relation
given by the disordered connectivity while staying agnostic of the
neuron model. To instantiate the approximation for the binary model
studied here, we must now provide knowledge about the input-to-output
relation $\rho[\boldsymbol{x}|\boldsymbol{h}]$.

\paragraph*{Binary neuron model}

We consider the binary neuron model, or kinetic Ising model, with
state $x_{i}\in\{-1,1\}$ \citep{Glauber63_294,Ginzburg94}. The states
of all neurons are updated asynchronously by independent Poisson processes
with rate $\tau^{-1}$ and an activation probability function $\Tbi:\,\R\to[0,1]$.
It is clear that the form of $\rho[\boldsymbol{x}|\boldsymbol{h}]$
depends on the realization of the update times, which constitute a
source of noise, or temporal stochasticity. The update sequence may
be thought of as another type of disorder in the sense that it breaks
the homogeneity of the time axis by selecting a set of time points
where the neuronal state can change. As with the random connectivity,
one may study the behavior of the system averaged over this disorder.
In this case, the probability of finding a neuron active at time $t$
\begin{equation}
p[x_{i}(t)=1|h_{i}]=\int_{-\infty}^{t}\,\frac{dt^{\prime}}{\tau}e^{-\frac{t-t^{\prime}}{\tau}}\,\mathrm{T}_{p}\left(h_{i}\left(t^{\prime}\right)\right)\label{eq:Time_evolution_p1}
\end{equation}
is given by the probability $\mathrm{T}_{p}\left(h_{i}\left(t^{\prime}\right)\right)$
to be activated at any prior update time point $t^{\prime}$ and the
survivor function $e^{-\frac{t-t^{\prime}}{\tau}}$ \citep{Cox62},
the probability that no further update happened since. While this
knowledge is far from knowing the complete probability functional
$\rho[\boldsymbol{x}|\boldsymbol{h}]$ across its infinite time dimension,
the information about this single time slice is sufficient to plug
into \prettyref{eq:saddle_point_R-1} and obtain, after taking a time
derivative, the mean-field equation
\begin{equation}
\tau\frac{d}{dt}R(t)+R(t)=\bar{g}\left\langle \T\left(h\right)\right\rangle _{h\sim{\cal N}\left(R(t),\Qs(t,t)\right)},\label{eq:ODE_mean}
\end{equation}
where
\begin{align}
\mathrm{T}(h) & =2\Tbi(h)-1.\label{eq:relation_Tp_T}
\end{align}
Details are provided in \prettyref{app:Deriv_MFeq_binary}.

In \eqref{eq:ODE_mean} only equal-time autocorrelations $Q(t,t)$
appear, because the dynamics is a Markov process; its evolution at
time $t$ depends only on the statistics at this very time point,
not on the prior history. Closing the equation is thus simple for
binary neurons, because, by $x_{i}\in\{-1,1\}$, their autocorrelation
is always $1$, so that $\Qs\left(t,t\right)=\frac{g^{2}}{N}\sum_{i}\left\langle x_{i}\left(t\right)x_{i}\left(t\right)\right\rangle =g^{2}$
when cross-correlations are negligible (see \prettyref{app:Model-independent-mean-field-the_appendix}).

To compute $Q(t,t+\Delta t)$ for binary neurons, we need more information
about $\rho[\boldsymbol{x}|\boldsymbol{h}]$, namely the joint probability
distribution over two time slices for a neuron:
\begin{equation}
\rho\left[x\left(t\right),x\left(s\right)|h\right]=\underbrace{\rho\left[x\left(t\right)|x\left(s\right),h\right]}_{\overset{t\searrow s}{\longrightarrow}\delta_{x\left(t\right),x\left(s\right)}}\rho\left[x\left(s\right)|h\right].\label{eq:prob_distr_t_s_with_cond}
\end{equation}
To construct $\rho\left[x\left(t\right)|x\left(s\right),h\right]$
for binary neurons, the basic idea is to iterate the $2\times2$ states
a neuron can assume at the points in time $s$ and $t$ and consider
all possible evolutions that match the respective initial and final
condition. From such a consideration, we derive $Q(\Delta t)=Q(t,t+\Delta t)$
for stationary dynamics in \prettyref{app:Derivation-of-autocorr-in-binary-1}
by again taking a time derivative of the saddle-point equation \prettyref{eq:saddle_point_Q-1},
yielding
\begin{align}
 & \tau\frac{d}{d\Delta t}Q\left(\Delta t\right)+Q\left(\Delta t\right)\label{eq:IDE_Q_stationary}\\
= & g^{2}\int_{0}^{\infty}\frac{dt^{\prime}}{\tau}e^{-\frac{t^{\prime}}{\tau}}\left\langle \T\left(h\right)\T\left(h^{\prime}\right)\right\rangle _{\left(h,h^{\prime}\right)\sim\mathcal{N}_{R,Q(0),Q(\Delta t+t^{\prime})}}.\nonumber 
\end{align}
This equation is the analogon of the integral equation (5.17) of \citet{VanVreeswijk98_1321}.
The advantage of the form \prettyref{eq:IDE_Q_stationary} compared
to the classical result is, as detailed in \prettyref{app:Derivation-of-autocorr-in-binary-1},
that by differentiating once more with respect to $\Delta t$ and
then using Price's theorem \citep{PapoulisProb4th}, it can be cast
into a Newtonian form

\begin{align}
\tau^{2}\ddot{Q}\left(\Delta t\right)= & -V_{R,Q(0)}^{\prime}\left(Q\left(\Delta t\right)\right),\label{eq:Eom_input_autocorr}\\
V_{R,Q(0)}\left(Q\right):= & -\frac{1}{2}Q^{2}+g^{2}\left\langle {\cal T}\left(h\right){\cal T}\left(h^{\prime}\right)\right\rangle _{\left(h,h^{\prime}\right)\sim\mathcal{N}_{R,Q(0),Q}},\label{eq:Eom_input_autocorr-Potential}
\end{align}
where $\mathcal{T}$ is a primitive of $\T$, which is $\partial_{h}\mathcal{T}(h)=\T(h),$
and $\mathcal{N}_{R,Q(0),Q}$ is the bivariate Gaussian with stationary
mean $R$ and covariance matrix $\left(\begin{array}{cc}
Q(0) & Q\\
Q & Q(0)
\end{array}\right)$. We exploit this result in \prettyref{sec:Same_stat_Different-mechanisms-of-chaos}
to construct rate models with exactly the same DMFT solution as a
binary network.

\subsection{Binary networks are always chaotic in the thermodynamic limit\label{sec:Binary-networks-always-chaotic-in-thermlim}}

In the setting of reservoir computing (\prettyref{fig:Transient-chaotic-dimensionality-summary}),
a particularly important measure for the classification performance
of a network is how the distance between two different dynamical states,
each caused by one stimulus, evolves over time. Tracking the evolution
of initially small differences between the states amounts to the characterization
of chaos \citep{Sompolinsky88_259,Bertschinger04_1413,Sussillo09_544,Toyoizumi11_051908,Kadmon15_041030,Schuecker18_041029}.
We assess chaos by studying the time evolution of two systems with
infinitesimally different initial conditions but identical connectivity
and identical realization of stochasticity, thus the same sequences
of update time points. Technically, this approach amounts to a replica
calculation, where one studies the network-averaged correlation between
the states of the two systems over time, an approach pioneered by
\citet{Derrida86_45}. Here we do not use the classical annealed approximation
of this original work, where the connectivity is redrawn in every
time step, but compute the full quenched averages, where the connectivity
is constant in time. The calculation leads to a dynamic mean-field
theory for the correlation between replicas.

In \prettyref{app:Replica-calculation-for-chaos-in-binary}, by an
approach analogous to the derivation of the ordinary differential
equation (ODE) for the autocorrelation \prettyref{eq:Eom_input_autocorr-Potential},
we obtain the evolution of the cross-replica equal-time correlation
$Q^{(12)}(t)$ in the binary network as
\begin{align}
 & \tau\frac{d}{dt}Q^{(12)}\left(t\right)\label{eq:full_ODE_crossreplicacorr_binary}\\
= & -Q^{(12)}\left(t\right)\nonumber \\
 & +g^{2}\,\left(1-\left\langle \left|\T\left(h^{(1)}\right)-\T\left(h^{(2)}\right)\right|\right\rangle _{h^{(1)},h^{(2)}}\right).\nonumber 
\end{align}
Here $(h^{(1)},h^{(2)})\sim\mathcal{N}(\b R,\b Q)$ is a measure of
a pair of Gaussian processes with means $\langle h^{(\alpha)}\rangle=R^{(\alpha)}=\left\langle \T\left(h\right)\right\rangle _{h\sim{\cal N}\left(R^{(\alpha)},g^{2}\right)}$,
the stationary solution of \prettyref{eq:ODE_mean} and covariance
matrix $\cum{h^{(\alpha)}h^{(\beta)}}{}=Q^{(\alpha\beta)}(t)$,
whose diagonal elements are each $Q^{(\alpha\alpha)}(t)=g^{2}$.

Since the two replicas are nearly perfectly correlated in the beginning,
we know that the correlation between a neuron and its ``copy'' in
the other replica is given by the autocorrelation at first, motivating
the ansatz
\[
Q^{(12)}\left(t\right)=Q(0)-\epsilon(t),\quad\epsilon(t)\geq0.
\]
 As shown in \prettyref{app:Replica-calculation-for-chaos-in-binary},
an expansion for small $\epsilon$ leads to the approximate equation
governing the evolution of $\epsilon(t)$
\begin{equation}
\tau\frac{d}{dt}\epsilon(t)=-\epsilon(t)+\frac{2}{\sqrt{\pi}}g^{2}\left\langle \T^{\prime}\left(h\right)\right\rangle _{h\sim\mathcal{N}\left(R,Q(0)\right)}\sqrt{\epsilon(t)},\label{eq:ODE_deviation_fully_correlated}
\end{equation}
which generalizes the result of \citet{VanVreeswijk98_1321} to arbitrary
activation functions. As was their conclusion for neurons with hard
threshold, we see from \prettyref{eq:ODE_deviation_fully_correlated}
that for any activation function with average positive slope and independent
of the parameters, the positive term $\propto\sqrt{\epsilon}$ is
always larger than the negative linear term for small $\epsilon$;
so an initial deviation between the replicas will grow, indicating
chaotic dynamics. Since the calculation becomes exact in the thermodynamic
limit, the conclusion is that infinitely large binary networks are
always chaotic, with formally infinite maximum Lyapunov exponent since
the slope of the right-hand side of \prettyref{eq:ODE_deviation_fully_correlated}
at $\epsilon=0^{+}$ is infinite, leading to an initial growth of
$\epsilon$ that is faster than exponential. More specifically, $\epsilon(t)\sim t^{2}$
for $\epsilon\ll1$, meaning that $\epsilon(0)=0+$ grows to a finite
value in finite time, as opposed to an exponential function. See \prettyref{app:Initial-growth-of-perturbations-in-binary}
for additional details. Since the slope of the activation function
appears only averaged over the input distribution, there is no qualitative
difference between different activation functions. In particular,
going from a stochastic activation function to the deterministic Heaviside
limit changes only the second term in \prettyref{eq:ODE_deviation_fully_correlated}
by a finite factor and thus does not qualitatively alter the chaotic
behavior. This result can also be understood by noting that for the
stochastic activation function, the function value at each update
is compared to a random number to decide the activity state. The comparison
is just like using a Heaviside function but with randomly drawn threshold
at each update.

\subsection{Transition to chaos in finite-size binary networks\label{sec:Transition-to-chaos-binary-finite-size}}

In contrast to the theoretical prediction, simulations of binary networks
in fact show parameter regimes with regular dynamics (\prettyref{fig:simulation-theory_chaos_comparison}).
Since the theory is only exact in the limit of infinite network size,
this behaviour suggests a finite size effect. But the result of the
replica calculation \prettyref{eq:ODE_deviation_fully_correlated}
does not rely on carrying out the $N\to\infty$ limit. Rather it is
expected to be a good approximation for finite, yet large networks
$N\gg1$. How can the theory be reconciled with the simulation?

\begin{figure}
\begin{centering}
\includegraphics{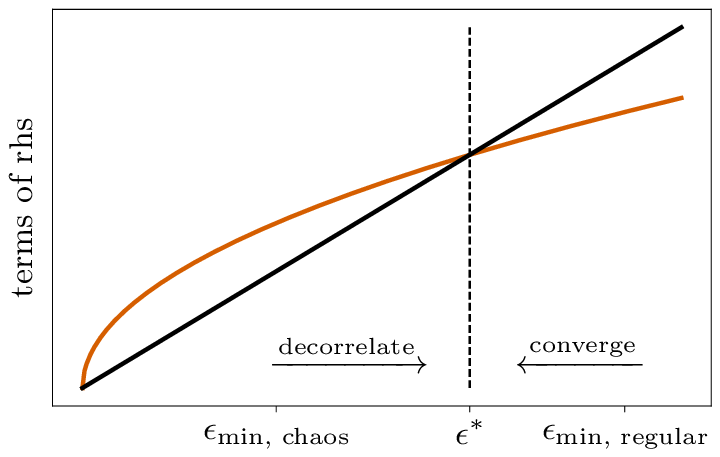}
\par\end{centering}
\caption{\textbf{Fixed-point average distance between replicas implies chaotic
subspace.} The square-root term $\propto\sqrt{\epsilon}$ (dark orange
curve) and the linear term $\propto\epsilon$ in \prettyref{eq:ODE_deviation_fully_correlated}
intersect and produce a fixed point $\epsilon^{\ast}$ for the covariance
$Q(0)-\epsilon$ between the two replicas. The resulting average
Hamming distance $d^{\ast}$ between states in the two copies of the
system is given by $\epsilon^{\ast}$ and \prettyref{eq:relation_hammingdist_epsilon}
as $d^{\ast}=N\epsilon^{\ast}/g^{2}$. Depending on whether $\epsilon_{\text{min}}$,
the minimum decorrelation due to a single flipped spin, is smaller
or larger than $\epsilon^{\ast}$, the replicas will either decorrelate,
or they will converge and forget the perturbation.\label{fig:decorr_stabilize_visualization}}
\end{figure}

\begin{figure*}
\begin{centering}
\includegraphics{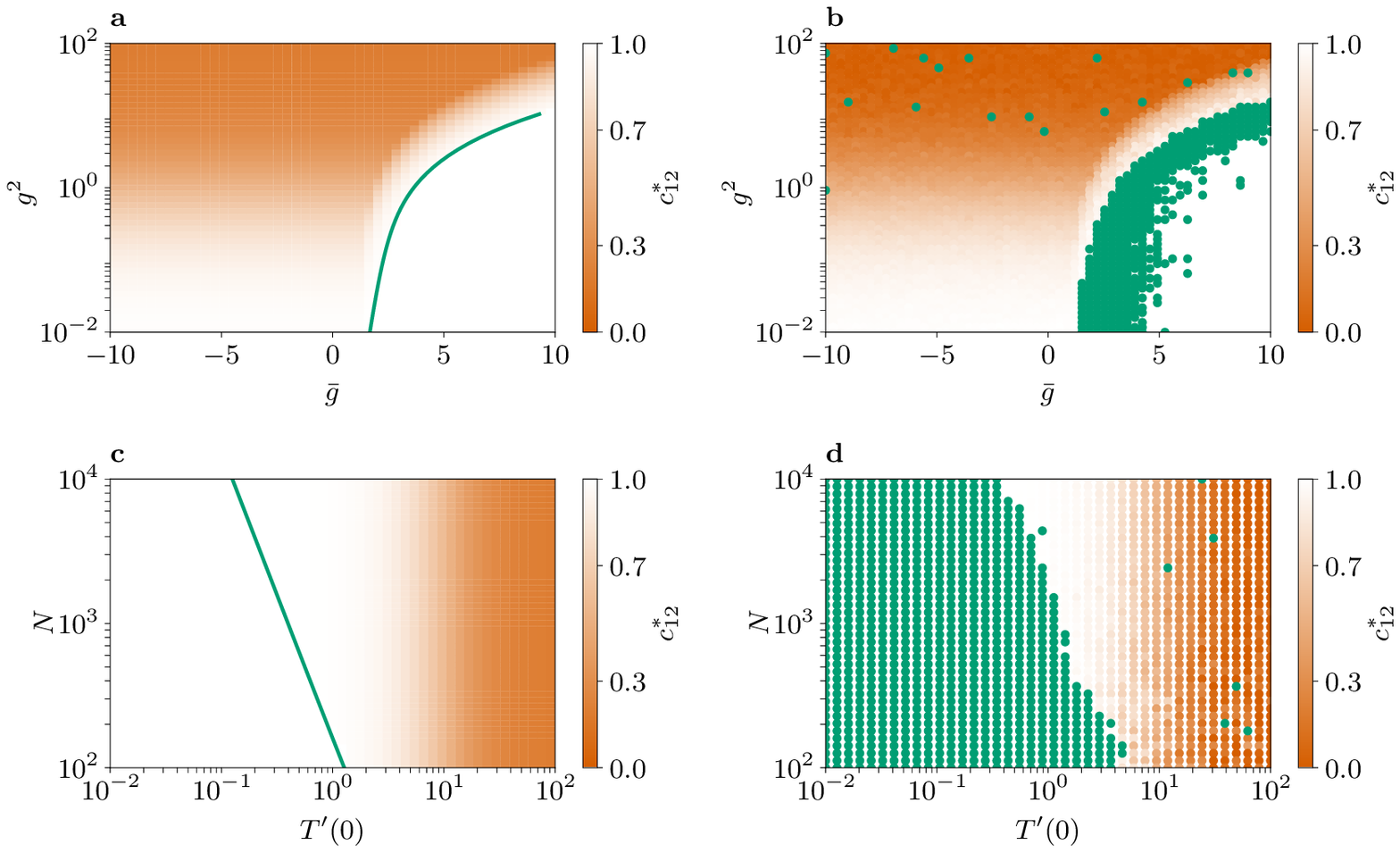}
\par\end{centering}
\caption{\textbf{Chaos transition and residual correlation in theory (a,c)
and simulations (b,d).} \textbf{a} Theoretical prediction of chaos
transition (green line, eq. \prettyref{eq:chaos_criterion_finite_size})
and residual correlation coefficient between replicas (dark orange
shading, $c_{12}^{\ast}\,=1-Q(0)^{-1}\epsilon^{\ast}$ and \prettyref{eq:residual_epsilon})
for varying mean $\bar{g}$ and variance $g^{2}$ of the connectivity.
Other parameters are $N=5000$, $\protect\T=\tanh$, $\tau=10\,\text{ms}$.
\textbf{b }As in (a) but each pixel is colored dark orange (chaotic)
or green (stable) according to a network simulation. Two identical
networks are evolved with identical random numbers, only one being
perturbed by flipping four spins, and after $T_{\text{sim}}=2500\,\text{ms}$
the correlation between the state vectors is computed. The few scattered
green dots in the chaotic regime are algorithmic artifacts where the
perturbation is unsuccessful, see \prettyref{app:Description-of-simulations}
for the perturbation method. \textbf{c} Theoretical prediction for
varying network size $N$ and slope $T'(0)$ of the activation function
$T(h)=\tanh(T'(0)h)$. Other parameters are $\bar{g}=0$, $g^{2}=0.01$
and $\tau=10\,\text{ms}$. \textbf{d} As in (c) but from network simulations,
with procedure as described for (b). \label{fig:simulation-theory_chaos_comparison}}
\end{figure*}

First, while the square-root term in \prettyref{eq:ODE_deviation_fully_correlated}
is always larger for sufficiently small $\epsilon$, there also exists
a point $\partial_{t}\epsilon^{\ast}\overset{!}{=}0$ where this
relationship reverses and the linear term starts to dominate: the
point where the right-hand side of \prettyref{eq:ODE_deviation_fully_correlated}
vanishes,
\begin{align}
\sqrt{\epsilon^{\ast}} & =\frac{2}{\sqrt{\pi}}g^{2}\av{\T^{\prime}(h)}h.\label{eq:residual_epsilon}
\end{align}
This point corresponds to a stable average distance between (partly)
decorrelated trajectories, as illustrated in \prettyref{fig:decorr_stabilize_visualization}.

Second, in a finite network of $N$ binary neurons, an infinitesimal
perturbation cannot be realized, since the smallest possible perturbation
is to flip a single spin at index $i_{\text{flip}}$. A single flip
implies for the minimally perturbed cross-replica correlation \prettyref{eq:saddle_replica}
\begin{align*}
Q(0)-\epsilon_{\text{min}} & =g^{2}\av{x^{(1)}x^{(2)}}{}\\
 & =g^{2}\frac{1}{N}\Big(\sum_{i=1}^{N}\left(x_{i}^{(1)}x_{i}^{(1)}\right)-\underbrace{x_{i_{\text{flip}}}^{(1)}x_{i_{\text{flip}}}^{(1)}}_{=1}+\underbrace{x_{i_{\text{flip}}}^{(1)}x_{i_{\text{flip}}}^{(2)}}_{=-1}\Big)\\
 & =Q(0)-2\frac{g^{2}}{N},
\end{align*}
so that 
\begin{equation}
\epsilon_{\text{min}}=2\frac{g^{2}}{N}.\label{eq:epsilon_min}
\end{equation}

Therefore, if $\epsilon_{\text{min}}>\epsilon^{\ast}$ the replicas
will tend toward more correlation. But as the only possible step below
$\epsilon_{\text{min}}$ is having zero different spins and thus perfect
correlation, the initial difference should tend to be completely forgotten,
resulting in regular dynamics. On the other hand, if $\epsilon_{\text{min}}<\epsilon^{\ast}$
an increase of the initial difference is possible.

Thus the chaos transition criterion in the finite binary network is
$\epsilon_{\text{min}}\overset{!}{\leq}\epsilon^{\ast}$ resulting
in
\begin{align}
1 & \apprle\sqrt{\frac{2}{\pi}}g\av{\T^{\prime}(h)}h\sqrt{N}.\label{eq:chaos_criterion_finite_size}
\end{align}
Because of the scaling with $\sqrt{N}$, it is clear that networks
with thousands or even only hundreds of neurons are only nonchaotic
if the connectivity is very weak, $g\apprle N^{-\frac{1}{2}},$ or
the dynamics is saturated (which gives a small $\av{T^{\prime}(h)}h$).
Also, $N\to\infty$ clearly recovers the limit of strictly chaotic
dynamics. For the special case of a Heaviside activation function
and vanishing mean connectivity $\bar{g}=0$, the network is always
chaotic, since $\av{H^{\prime}(h)}{h\sim\mathcal{N}(0,g^{2})}=\sqrt{2/(\pi g^{2})}$
results in $\pi/2\le\sqrt{N}$, which is certainly true for typical
network sizes.

The predicted transition and the residual correlation $\epsilon^{\ast}$
fit those observed in simulations quite well (\prettyref{fig:simulation-theory_chaos_comparison}).
The dependence of the transition on the positive mean connectivity
$\bar{g}$ in the upper panels arises because the network settles
in a state with nonzero mean activity that depends on $\bar{g}$;
it selects one of the two degenerate states in this bistable, ``ferromagnetic'',
regime. The symmetry with respect to a global sign flip of the activity
is spontaneously broken. In this state neurons show a very small average
slope $\av{T^{\prime}(H)}H$, thus shifting the point of transition
to larger $g$ with increasing $\bar{g}$. The predicted residual
correlation is independent of $N$ (\prettyref{fig:simulation-theory_chaos_comparison}c,
compare \prettyref{eq:residual_epsilon}), while the chaos transition
depends on $N$ (compare \prettyref{eq:chaos_criterion_finite_size}).

We obtain the same criterion \prettyref{eq:chaos_criterion_finite_size}
through a less general, but more intuitive perspective by analyzing
the probability that, given a single-spin difference, the difference
in inputs is such that during the next updates, another neuron will
also be updated to a ``wrong'' state (see \prettyref{app:Fluxtubes-in-binary-networks}).
This view provides an expression for the average rate of decorrelation
caused by an initial single spin flip. Requiring this rate to be unity,
we obtain the same chaos transition criterion as \prettyref{eq:chaos_criterion_finite_size}.
The approach is inspired from and very similar to calculating the
divergence rate of flux tubes in spiking networks \citep{Touzel19}.
Such flux tubes are stable local environments of a phase-space trajectory,
while the network is globally unstable. Thus, the phase space can
be partitioned into tubes which diverge from each other, while perturbations
within a tube decay. Indeed, the binary network has relatively trivial
flux tubes in the input phase space given by those regions that result
in the same updated state.

Note that the chaos transition shown in \prettyref{fig:simulation-theory_chaos_comparison}\textbf{d}
in simulations happens at slightly larger slopes $T^{\prime}(0)$
than predicted by \prettyref{eq:chaos_criterion_finite_size}. Considering
the cascade of spin flips evoked by the initial perturbation provides
an explanation: If the average proliferation rate of spin flips per
time constant is only slightly above one, the cascade triggered by
a single flipped spin still has a large probability of dying out.

\textcolor{brown}{}

\subsection{Dynamics in binary networks is governed by a chaotic submanifold\label{sec:dynamics-submanifold}}

The chaos in binary networks found in \prettyref{sec:Binary-networks-always-chaotic-in-thermlim}
causes nearby trajectories to diverge at first. Because of the fixed
point value $\epsilon_{\ast}$ of the residual correlation found in
\prettyref{sec:Transition-to-chaos-binary-finite-size}, however,
the network states of the replicas do not decorrelate completely.
Instead, any pair of trajectories has an average maximal distance
determined by $\epsilon^{\ast}$. This limited distance is a result
of the two trajectories evolving by the same network connectivity
and update sequence. Also trajectories that are very far apart will
converge to this residual correlation. The fixed-point distance $\epsilon^{\ast}$
is thus a representative of the average distance between any two trajectories
in the long-time limit. The corresponding Hamming distance $H^{(12)}=\frac{1}{4}||\mathbf{x}^{(1)}-\mathbf{x}^{(2)}||^{2}$,
that is, the number of different spins between a pair of binary states,
is given by
\begin{align}
H^{(12)}(t) & =\frac{N\,\epsilon(t)}{2g^{2}},\label{eq:relation_hammingdist_epsilon}
\end{align}
where we use $Q^{(12)}=\frac{g^{2}}{N}\big(N-2H^{(12)}\big)$ and
the prefactor of $H^{(12)}$ arises because every flipped spin causes
a decrease by $2$ (from $+1$ to $-1$). Even though the $H^{(12)}$
spin flips distinguishing two trajectories can in principle be distributed
across any of the $N$ neurons, the subspace spanned by the set of
possible trajectories has an approximate dimensionality of 
\begin{align}
d(t) & \simeq2\,H^{(12)}(t)\simeq\frac{N}{g^{2}}\epsilon(t).\label{eq:d_2_H}
\end{align}
This relation can be understood by considering two independently drawn
binary random vectors of dimension $d$ that have, on average, the
distance $H^{(12)}=d/2$, because the average distance between any
pair of spins that take the values $x^{(1)},x^{(2)}\in\{-1,1\}$ with
equal probabilities is $\langle(x^{(1)}-x^{(2)})^{2}\rangle/4=1/2$.
Therefore, in the following, we quantify dimensionality via the Hamming
distance $H^{(12)}$ using \prettyref{eq:d_2_H}.

Thus, if $\epsilon_{\text{min}}>\epsilon^{\ast}$, then $H^{(12)}(\infty)<1$
and the set of long-term trajectories contains only a single trajectory,
thus constituting a limit cycle (although the return time is astronomically
large \citep{Hwang19_053402}). Irrespective of the initial state,
the network is attracted to a stereotypical trajectory; the dynamics
is regular. This situation arises for very weak coupling.

If $\epsilon_{\text{min}}<\epsilon^{\ast}$, then $H^{(12)}(\infty)>1$
and there are many trajectories that constitute the attractive subspace.
 The evolution within the space is chaotic, because for any pair
of states with an initial distance $\epsilon<\epsilon^{\ast}$ the
distance increases; thus small differences are amplified. A set of
trajectories that initially spans a low-dimensional subspace is thus
expanded into a higher-dimensional space. For long times, however,
any two states differ in only typically $d(\infty)/2$ of their neurons.
This limiting dimensionality grows proportional to $Ng^{2}$ as seen
by inserting \prettyref{eq:residual_epsilon} into \prettyref{eq:relation_hammingdist_epsilon}
\begin{align}
d^{\ast} & =Ng^{2}\,\left(\frac{2}{\sqrt{\pi}}\av{\T^{\prime}(h)}h\right)^{2}.\label{eq:d_star}
\end{align}
The time evolution when starting with a set of trajectories with dimensionality
$d(0)$ is given by
\begin{align}
d(t) & =\big(\sqrt{d_{\ast}}-\big(\sqrt{d_{\ast}}-\sqrt{d(0)}\big)\,e^{-\frac{t}{2\tau}}\big)^{2},\label{eq:d_t_exact}
\end{align}
obtained by integrating \prettyref{eq:ODE_deviation_fully_correlated}
(see \prettyref{app:Initial-growth-of-perturbations-in-binary}),
and shown in \prettyref{fig:Discrete-chaos-versus-rate-chaos}\textbf{a}.
This explicit solution shows that the expansion happens very quickly
on a timescale of $2\tau$, where $\tau$ is the average time to have
one update per neuron, and then converges to the residual value $d^{\ast}$
\prettyref{eq:d_star} for long times. This exclusive dependence on
$\tau$ can intuitively be understood from the right-hand side of
\prettyref{eq:chaos_criterion_finite_size}, which can be interpreted
as the average number of flips $n_{\text{spawns}}$ caused by an initial
spin flip within one time constant (as obtained in \prettyref{app:Fluxtubes-in-binary-networks}).
Hence, using \prettyref{eq:residual_epsilon}, one has 
\begin{align}
\frac{d^{\ast}}{2} & =n_{\text{spawns}}^{2},\label{eq:decorrelation_speed-1}
\end{align}
so that after two time constants have passed, the residual correlation
would be reached if the functional form of initial decorrelation would
be extrapolated to later times, neglecting saturating terms (see \prettyref{fig:Discrete-chaos-versus-rate-chaos}\textbf{a}
and \prettyref{app:Initial-growth-of-perturbations-in-binary}).
But because the residual correlation limits the spread of the cascade
of flips, in a similar way as the population size limits the growth
of an epidemic \citep{Turner1976_367}, the growth slows down and
asymptotically approaches the residual correlation.

\subsection{Same statistics, different chaotic dynamics in continuously and discretely
coupled networks\label{sec:Same_stat_Different-mechanisms-of-chaos}}

Having quantified how binary networks with discrete signaling separate
different states, as required to understand classification in reservoir
computing (\prettyref{fig:Transient-chaotic-dimensionality-summary}),
we now turn to the well-established alternative of units with continuous-valued
activity and signaling, commonly referred to as ``rate models''
and typically employed in artificial neuronal networks. Concretely,
we consider the coupled set of stochastic differential equations \citep{Sompolinsky88_259,Schuecker18_041029}
\begin{align}
\tau\partial_{t}\b h= & -\b h+\b J\,\mathrm{T}\left(\b h\right)+\sqrt{\tau}\b{\xi}\label{eq:rate_SCS_form}
\end{align}
with the activation function $\mathrm{T}:\,\R\to[-1,1]$ given by
\eqref{eq:relation_Tp_T}, timescale $\tau$ and a white noise process
$\xi$ with $\av{\xi_{i}(t)\xi_{j}(s)}{}=\sigma_{\xi}^{2}\,\delta(t-s)\delta_{ij}$.
Chaos in such networks has been intensely studied \citep{Sompolinsky88_259,Kadmon15_041030,Schuecker18_041029}.

\begin{figure}
\centering{}\includegraphics[width=1\columnwidth]{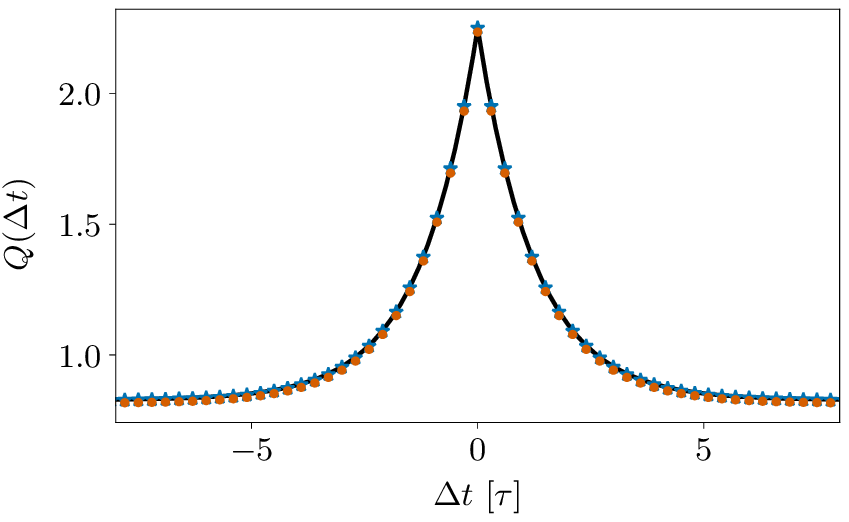}\caption{\textbf{Matched second order statistics in binary and rate networks.}
\textbf{a} Autocorrelation functions in simulations of binary (dark
orange dots) and rate networks (blue stars). Theoretical curve (black)
given by the solution of \prettyref{eq:Eom_input_autocorr-Potential}.
Noise amplitude of the rate network chosen by \prettyref{eq:Initial_condition_Q_dot_binary}
to obtain matched statistics. Other parameters: $N=5000$, $g=1.5$,
$\bar{g}=0$, $\tau=1\,\mathrm{ms}$ and $\protect\T(h)=\text{tanh}(h-\Theta)$
with $\Theta=1.173$ such that $\left\langle x\right\rangle _{\mathcal{N}(R,Q)}\approx-0.5$
according to the stationary solution of \prettyref{eq:ODE_mean} (see
\prettyref{app:Description-of-simulations} for details). We simulate
a single realization of the binary network for 5000 ms and average
over five realizations of the rate network running for 1000 ms each.
We average over neurons in both cases.\label{fig:rate_binary_autocorr_match}
}
\end{figure}

We show in \prettyref{app:Equivalence-of-mean-field-rate-binary}
that the model-independent field theory applied to this stochastic
rate model yields the same set of self-consistency equations for the
first- \prettyref{eq:ODE_mean} and second-order statistics \prettyref{eq:Eom_input_autocorr}
as the binary model; also the conditions on $Q_{\infty}\coloneqq\underset{\tau\rightarrow\infty}{\lim}Q\left(\tau\right)$
agree. In contrast to the binary network, however, where the initial
value $Q(0)=g^{2}$ is known, \eqref{eq:Eom_input_autocorr} must
be solved with an initial condition for the slope $\dot{Q}(0+)$.
This slope is determined by the variance $\sigma_{\xi}^{2}$ of the
noise in \eqref{eq:rate_SCS_form}. Demanding identical mean-field
solutions for the two neuron types, the variance of the noise follows
as (see \prettyref{app:App_Equivalence_binary_rate_matching_init})

\begin{align}
\sigma_{\xi}^{2} & =\frac{2}{\tau}\,\sqrt{2\left(V_{g^{2}}(Q_{\infty})-V_{g^{2}}(g^{2})\right)}.\label{eq:Initial_condition_Q_dot_binary}
\end{align}
Equation \prettyref{eq:Initial_condition_Q_dot_binary} tells us that,
given a pair of equivalent activation functions \prettyref{eq:relation_Tp_T},
moments of connectivity $\bar{g}$, $g^{2}$, and timescale $\tau$,
asynchronously updated binary networks are statistically equivalent
in DMFT approximation to rate networks with appropriately chosen Gaussian
white noise input. This result is confirmed in simulations by comparing
the autocorrelation functions averaged across many neurons in \prettyref{fig:rate_binary_autocorr_match}.
The good agreement between the autocorrelation that is averaged over
all neurons in a network with a single random realization of the coupling
matrix and the theoretical curves, which describe ensembles of networks
averaged over many realizations of the random couplings, moreover
shows that these quantities are self-averaging.

\subsubsection*{Condition for chaos in rate networks}

Having established their equivalence on the level of statistics, we
now compare the chaotic evolution of binary and rate networks. As
its binary counterpart, the rate neuron model can be studied in a
replica calculation in dynamical mean-field approximation, which yields
the equation of the cross-replica time-lagged covariance of the form
\citep{Schuecker18_041029,Kadmon15_041030}
\begin{align}
(\partial_{t}+1)(\partial_{s}+1)Q^{(12)}(t,s) & =g^{2}\,f_{\T}(Q_{0},Q^{(12)}),\label{eq:crossreplica_variance_evolution_rate-1}
\end{align}
with $f_{\T}(Q_{0},Q^{(12)})=\langle\T(x_{1})\T(x_{2})\rangle$ and
the average is taken with respect to $(x_{1},x_{2})\sim\mathcal{N}\Big(0,\big(\begin{array}{cc}
Q_{0} & Q^{(12)}\\
Q^{(12)} & Q_{0}
\end{array}\big)\Big)$. The approximation for small differences $Q^{(12)}=Q_{0}-\epsilon$,
to linear order in $\epsilon$, is
\begin{align}
(\partial_{t}+1)(\partial_{s}+1)\,\epsilon(t,s) & =g^{2}f_{\T'}(t,s)\,\epsilon(t,s),\label{eq:epsilon_evolution_rate-1}
\end{align}
which is solved by
\begin{align}
\epsilon(t,t) & =\epsilon(0,0)\,e^{\lambda_{\max}(g)\,t},\label{eq:Lyapunov}
\end{align}
where $\lambda_{\mathrm{max}}(g)$ is the largest Lyapunov exponent
that follows from an eigenvalue problem, see (\citet[their eqs. 10-11]{Sompolinsky88_259},
\citet[their eq. 17]{Schuecker18_041029}). The linear stability analysis
in \prettyref{eq:epsilon_evolution_rate-1} leads to the criterion
for the chaos transition \citep[their eq. 20]{Schuecker18_041029}
\begin{align}
g^{2}\,\langle T(h)T(h)\rangle_{h\sim\N(R^{\ast},Q_{0})}-Q_{0} & \geq0.\label{eq:Rate_chaos_cond_from_matched_binary}
\end{align}

\subsubsection*{No chaos in rate networks with matched statistics}

Applying criterion \eqref{eq:Rate_chaos_cond_from_matched_binary}
to a network of rate neurons with the noise matched to its binary
counterpart via \prettyref{eq:Initial_condition_Q_dot_binary}, we
obtain $Q_{0}=g^{2}$ by construction. Using that $\langle T(h)T(h)\rangle\leq1$
because of $|T|\le1$, we observe that the condition \prettyref{eq:Rate_chaos_cond_from_matched_binary}
cannot be fulfilled. The dynamics is therefore always in the regular
regime because the frozen noise of amplitude given by \prettyref{eq:Initial_condition_Q_dot_binary}
is so large that it drives the dynamics and suppresses chaos. Only
asymptotically the chaos transition is approached for an infinite
slope of the activation function, $T^{\prime}\to\infty$, or equivalently
$g^{2}\to\infty$.

Everything else being identical, the only difference between the
two models is the type of signals exchanged between units, either
being discrete or continuous. This demonstrates that chaos in binary
networks is intrinsically caused by the discrete signaling. Formally,
the difference between the two forms of signaling here shows up in
the effective noise $\xi$: In the rate network, the realization of
this noise is identical across the two replicas, because it represents
the random realizations of the discrete variables of the binary network
whose statistics we want to match. In the binary network, this noise
itself changes, because it is intrinsically generated by the discrete
switching dynamics, so that the realization is not external and frozen,
but depends on the microscopic state.

\subsection*{Qualitative differences of chaos between rate and binary networks}

In the following, we give up on matching the statistics between rate
and binary networks to discuss the qualitative differences of the
respective chaotic dynamics.

\subsubsection*{Residual correlation}

The first qualitative difference concerns the residual correlation
of the replicas. Since there is no term $\propto\sqrt{\epsilon}$
in \prettyref{eq:epsilon_evolution_rate-1}, there is no residual
correlation for small $\epsilon$. Furthermore, equation \prettyref{eq:crossreplica_variance_evolution_rate-1}
is also valid for small $Q^{(12)}$ and shows that the completely
decorrelated state $Q^{(12)}=0$ is always a fixed point: the expectation
value factorizes, and for any point symmetric $\T$ the right-hand
side vanishes. In the thermodynamic limit, the residual correlation
in rate networks is thus zero for any $g>1$. A network that has been
infinitesimally perturbed eventually has a state that is completely
uncorrelated to the unperturbed system.

Trajectories in rate networks of finite size, in fact, show very small
residual correlation closely beyond the edge of chaos $g^{2}\gtrsim1$.
However, as already noted by \citet{Sompolinsky88_259}, the transition
at $g=1$ is not completely sharp in finite-size networks \citep[see also][]{wainrib13_118101}.
For larger networks, however, the residual correlation approaches
zero; this is in contrast to binary networks, which in otherwise identical
settings have a finite residual correlation \prettyref{eq:d_star}
even in the large-$N$ limit.

This qualitative difference is shown in \prettyref{fig:Discrete-chaos-versus-rate-chaos}:
In binary networks the decorrelation between the original and the
perturbed system $d(t)/N=1-Q^{(12)}(t,t)/Q_{0}$ in the long-time
limit saturates below unity, on a level that depends on the coupling
$g$ by Eq. \prettyref{eq:d_star} and is bounded by $8/\pi^{2}\simeq0.81$
in the limit $g\to\infty$. The quantity $d(t)/N$ can also be interpreted
as the relative dimensionality of the explored space. In rate networks
with otherwise identical parameters, the relative decorrelation reaches
unity independent of $g$. The decorrelation in rate networks cannot
be interpreted in terms of dimensionality, however. Indeed, a recent
work demonstrates a structured chaotic attractor in such networks
\citep{Engelken20_02427}.

\subsubsection*{Transient of decorrelation}

\begin{figure}
\begin{centering}
\includegraphics{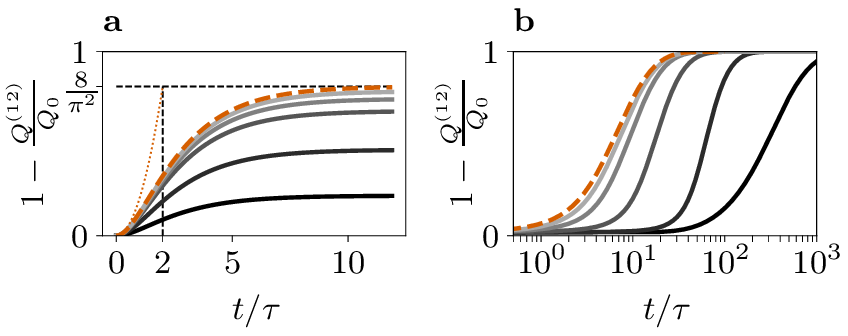}
\par\end{centering}
\caption{\textbf{Discrete chaos versus continuous chaos. a }Evolution of the
decorrelation between replicas in the binary network \prettyref{eq:d_t_exact}
with $T=\tanh$. Increasing coupling strength from dark to light gray,
$g=\{0.5,1,2,3,5\}$. Dark orange curves: limit $g\to\infty$ or Heaviside
$T$ (dashed); quadratic solution for nonsaturated growth derived
in \prettyref{app:Initial-growth-of-perturbations-in-binary} (dotted).
\textbf{b }Evolution of decorrelation between replicas in the rate
network \prettyref{eq:rate_SCS_form} obtained by numerical solution
of \prettyref{eq:epsilon_evolution_rate-1} (details in \prettyref{app:Initial-growth-of-perturbations-in-binary-1}).
Color code as in a, but using parameters $g=\{1.1,1.5,2.5,5.10\}$
and initial decorrelation of $2\%$ instead of $0\%$.\label{fig:Discrete-chaos-versus-rate-chaos}}
\end{figure}

The second qualitative difference concerns the transient of decorrelation.
The solution \prettyref{eq:d_t_exact} shows that the characteristic
timescale of decorrelation is $2\tau$, where $\tau$ is the average
interval between two state changes of a neuron, exposing that the
microscopic state drives the chaotic evolution. Decorrelation slows
down only mildly for weaker coupling $g$, as shown in \prettyref{fig:Discrete-chaos-versus-rate-chaos}a.
Moreover, it has a finite slope shortly after the infinitesimal perturbation
of the system, reflecting the infinite Lyapunov exponent.

In rate networks, the maximal Lyapunov exponent $\lambda_{\mathrm{max}}(g)$
is finite and depends continuously on the coupling $g$. Decorrelation
therefore starts with a vanishing speed for infinitesimal perturbations,
well described by the exponential behavior \prettyref{eq:Lyapunov},
as shown in \prettyref{fig:Discrete-chaos-versus-rate-chaos}b. The
time to reach a given level of decorrelation, moreover, strongly depends
on the coupling strength $g$, corresponding to a critical slowing-down
at the transition to chaos, which does not occur in binary networks.

\subsection{Computation by transient chaotic dimensionality expansion\label{sec:Computation-by-transient-dim-expansion}}

\begin{figure*}
\begin{centering}
\includegraphics{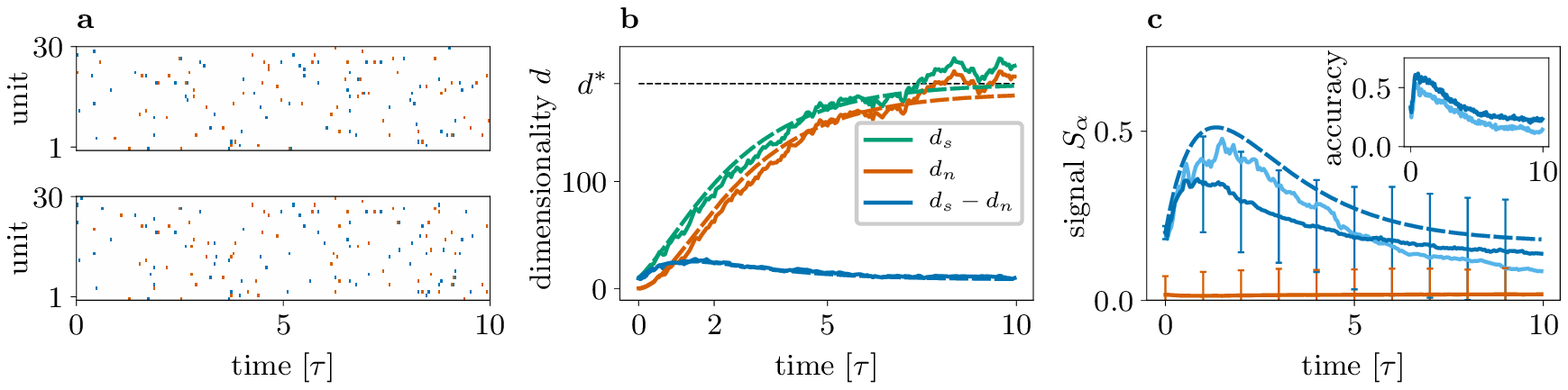}
\par\end{centering}

\caption{\textbf{Transient dimensionality expansion of stimulus representation
by a chaotic binary network.} \textbf{a }Time evolution of first $50$
neurons for two different patterns of initial states; up transitions
in red, down in blue. Initial state of first $L=10$ neurons set to
one of the $P=50$ fixed, random patterns. The initial state of the
remaining neurons is kept constant. \textbf{b }Evolution of signal
and noise subspace dimensionality. Dimensionality $d_{\mathrm{s}}(t)$
given by \prettyref{eq:d_2_H} explored by the network across different
patterns (green; solid curve: using averaged simulated distances across
all pairs of patterns; dashed curve: theory \prettyref{eq:d_t_exact}).
Dimensionality $d_{\mathrm{n}}(t)$ explored across different noisy
realizations of a pattern (dark orange; solid curve: using averaged
simulated distances across all pairs of $20$ realizations per pattern;
dashed curve: theory \prettyref{eq:d_t_exact}). Noisy realizations
of patterns have Gaussian noise with standard deviation $\sigma=0.3$
added to each of the $L$ entries of the initial pattern state. Difference
between signal and noise dimensionality (blue, theory is dashed).
\textbf{c} Linear readout $S_{\alpha^{\prime}}(t)=w_{\alpha^{\prime}}(t)^{\protect\T}(x_{\alpha}(t)+\xi_{\text{pre}})+\xi_{\text{post}}$
trained for each time point $t$ to detect stimulus identity by minimizing
the quadratic error $\sum_{\alpha}(S_{\alpha^{\prime}}-\delta_{\alpha\alpha^{\prime}})^{2}$
for the correct stimulus $\alpha=\alpha^{\prime}$ (blue) and nonmatching
stimuli $\alpha\protect\neq\alpha^{\prime}$ (dark orange). Error
bars show the variability across patterns and noise realizations,
excluding $\xi_{\text{post}}$. Theoretical prediction \prettyref{eq:signal_approximation_final}
(dashed blue). Readout by the theoretical weight vector \prettyref{eq:w_star-2}
using the approximation \prettyref{eq:uncorrelated_patterns-1} (light
blue). Inset: classification accuracy of the initial input stimulus
based on choosing the readout with largest signal for trained readouts
(blue) and approximate readouts (light blue), including $\xi_{\text{post}}$.
The approximate readout vectors yield a higher average signal, but
also the variance is higher (not shown) resulting in slightly worse
classification accuracy. Other parameters: $N=500$ neurons, coupling
strength $g=0.8$, prereadout noise $\sigma_{\xi\text{,pre}}=0.1$
and postreadout noise $\sigma_{\xi\text{,post}}=0.1$ as detailed
in \prettyref{app:Noisy-binary-pattern-task}. The training set comprised
$100$ noisy realizations of each pattern, and the test set $20$.
All theoretical curves are corrected for the probability that a noise
realization does not leave the original flux tube, see \prettyref{app:Nonzero-plateau-of-signal}.\label{fig:transient_dim_expansion_binary}}
\end{figure*}

We now return to the question how the separation of trajectories by
the chaotic dynamics of binary networks affects computation in a setting
of reservoir computing (\prettyref{fig:Transient-chaotic-dimensionality-summary}).
We investigate the network performance in a pattern classification
task: We consider $P$ fixed patterns, numbered by the index $1\leq\alpha\leq P$,
each given by a randomly drawn binary vector of length $L$. Noisy
realizations of a pattern are then created by adding Gaussian independent
noise of variance $\sigma^{2}$ to each entry of the original pattern,
creating $P$ classes of noisy pattern realizations. The network is
prepared at $t=0$ in a fixed initial state consistent with its stationary
statistics. A noisy pattern is presented to the network as the initial
state of a (fixed) subset of $L$ of the $N$ neurons. The corresponding
network state is denoted as $x_{\alpha}(t)$. At each time $t$, we
train one linear readout $S_{\alpha^{\prime}}(t)=w_{\alpha^{\prime}}(t)^{\T}x_{\alpha}(t)$
per pattern class $\alpha^{\prime}$ by linear regression to provide
the output $S_{\alpha^{\prime}}(t)=1$ if the $\alpha^{\prime}$-th
pattern has been presented ($\alpha=\alpha^{\prime}$) and $0$ else
($\alpha\neq\alpha^{\prime}$, see \prettyref{app:Linear-regression}
for details). Thus we have $P$ readouts, one for detecting each of
the presented patterns (one-hot encoding). Classification is performed
by selecting the strongest readout signal. Additional noise sources
are present at the readout and classification to ensure robustness.
The setup, training and following theory are detailed in \prettyref{app:Noisy-binary-pattern-task}.

Clearly, the set of possible trajectories resulting from the different
initial-state preparations has dimensionality $d_{s}(0)=L$ at $t=0$.
The linear separability of pattern classes is thus initially low,
if $\begin{gathered}L\ll P\ll N\end{gathered}
$. From \prettyref{sec:dynamics-submanifold} we know that the chaotic
dynamics will quickly increase the dimensionality of the state space
that encodes the patterns, eventually approaching that of the chaotic
submanifold $d_{s}(\infty)=d^{\ast}$. To explain the effect on the
separability of patterns and the classification performance, we must
distinguish between the dimensionality of the total set of trajectories
(including all patterns and their noisy variations), referred to as
the signal dimensionality $d_{s}(t)$, and the dimensionality of the
set of trajectories given by noisy variations of a single pattern,
referred to as the noise dimensionality $d_{n}(t)$. Let us at first
neglect the noise. With the increase of $d_{s}(t)$ also the linear
separability of patterns increases. From the property of the linear
regression this means that the average readout signal $S_{\alpha}(t)$
of the correct pattern class $\alpha$ increases; if network responses
were pairwise orthogonal, which to good approximation is satisfied
in the high-dimensional signal subspace, the maximal attainable signal
would be
\begin{align*}
\hat{S}_{\alpha}(t) & =\frac{d_{s}(t)}{P},
\end{align*}
as shown in \prettyref{app:Approximation-orthogonal-patterns}.

However, also the noise dimensionality $d_{n}(t)$, spanned by all
noisy realizations of the same pattern, increases in the same way
as $d_{s}(t)$ due to the chaotic dynamics, as shown in \textcolor{brown}{\prettyref{fig:transient_dim_expansion_binary}}\textbf{\textcolor{black}{b}}\textcolor{black}{.}
But $d_{s}(t)$ has a head start because noisy realizations of one
pattern are more similar to each other than to other patterns. Now
let us assume that different noise realizations cause different responses
that lie entirely within and are uniformly distributed across the
signal subspace. This means that the noise randomly flips a number
of spins that encode the pattern and thus effectively reduces the
dimensionality of the space that faithfully encodes the signal. The
effective dimension of the space that is available to represent the
signal is then
\begin{align}
\Delta d(t) & =d_{s}(t)-d_{n}(t).\label{eq:def_delta}
\end{align}
 The expected signal is then given by
\begin{align}
S_{\alpha}(t) & =\frac{d_{s}(t)-d_{n}(t)}{P}\label{eq:signal_approximation_final}\\
 & =\frac{d_{s}(t)}{P}\,\Big(1-\frac{d_{n}(t)}{d_{s}(t)}\Big).\nonumber 
\end{align}
This approximate expression overestimates, but captures quite well
the overall shape of the average readout signal shown in \prettyref{fig:transient_dim_expansion_binary}\textbf{c}:
Initially, the signal rises in relation to the ratio of dimension
of representation space and number of patterns, but ultimately, the
signal declines, because the dimensionality spanned by the noise approaches
that spanned by the signal. As seen in the inset of \prettyref{fig:transient_dim_expansion_binary}\textbf{c},
the classification accuracy mirrors the behavior of the average readout
signal. This transient increase of the linear separability of the
pattern classes is rooted in the property of the system that the signal
dimensionality initially rises faster than the noise dimensionality.

\begin{figure}
\begin{centering}
\includegraphics{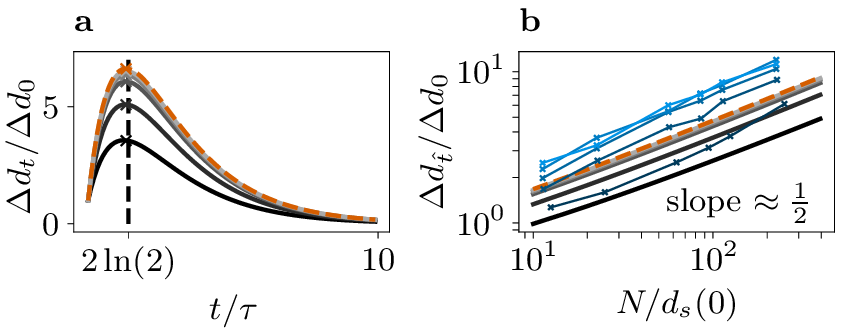}
\par\end{centering}
\caption{\textbf{Optimal expansion within $2\ln2$ neuronal activations.} \textbf{a}
Effective dimension $\Delta d(t)$ \prettyref{eq:def_delta} relative
to initial $\Delta d(0)$ for vanishing initial noise amplitude $d_{n}(0)\searrow0$.
Maxima at $\hat{t}=2\tau\,\ln\frac{1-(1-\delta)^{2}}{\delta}$, $\delta=\sqrt{d_{s}(0)/d_{\ast}}$
\prettyref{eq:t_hat_exact} marked by crosses. The time to maximum,
for weak initial stimuli $\delta\ll1$, is well approximated by $\hat{t}\simeq2\ln2\,\tau$
(vertical dashed line). From dark to light gray: $g=0.5,1,2,3,5$
and $\tanh$ gain function; dashed dark orange curve: Heaviside gain
function, limit $g\to\infty$. Initial parameter $N/d_{s}(0)=200$.
\textbf{b} Maximal effective dimension $\Delta d(\hat{t})$ relative
to initial $\Delta d(0)$ as a function of $N/d_{s}(0)$. Same color
code as a. In blue, corresponding simulation results, averaged over
ten connectivity seeds each, and corrected for the probability that
a noise realization does not leave the original flux tube, see \prettyref{app:Nonzero-plateau-of-signal}.
Values of $g$ from dark to light as in a. \label{fig:Optimal-expansion}}
\end{figure}

For small initial $d_{s}(0)$ and in the limit of vanishing noise
$d_{n}(0)\searrow0$ the maximum of $\Delta d=d_{s}-d_{n}$, and therefore
$S_{\alpha}$, is reached at $\hat{t}\simeq2\ln2\,\tau\simeq1.39\,\tau$
(details given in \prettyref{app:Initial-growth-of-perturbations-in-binary},
Eq. \prettyref{eq:t_hat}); the peak time depends only weakly on $g$,
as shown in \prettyref{fig:Optimal-expansion}\textbf{a}. The improvement
of the separability due to the transient expansion, the maximum $\Delta d(\hat{t})$
compared to its initial value at $t=0$ is given by \prettyref{eq:improvement_signal}
\begin{align*}
\frac{\Delta d(\hat{t})}{\Delta d(0)} & \simeq\frac{1}{2}\sqrt{\frac{d_{\ast}}{d_{s}(0)}}+\frac{1}{4}\\
 & \stackrel{g\to\infty}{=}\frac{1}{\pi}\,\sqrt{\frac{2N}{d_{s}(0)}}+\frac{1}{4},
\end{align*}
where the latter expression is the limit of $g\to\infty$ of the former.
The improvement of the signal scales with $N^{\frac{1}{2}}$ and $d_{s}(0)^{-\frac{1}{2}}$,
as shown in \prettyref{fig:Optimal-expansion}\textbf{b}. The theory
slightly underestimates the maximum of $\Delta d$ in simulations,
but captures the scaling relation, predicting a slope of $\approx1/2$.
So the peak classification accuracy can be improved by larger networks.

Note that in \prettyref{fig:transient_dim_expansion_binary}\textbf{c},
neither in theory nor in the simulation the average signal drops to
zero for $t\to\infty$, in \textbf{b} the noise curve saturates to
a smaller value than that of the signal; this is because not all noise
realizations leave the flux tube of the original pattern, so that
the average distance between realizations remains smaller than that
between patterns (analyzed in \prettyref{app:Nonzero-plateau-of-signal}).

\subsection{Generalization to other network models: Transient chaotic SNR amplification
\label{sec:Generalization-SNR-amplificaiton}}

The computational effect described in the last section relies on two
factors: First, a high-dimensional space in which trajectories are
nonlinearly embedded, and second, the decorrelation dynamics eq. \prettyref{eq:ODE_deviation_fully_correlated}
and \prettyref{fig:transient_dim_expansion_binary}\textbf{b} that
causes a small deviation to initially grow slower than a larger deviation.
In the setting of a classification task, this mechanism thus enhances
the signal-to-noise ratio (SNR).

The first factor is a general feature of all nonlinear neuronal networks.
The second factor, we conjecture, should also be a typical property
of chaotic networks, because the expansion of distances between three
arbitrary trajectories should imply the largest distance to grow faster
(in absolute terms) than the two smaller distances, as by a triangle
inequality. Therefore, we expect the transient disentanglement of
pattern classes to be a general phenomenon in strongly chaotic neuronal
networks. To substantiate this claim, we demonstrate this effect in
rate networks, and, in a proof-of-principle manner, in a spiking leaky-integrate-and-fire
(LIF) network and a long-short-term-memory (LSTM) network \citep{Hochreiter97}.

\subsubsection*{Rate networks}

A network of stochastic rate units with first- and second-order statistics
matched to those of a binary network, as in \prettyref{sec:Same_stat_Different-mechanisms-of-chaos},
has a classification performance close to zero, as shown in \prettyref{fig:rate_chaos_B}\textbf{a}.
This is because the matched stochastic rate network is not chaotic;
trajectories for different presented patterns converge. A small readout
noise thus destroys classification accuracy.

However, how is the performance in the chaotic regime of a rate network?
For this we no longer consider matched statistics and from here on
again use rate networks without effective noise and $g^{2}>1$, which
are chaotic \citep{Sompolinsky88_259}.

\begin{figure}
\begin{centering}
\includegraphics{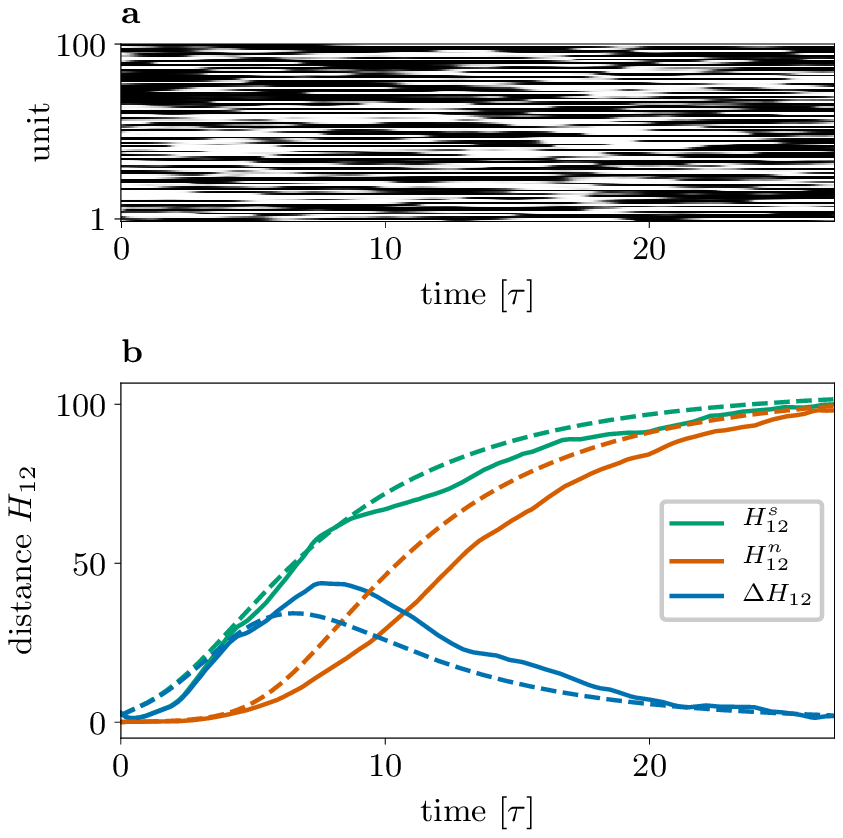}
\par\end{centering}
\caption{\textbf{Evolution of trajectories in a chaotic rate network.} \textbf{a}
Chaotic network activity for $N=250$, $g=5.8$, $\sigma_{\text{eff}}^{2}=0$
and $\protect\T(h)=\tanh(h)$. Gray scale shows the activity of each
neuron between $-1$ (white) and $1$ (black).\textbf{ b} Time evolution
of average distances between patterns (green) and between noisy realizations
of a pattern (dark orange). Difference between signal and noise distances
(blue). Numerical solutions of \prettyref{eq:epsilon_evolution_rate-1}
(dashed), for details see \prettyref{app:Initial-growth-of-perturbations-in-binary-1}.
Network as in (a) and $P=50$, $L=10$, $\sigma=0.3$.\label{fig:rate_chaos_A}}
\end{figure}

Similar to binary networks, finite-size chaotic activity (\prettyref{fig:rate_chaos_A}\textbf{a})
in rate networks\textbf{ }shows a transiently smaller difference between
noisy realizations of a pattern than between patterns in the classification
task, as shown in \prettyref{fig:rate_chaos_A}\textbf{b}. This also
follows from \prettyref{eq:epsilon_evolution_rate-1} and \prettyref{eq:Lyapunov}
since the acceleration of the decorrelation is smaller for smaller
initial $\epsilon$, making the noise distance grow slower than the
signal distance.

Decorrelation in the rate network, however, can take many neuronal
time constants if the network is in the mildly chaotic regime (\prettyref{fig:rate_chaos_A}\textbf{b}).
The timescale sensitively depends on the recurrent coupling strength,
as shown in \prettyref{fig:Discrete-chaos-versus-rate-chaos}\textbf{b}.
This can be understood in terms of the decay constant of the time
lagged autocorrelation function: In a noiseless rate network, the
timescale of the autocorrelation diverges at the transition to chaos
\citep{Sompolinsky88_259}.

In the rate network classification performance jumps to a high value
already after the first time step (\prettyref{fig:rate_chaos_B}\textbf{c}).
The reason is that all neurons' states are immediately nonlinearly
affected by the input pattern. The stimulus is thus immediately projected
nonlinearly into an $N$-dimensional representation that allows linear
classification. This becomes apparent by considering that in the time
step after stimulus presentation, the input to the network contains
a term $\propto\frac{\delta t}{\tau}\,\sum_{j}J_{ij}\,T(h_{j}(0))$.

\begin{figure}
\begin{centering}
\includegraphics{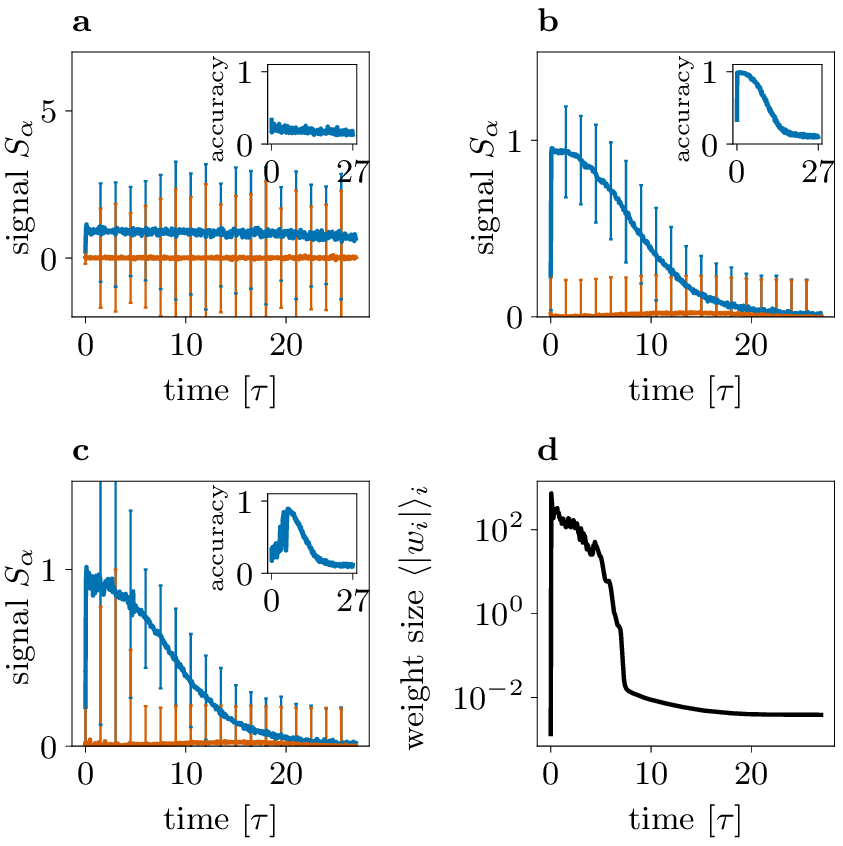}
\par\end{centering}
\caption{\textbf{Stimulus representation in a rate network.} \textbf{a},\textbf{b},\textbf{c
}Signal of correct readout (blue), signal of wrong readouts (dark
orange), and classification accuracy (inset). \textbf{a} Including
noise $\sigma_{\xi}^{2}$ matched to the statistics of a binary network
according to \prettyref{eq:Initial_condition_Q_dot_binary}. Parameters
otherwise as in \prettyref{fig:rate_chaos_A}. Because the frozen
effective noise suppresses chaos, trajectories for different stimuli
converge toward the same state and a small readout noise $\sigma_{\text{readout}}=10^{-4}$
results in classification at chance level. \textbf{b} Removing the
noise in the network, and no readout noise. The network is chaotic
as in \prettyref{fig:rate_chaos_A}, and classification accuracy jumps
to $1$ in the first time step. \textbf{c} Including readout noise
$\sigma_{\text{readout}}=10^{-4}$ impedes classification accuracy
until trajectories are sufficiently separated. \textbf{d} Average
norm of individual readout weights for the setting in (b). Since trajectories
are very close in the initial time period (compare also \prettyref{fig:rate_chaos_A}b),
initial readout weights are very large, explaining the sensitivity
to readout noise in (c). \label{fig:rate_chaos_B}}
\end{figure}
 But even though the dimensionality is immediately $N$ dimensional,
the amplitude grows continuously with time and is thus very small
at first, $\propto\delta t$. This behaviour exposes the qualitative
difference in the interpretations of $\epsilon$: In the binary network,
there is a direct link between $\epsilon$ and the dimensionality
of the signal space, while in the rate network $\epsilon$ is a measure
of the Euclidean distance between trajectories and is only very indirectly
related to the number of dimensions across which this distance is
distributed. Hence, the difference between inter- and intrapattern
distances in \prettyref{fig:rate_chaos_A}\textbf{b} peaks at a later
time point than the classification accuracy in \prettyref{fig:rate_chaos_B}\textbf{b},
because the rate network performs a transient signal amplification
rather than a transient dimensionality expansion as in the binary
network case.

Even though the dimensionality of the representation immediately after
stimulus presentation equals $N$, distances between stimuli in the
new directions are small at first. The classification in the rate
network therefore relies on fine-tuned and very large readout weights
in the beginning (\prettyref{fig:rate_chaos_B}\textbf{d}). Therefore,
the initial classification accuracy is severely impaired by adding
even weak noise to the readout (\prettyref{fig:rate_chaos_B}\textbf{c}).
This addition of noise, in turn, results in a peak of the accuracy
predictable by the theoretical peak in signal amplification (\prettyref{fig:rate_chaos_A}\textbf{b}).

In summary, rate networks with continuous signaling perform a transient
amplification of the signal-to-noise ratio, rather than a transient
dimensionality expansion. Compared to binary networks with discrete
signaling, the resulting empirical differences are the strong dependence
of the decorrelation timescale on the coupling strength, the existence
of a minimal coupling strength required for amplification, the absence
of the residual correlation, and the initially high sensitivity to
readout noise.

\begin{figure*}
\includegraphics{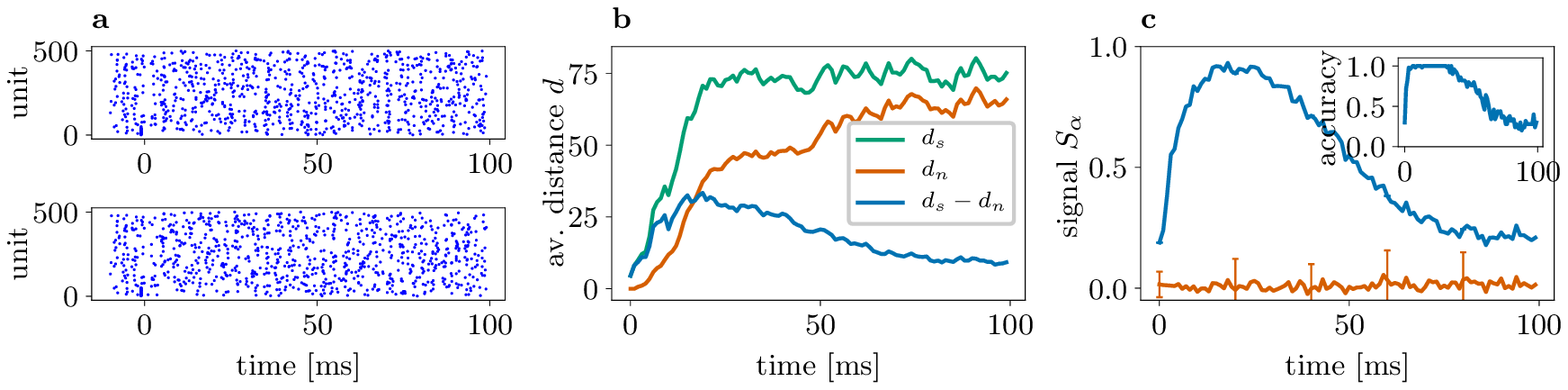}

\caption{\textbf{Transient chaotic dimensionality expansion in a spiking network.
a }Raster plots of spiking activity in the network for two different
patterns; average firing rate per neuron $\nu\simeq19\,\mathrm{Hz}$.
Parameters $N=500$, in-degree $K=125$, weights $J_{ii}=-1.0\,\mathrm{mV}$,
and membrane time constant $\tau_{m}=10\,\mathrm{ms}$; additional
detail given in \prettyref{app:Description-of-simulations}.\textbf{
b }Evolution of average Euclidean distance between different patterns
(green), noisy realizations of a pattern (dark orange), and the difference
of the two (blue). \textbf{c }Signal of matching readout (blue), nonmatching
readouts (dark orange), and classification performance (inset). \label{fig:LIF-SNR-amplification}}
\end{figure*}
\begin{figure*}
\includegraphics{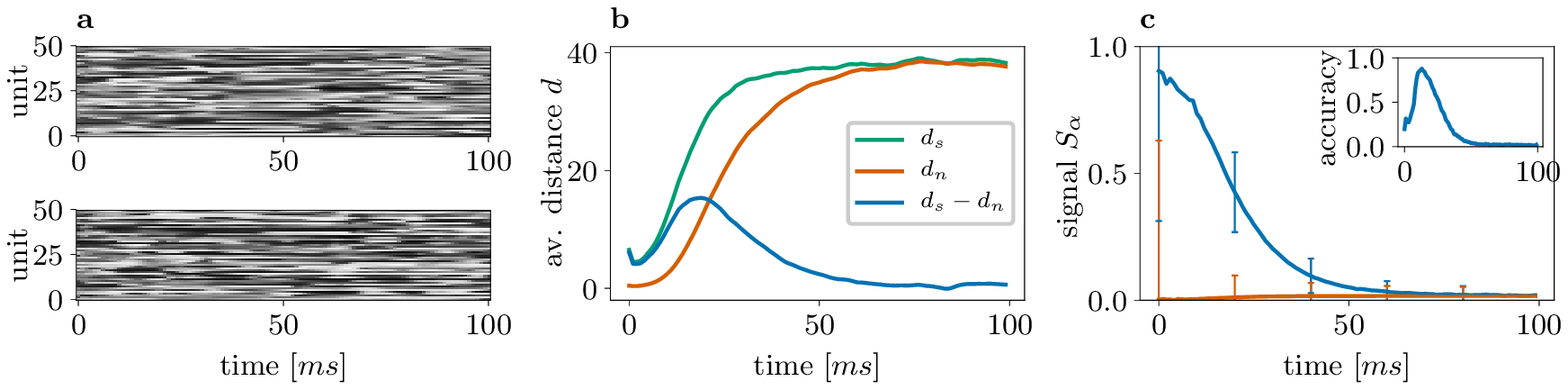}

\caption{\textbf{Transient chaotic SNR amplification in a LSTM network. a }Chaotic
network activity of the first $50$ units for two different patterns.
Parameters $N=200$ and $g=5.8$ for the hidden to input-, forget-
and cell-gates weights and all other weights using a standard $\mathcal{U}(-N^{-\frac{1}{2}},N^{-\frac{1}{2}})$
distribution, with additional detail given in \prettyref{app:Description-of-simulations}.
Gray scale shows the activity of each hidden unit between $-1$ (white)
and $1$ (black).\textbf{ b }Evolution of average Euclidean distance
between different patterns (green), noisy realizations of a pattern
(dark orange), and the difference of the two (blue). The short initial
decay in pattern distances is likely due to a short networkwide contraction
of the dynamics after pattern presentation, caused by the interaction
of the gating variable states and the $L$ changed hidden states.
\textbf{c }Signal of correct readout (blue), incorrect readouts (dark
orange) and classification performance (inset).\label{fig:LSTM-SNR-amplification}}
\end{figure*}

\subsubsection*{Spiking networks}

To demonstrate that the same computational effect translates to spiking
networks which are furthermore not all-to-all connected, we consider
the same task in a purely inhibitory network of LIF neurons with fixed
in-degree in the asynchronous-irregular firing state. Trajectories
in these networks are known to have a small stable local environment
(flux tube) but exhibit chaos for perturbations leaving the flux tube
\citep{Monteforte12_041007,Touzel19}, just as the binary networks
considered in this manuscript. In the binary networks, the flux-tube
borders are given by input perturbations that are just sufficient
to cause a single spin to flip. By binning the spike trains of the
LIF network with a bin width equal to the membrane time constant,
we obtain approximately binary vectors if the bin width is small compared
to the inverse firing rate. Therefore we can use the same training
and analysis procedure as for the binary networks, and also interpret
distances in terms of dimensionality. Further details are given in
\prettyref{app:Description-of-simulations}. The LIF network dynamics
and performance show the same features found in binary networks: Distances
between states quickly grow toward a residual correlation (\prettyref{fig:LIF-SNR-amplification}).
Dimensionality expansion takes place on a timescale within which each
neuron fires only a single spike or less on average ($\nu^{-1}\simeq55\,\mathrm{ms}).$
The distance between pattern classes initially grows faster than the
noise distance, \prettyref{fig:LIF-SNR-amplification}\textbf{b},
and the readout signal and classification performance show a corresponding
transient peak, \prettyref{fig:LIF-SNR-amplification}\textbf{c}.
As in the binary network, some noise realizations do not leave the
flux tube of the unperturbed pattern, causing a reduced long-term
average noise distance and a nonzero plateau of the residual classification
performance. 

\subsubsection*{LSTM networks}

Finally, to demonstrate the existence of the computational effect
in a powerful specialized machine-learning architecture, we consider
the same task in a recurrent LSTM network \citep{Hochreiter97}. The
architecture is similar to the rate networks we consider, but contains
a large number of additional dynamical ``gating'' variables which
control when, where and by how much the cell states interact with
each other, resulting in considerably more complex dynamics than in
a rate network with fixed coupling matrix. We us a vanilla pytorch
implementation, choosing the initialization parameters such that the
network exhibits spontaneous chaotic fluctuations over a moderate
range of timescales, as seen in \prettyref{fig:LSTM-SNR-amplification}\textbf{a}.
Details about the parameters and task implementation are given in
\prettyref{app:Description-of-simulations}. In \prettyref{fig:LSTM-SNR-amplification}\textbf{b}
and \textbf{c}, we see a behavior in close analogy to that shown in
\prettyref{fig:rate_chaos_A}\textbf{b} and \prettyref{fig:rate_chaos_B}
for rate networks. Signal and noise distances increase differentially
fast, and there is a pronounced transient peak in the classification
accuracy. Other than in the binary and LIF networks, which use discrete
signaling, the rate and LSTM networks do not have locally stable flux
tubes and no residual plateau in the classification performance.

\section{Discussion\label{sec:Discussion}}

This manuscript compares the effect of discrete and continuous signaling
on the dynamics and function of neuronal networks. Focusing on binary
classification as a fundamental computation, it addresses the question
how the temporal dynamics can be used to represent stimuli. Separating
representations translates into asking how state trajectories diverge
or converge if different stimuli are presented. Technically this amounts
to quantifying chaos in such networks.

A model-independent path-integral approach enables comparisons across
models. We find that the dynamic mean-field theory is of identical
structure for networks of binary units and for continuous rate networks.
In binary networks we discover a network-size-dependent transition
to chaos and the existence of a chaotic submanifold. We elucidate
the qualitative differences to chaos in rate networks in terms of
the mechanism causing chaos, timescales, and parameter regimes.

Applied to classification, chaotic dynamics causing a relative dimensionality
expansion of representations leads to a mechanism of fast and transient
computation in binary networks with discrete signaling. We describe
a generalization of this effect as a transient signal-to-noise amplification
in chaotic rate networks with continuous signaling.

The remainder of the discussion puts these results into context of
the literature, mentions limitations, and provides an outlook.

\subsection{Differences and similarities across neuron models\label{sec:Differences-and-similarities-across-neuron-models}}

\subsubsection*{Transition to chaos in binary networks at finite size}

We demonstrate that there is a transition to chaos in finite-size
binary networks, described by a field-theoretical replica calculation.
Our results are consistent with works on sparse random boolean networks
with synchronous update showing a chaos transition for in-degree $K=2$
\citep{Derrida86_45,Kauffman93}, which relates here to the transition
at $N\approx2$ for the Heaviside activation function. \citet{Derrida86_45}
approximate the disorder by annealed averages and use synchronous
update, while we compute the quenched averages and employ asynchronous
update. In mean-field theory, the in-degree in sparse networks plays
a role similar to the network size in dense networks. Other works
that investigated the edge of chaos numerically in discretely coupled
networks have also found small in-degrees as critical coupling \citep{Bertschinger04_1413,Legenstein07_323,Snyder12}.

\subsubsection*{Correspondence of DMFT in binary and rate networks}

The model-independent field theory presented here exposes a one-to-one
match of the stationary activity statistics in dynamical mean-field
approximation of binary and rate networks. Exposing identities between
neuron models is useful to see if and how the results generalize.
Steps in this direction where already taken in \citet{Grytskyy13_131},
who showed that weak pairwise correlations can be explained by linearizing
LIF neurons, Hawkes processes, and binary neurons, mapping them to
noisy linear rate models. The results presented here are more general
since they apply not only to the linearization of the models but hold
for the nonlinear behavior as well. The equivalence of time-lagged
autocorrelations is shown here for stationary statistics; for nonstationary
dynamics also the effective noise strength should vary as a function
of time \citep[chap. 6.4]{Kuehn20_diss}.

\subsubsection*{Assumptions on connectivity}

The assumption of Gaussian connectivity $J_{ij}\sim\mathcal{N}(\frac{\bar{g}}{N},\frac{g^{2}}{N})$
straightforwardly generalizes to other connectivities, as long as
higher than second cumulants are suppressed by powers of $N^{-1}$.
Scaling the mean connectivity as $\bar{g}/N$ yields a consistent
approximation in $1/N$. Sparse connectivity, however, typically leads
to a scaling $\bar{g}/\sqrt{N}$. Formally, a consistent treatment
therefore requires Gaussian fluctuations of the mean activity field
$\av{\mathcal{R}^{2}}{}\sim1/N$, which is possible in the presented
framework. Such fluctuations are, however, suppressed by negative
feedback \citep{Tetzlaff12_e1002596} in the inhibition-dominated
(balanced) regime $\bar{g}<0$. For multiple populations, the DMFT
equations acquire population indices, but stay structurally the same
(cf. \citep{VanVreeswijk98_1321} for binary neurons and \citep{Kadmon15_041030,Aljadeff15_088101}
for rate neurons). Scale-free distributions of weights can violate
the assumptions and require a different approach \citep{Kusmierz19_05780}.

\subsubsection*{Relation of the model-independent path integral formulation to earlier
work}

The seminal work by \citet{Sompolinsky88_259} on rate neurons used
statistical field theory \citep{Crisanti18_062120}, and the work
by \citet{VanVreeswijk98_1321} on binary neurons relied on a disorder
average of the master equation \citep{Glauber63_294,Ginzburg94}.
The statistical field theory that we develop here captures both model
classes, and is similar to the Martin-Siggia-Rose-de Dominicis-Janssen
(MSRDJ) formalism \citep{Martin73,dedominicis1976_247} for rate neurons
\citep[reviewed e.g. in][]{Chow15,Hertz16_033001,Helias20_970}. In
particular, this formulation exposes the identical structure of the
mean-field approximations.

Binary networks with asymmetric connectivity show nonequilibrium dynamics,
so that the Ising Hamiltonian cannot be used. Instead, complete information
about the system dynamics needs to be captured. Full information is
supplied by the master equation, for which an established approach
is the Doi-Peliti formalism \citep{Doi76_1465,Peliti85_1469}. The
fields in the latter approach, however, have no intuitive physical
interpretation, even though they allow the construction of mean-field
equations and fluctuation corrections \citep{Buice07_051919}. Closer
to our method are the approaches by \citet{Sommers87_1268,Andreanov2006_030101,Lefvre2007_07024},
which can be obtained as special cases from our formulation.

\subsubsection*{Different chaotic dynamics in binary and rate networks}

Qualitative differences between chaos in binary and rate networks
can be summarized as follows: i) Rate networks with activity statistics
matched to that of binary networks are nonchaotic. This shows that
discrete signaling provides a different mechanism that drives chaos
in binary networks, and that the mechanism causing chaos in rate networks
is not effective in binary networks. A marginally chaotic solution
is approached in matched rate networks when sending the activation
function to the Heaviside limit. This is consistent with the finding
that rate networks are always chaotic if the activation function has
an infinite slope \citep{Kadmon15_041030}. ii) In the limit of large
numbers of neurons, (now unmatched) rate networks have a critical
coupling strength beyond which they transition to chaos, while binary
networks are always chaotic in this limit. At finite network sizes,
binary networks have a size-dependent, critical coupling strength,
which is typically very low. iii) Decorrelation of trajectories in
binary networks is generally faster than in rate models. In binary
networks it takes place on a timescale given by the interval between
state changes of individual neurons and is only mildly affected by
the network coupling. In rate models, it strongly depends on the coupling
strength, showing a critical slowing-down at the transition to chaos.
Stochasticity gradually smooths out this divergence \citep{Schuecker18_041029}.
iv) Trajectories in binary networks decorrelate only up to a residual
correlation, while those in rate networks completely decorrelate.
v) Binary networks have an infinite Lyapunov exponent, so that decorrelation
starts off with a finite slope even for infinitesimal initial perturbations.
In rate networks, the initial decorrelation is an exponential function,
whose slope therefore vanishes for infinitesimal perturbations.

\subsubsection*{Origin of the difference}

In the rate network, the noise is external and frozen, but a binary
network's noise realization depends acutely on the initial value of
the system. Thus, perturbing the initial value also changes the noise
realization. In particular, due to the thresholding operation that
produces the discrete signal, a tiny perturbation in the input can
cause a flip of the neuron, and consequently a macroscopic change
of the network state; the probability that this change happens increases
with the number of targets that receive this perturbation, and thus
with network size; this increase is due to the strong synapses $\abs{J_{ij}}\propto N^{-\frac{1}{2}}$
(for $\cum{J_{ij}^{2}}{}\sim gN^{-1}$). For large networks the growth
of the perturbation corresponds to a macroscopic change in the noise
realization.

\begin{figure}
\begin{centering}
\includegraphics{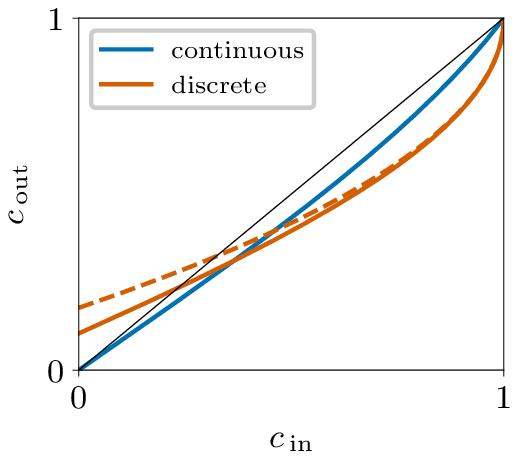}
\par\end{centering}
\caption{\textbf{Correlation transmission by a pair of neurons.} Correlation
coefficient $c_{\mathrm{out}}$ between outputs as a function of the
correlation coefficient $c_{\mathrm{in}}$ between inputs. Discrete
signaling (dark orange, \prettyref{eq:cout_disc}); continuous signaling
(blue, \prettyref{eq:cout_cont}). Approximation for discrete signaling
in the limit $c_{\mathrm{in}}\to1$ (dashed dark orange) shows a behavior
of the form $\propto-(1-c_{\mathrm{in}})^{\frac{1}{2}}$, so the slope
diverges as $\propto(1-c_{\mathrm{in}})^{-\frac{1}{2}}$. For comparison:
identity mapping with unit slope (black line).\label{fig:Correlation-transmission-by-pair-binary-2-1-1}}
\end{figure}

The presented replica calculation provides a complementary explanation
for the qualitative difference between binary and rate neurons. The
question of a chaos transition is reduced to studying how correlations
are transferred from the inputs of a pair of neurons to their outputs:
The change of the correlation between replicas is proportional to
the mismatch between the correlation at time $t$, the second line
on the right of Eq. \prettyref{eq:full_ODE_crossreplicacorr_binary},
and the correlation transmitted through a pair of neurons and connectivity,
given by the third line, a function $c_{\mathrm{out}}(c_{\mathrm{in}}).$
Such a transmission curve $c_{\mathrm{out}}(c_{\mathrm{in}}$) is
shown for discrete signaling and for continuous signaling in \prettyref{fig:Correlation-transmission-by-pair-binary-2-1-1}.
Clearly, if $c_{\text{out}}(c_{\text{in}})<c_{\text{in}}$ for a $c_{\mathrm{in}}$
close to perfect correlation just below unity, the correlation decreases
over time, the dynamics is chaotic; if $c_{\text{out}}(c_{\text{in}})>c_{\text{in}}$,
the correlation regenerates, the dynamics is regular. While the slope
$c_{\mathrm{out}}^{\prime}(c_{\mathrm{in}}\to1$) for discrete signaling
diverges as $\propto(1-c_{\mathrm{in}})^{-\frac{1}{2}}$, it stays
finite for continuous signaling. The infinite slope for networks with
discrete signaling leads to an infinite Lyapunov exponent for $N\to\infty$.
For continuous signaling, the slope is finite as long as the slope
of the activation function is bounded, resulting in finite Lyapunov
exponents. This analytical view of the different transitions to chaos
in continuous and discrete networks is also consistent with numerical
findings \citep{Busing10_1272}.

Correlation transmission by pairs of neurons is well studied experimentally
and theoretically \citep[e.g. ][]{Tetzlaff04_117,Shea-Brown08,Tchumatchenko10_058102,Tchumatchenko11_review}.
A diverging slope of the correlation-transmission curve, here shown
for binary neurons, has also been demonstrated for spiking neurons
without reset \citep{Tchumatchenko10_058102} and for the LIF model
\citep{Tetzlaff03_949,sheabrown07,Shea-Brown08,SchultzeKraft13_e1002904,Deniz2017_012412},
suggesting that these model classes behave similarly with regard to
the transition to chaos.

\subsubsection*{Flux tubes in binary and spiking networks}

There is a tight link between the replica calculation and chaos in
spiking networks examined in terms of the divergence rate between
flux tubes \citep{Monteforte12_041007}. Flux tubes are neighboring
portions of the phase space within which perturbations of the state
do not cause a global change of subsequent activity \citep{Touzel19}.
Binary neurons are formally simpler than spiking models, because one
can investigate changes of network states directly instead of analyzing
spike patterns (\prettyref{app:Fluxtubes-in-binary-networks}), binary
neurons do not have additional internal degrees of freedom, such as
the membrane potential, and their activation times are given by predetermined
update times. Therefore perturbations inside a flux tube are not forgotten
exponentially as in LIF neurons, but instantly. Just like in LIF networks,
the distance to a flux-tube boundary shrinks with network size as
$\sim1/N$. The divergence rate between flux tubes scales as $\sim\sqrt{N}$,
opposed to $\sim N$ in the LIF network \citep{Monteforte12_041007}.
Both are consistent with an infinite Lyapunov exponent for $N\to\infty$.
Chaotic spiking activity has also been investigated in \citet{Lajoie13_052901}
and \citet{Lajoie14_00123}, who showed that chaotic quadratic-integrate-and-fire
networks exhibit a reduced spike pattern entropy, indicating that
they explore only a lower-dimensional manifold in phase space.\textcolor{brown}{}

\subsection{Computation in the chaotic regime}

\subsubsection*{Transient chaotic SNR amplification}

We find that strongly chaotic neuronal networks of different types
invariably exhibit a transiently improved separability of low-dimensional
inputs. This is at first sight surprising, because chaotic dynamics
amplifies noise as well as informative differences. However, the variability
within a class (noise) is typically smaller than the variability across
classes. We find the latter to be amplified more strongly than the
former, thus improving the linear separability of the classes. We
argue that this relative amplification is a general effect in nonlinear,
high-dimensional chaotic systems. There is an analogy to astrophysics:
Space in the Universe is locally expanding everywhere. Thus all points
are drawn apart, forming diverging trajectories. As a result of the
ubiquitous expansion, galaxies move apart ever faster the greater
the distance between them, as described by the Hubble constant. In
the same way, trajectories in the network that are farther apart (different
classes) separate faster than trajectories that are initially closer
(noisy realizations within the same pattern class). Unlike the Universe,
however, the network state space is higher dimensional and inherently
expands along highly curved directions. Therefore the faster expanding,
larger differences are also more strongly affected by the nonlinearity
and are more quickly embedded into the surrounding higher dimensions.
Also unlike the Universe, the state space volume of the network is
finite and constant, so some directions must shrink to conserve the
total volume. Trajectories therefore do not diverge indefinitely but
reach a stable average distance determined by the volume of the limiting
chaotic attractor. They continue to be mixed by the expanding and
shrinking dynamics, such that information about their initial distance
relations is eventually forgotten and classification performance subsides.

\subsubsection*{Elucidation by dimensionality in binary networks}

This qualitative picture is made concrete in binary networks: Their
discrete state space allows an interpretation of the growing average
distances in terms of dimensionality. This allows us to express the
improvement of classification accuracy in terms of the difference
between the dimensionality of the representation of the signal $d_{s}$
and the number of dimensions corrupted by noise $d_{n}$. Their temporal
evolution follows stereotypic decorrelation curves obtained from a
replica calculation. Dimensionality and separability are linked, because
each dimension allows the linear separation of two additional random
features \citep{Cover1965_326,Gardner88_271}. The effect can also
be viewed as a dynamical version of the usually statically applied
kernel trick \citep{Vapnik98}.

\subsubsection*{Relation of chaos and computational power}

It has been argued that close to the edge of chaos, networks show
an optimal trade-off between separation of stimuli and generalization
\citep{Bertschinger04_1413,Sussillo09_544,Toyoizumi11_051908,Legenstein07_323}.
Our analysis provides a time-dependent perspective on this hypothesis.
Indeed, the signal and noise dimensionalities that approximate the
classification performance can be compared to the Kernel Quality and
Generalization Rank \citep[VC dimension,][]{Vapnik98}, introduced
by \citet{Legenstein07_323}. Our results show that binary networks
generally have a short memory lifetime. But in contrast to rate networks,
it also does not reduce strongly when moving deeper into the chaotic
regime. For tasks that require only short memory, performance can
benefit from the increased separation even deep in the chaotic regime,
in particular because the peak of the informative dimensionality $\max(d_{s}-d_{n})$
increases with network size. This is in line with the results of \citet{Snyder12},
where for small readout delay, performance stays high when increasing
the in-degrees (e.g. their figure 3 a,b). Overall, these observations
raise the question, whether the effect could be combined with a prolonged
memory lifetime, for example by heterogeneous time constants of neurons
or synapses, clustered connectivity \citep{LitvinKumar12_e1002667},
or by feeding the readouts back into the network.

\subsubsection*{Other related works}

Recurrent networks can be transformed to deep feed-forward networks
with weight sharing by ``unrolling'' them in time. Successive layers
then correspond to adjacent time steps in the recurrent network. Chaotic
iterative maps were shown to yield an increase of the dimensionality
of representations toward deeper layers \citep{Poole16_3360}. Training
specifically on low-dimensional representations yields facilitated
feature generation by dimensionality expansion in early layers and
feature selection and generalization by dimensionality suppression
in later layers \citep{Recanatesi2019_00443}. \citet{Farrell19_biorxiv}
investigated (continuous) recurrent networks in a classification task
similar to ours, also considering the strongly chaotic regime. They
focused on late-time compression of the representation by training
and found that chaos benefits learning of the task, for which our
results provide a principled explanation. Their results provide clues
on how training interacts with the random connectivity, an interesting
avenue for future work.

The mechanism of transient computation we investigate coexists with
nonnormal amplification \citep{Hennequin_12,Kerg19_13613,Bondanelli20_1,Tarnowski20_arxiv},
which is caused by effective feed-forward structures embedded in nonorthogonalizable
coupling matrices. Non-normal amplification is especially strong
in the chaotic regime \citep{Hennequin_12}. The here described mechanism,
however, also applies to normal matrices. Finally, recent developments
in statistical mechanics of computation in neuronal networks are reviewed
by \citet{Bahri20_501}.

\subsubsection*{A mechanism of fast computation in spiking networks}

The peak classification performance in binary reservoirs is reached
on the scale of a single neuronal time constant after stimulus onset.
This scale also holds approximately in the LIF network. However, the
peak time can likely be even shorter for higher in-degrees or firing
rates, as these influence the divergence rate of flux tubes in such
networks \citep{Monteforte12_041007}. The mechanism may explain how
computation can spread rapidly through the hierarchical networks of
the brain: The time window corresponds to one activation per contributing
neuron on average, allowing the fast feed-forward processing latencies
of $\apprle50\,\text{ms}$ per stage measured experimentally in cortical
areas \citep{Thorpe96,Hung05_863}. This perspective suggests how
networks that employ discrete communication may compute rapidly on
the basis of a few spikes rather than requiring a prolonged averaging
over time. A possible impediment to computation in a chaotic system
is its sensitivity to initial conditions, which, apart from the input
patterns, are kept fixed in this manuscript. One solution would be
a mechanism which quenches variability at appropriate times. Another
possibility was exposed by \citet{Lajoie16_e1005258}, who showed
that chaotic spiking networks can reliably encode inputs despite changing
initial conditions, because each input confines the chaotic activity
to a different manifold.

\subsubsection*{Experimental evidence and predictions}

The olfactory system is a potential candidate to rely on the transient
computational mechanism we describe, because it is specialized on
classification of patterns without a temporal component. In the vertebrate
olfactory bulb, an odor activates a comparably low-dimensional pattern
of glomeruli, the input layer to a higher dimensional recurrent network
that needs to separate representations to enable classification of
odor identity by subsequent processing stages. The insect antennal
lobe shares this basic organization. Recordings in the olfactory systems
in zebrafish \citep{Friedrich01_889}, locust \citep{Mazor05_661},
and rats \citep{Cury10_570} show a representation of stimuli that
is consistent with the here found mechanism of transient dimensionality
expansion.

In zebrafish, the activities of mitral cells in the olfactory bulb
show a high correlation for similar odors shortly after stimulus presentation.
Subsequently they decorrelate on a timescale of $\sim800\,\mathrm{ms}$,
reaching a residual correlation of about 40\% \citep[Fig 2E]{Friedrich01_889}.
The discriminability of these similar odors by a linear readout from
the mitral cells improves within the same time span to nearly error-free
classification \citep[Fig 2I]{Friedrich01_889}. In the process,
the population statistics stays approximately constant. These features
are in line with transient dimensionality expansion, except that the
classification accuracy does not decline again after the improvement.
However, this missing decline could be caused by feedback stabilizing
the representation after recognition, or be related to the sustained
presentation of the odor stimulus. Recordings in the locust antennal
lobe show qualitatively similar behavior \citep{Mazor05_661}. In
particular, these experiments report a decoding accuracy that is highest
within the transient phase. In the rat olfactory bulb, inhalation
also triggers a fast decorrelation transient of $\sim100\,\mathrm{ms}$,
during which odor identity is encoded in the instantaneous spike pattern,
and decoding accuracy rapidly peaks after $\sim70\,\mathrm{ms}$ before
declining to a lower level \citep{Cury10_570}. Future work should
systematically investigate if the dynamics in these biological systems
is in fact chaotic, for which suitable analysis methods are available
\citep{Grassberger83_1983,Toker20_jan}. Also, the analysis of inter-
and intraclass distances and classification by linear readouts is
applicable to experimental spiking data. The computational mechanism
we describe needs a reliable initial state from which trajectories
diverge. In the rat olfactory system, this reset could be tied to
inhalation onset. In cortex, stimuli seem to quench the variability
of spontaneous activity to evoke relatively low-dimensional responses
\citep{Celletti96_387,Churchland10,Mazzucato16,Gao17_4262}. To check
for chaotic dynamics, one could therefore analyze the growth of intertrial
variability during and after stimulation offset. We finally list concrete
testable predictions for neural systems that implement classification
by transient chaotic dimensionality expansion:
\begin{enumerate}
\item Variability is small or quenched at stimulus onset, then transiently
increases and reaches a stable value.
\item Not only the interclass distances, but also the intraclass (noise)
distances increase, although initially slower.
\item Decoding accuracy based on linear readouts trained at each time point
shows a peak, and this peak occurs before the distances saturate.
\end{enumerate}

\textcolor{brown}{}

\subsection*{Acknowledgements}

We thank Jonathan Kadmon for suggesting the analogy to the Hubble-Lema\^itre
law. This work is partially supported by the Helmholtz young investigator's
group VH-NG-1028, the European Union's Horizon 2020 research and innovation
programme under grant agreement No.\ 785907 (Human Brain Project
SGA2), the Exploratory Research Space (ERS) seed fund neuroIC002 {[}part
of the Deutsche Forschungsgemeinschaft (DFG), German Research Foundation
excellence initiative{]} of the RWTH University and the JARA Center
for Doctoral studies within the graduate School for Simulation and
Data Science, funded by the Deutsche Forschungsgemeinschaft (DFG,
German Research Foundation) - 368482240/GRK2416, funded by the Human
Frontier Science Program RGP0057/2016 grant, and funded by the Excellence
Initiative of the German federal and state governments (G:(DE-82)EXS-PF-JARA-SDS005).

\appendix

\section*{Appendices}

\subsection{Model-independent mean-field theory for random networks\label{app:Model-independent-mean-field-the_appendix}}

This section presents a self-contained derivation of the model-independent
mean-field theory for networks with Gaussian random connectivity $J_{ij}\stackrel{\text{i.i.d.}}{\sim}\mathcal{N}\left(\frac{\bar{g}}{N},\frac{g^{2}}{N}\right)$.
The $N$ neurons have inputs $\b h(t)=\left(h_{1}(t),...,h_{N}(t)\right)$
and outputs $\b x(t)=\left(x_{1}(t),...,x_{N}(t)\right)$ and the
neuronal dynamics is described by the conditional probability functional
$\rho[x_{i}|h_{i}]$. For deterministic neurons, where $x_{i}=f[h_{i}]$
is some causal functional of the input, one may set $\rho[x_{i}|h_{i}]=\delta[x_{i}-f[h_{i}]]$.
We use vectorial notation to denote
\begin{align}
\rho[\boldsymbol{x}|\boldsymbol{h}] & =\prod_{i=1}^{N}\rho[x_{i}|h_{i}],\label{eq:rho_x_given_h}
\end{align}
because, given their inputs $\{h_{i}\}$, neurons are otherwise pairwise
independent. The probability functional $\rho[x_{i}|h_{i}]$ is assumed
to be strictly causal, which is $x_{i}(t)$ is independent of $h_{i}(s>t)$;
a more explicit notation would be $\rho[x_{i}(\circ+)|h_{i}(\circ)]$,
denoting that the time-argument $x_{i}(t+\epsilon)$ must be infinitesimally
advanced by $\epsilon>0$ compared to the argument of $h_{i}(t)$
for $\rho$ to depend on $h$.

The joint statistics of input and output is then
\begin{equation}
\rho[\boldsymbol{x},\boldsymbol{h}]=\rho[\boldsymbol{x}|\boldsymbol{h}]\,\rho[\boldsymbol{h}].\label{eq:joint_prob_input_output_app}
\end{equation}
The distribution of the inputs $\bh$ is given as the marginalization
over $\bx$ as
\begin{align}
\rho[\bh]= & \int\D\bx\,\rho[\bx,\bh]\label{eq:marginalization_x}\\
= & \int\D\bx\,\rho[\boldsymbol{x}|\boldsymbol{h}]\,\rho[\boldsymbol{h}].\nonumber 
\end{align}
The connectivity $\bJ$ couples the outputs $\bx$ of the neurons
to the input $\bh$ as
\begin{align*}
\bh(t) & =\bJ\,\bx(t).
\end{align*}
So in the marginalization \prettyref{eq:marginalization_x} over $\bx$
we need to set

\begin{align}
\rho[\bh(\circ)|\bx(\circ)] & =\delta\big[\bh-\bJ\,\bx\big]\label{eq:rho_h}\\
= & \int\D\bhh\,\exp\big(\bhh^{\T}\bh\big)\,\exp\big(-\bhh^{\T}\bJ\,\bx\big),\nonumber 
\end{align}
where the path-integral measure is $\int\D\bhh=\prod_{t}\int_{-i\infty}^{i\infty}\frac{d\hat{h}(t)}{2\pi i}$
and the inner product is meant as $\bhh^{\T}\bh=\sum_{i=1}^{N}\int_{-\infty}^{\infty}dt\,\hat{h}_{i}(t)h_{i}(t)$.
By connecting the outputs back to the inputs, \prettyref{eq:joint_prob_input_output_app}
may seem to take a circular structure like $\rho[\boldsymbol{x},\boldsymbol{h}]=\rho[\boldsymbol{x}|\boldsymbol{h}]\rho[\boldsymbol{h}|\boldsymbol{x}]$.
But since the first conditional probability is causal, and the second
couples only equal time points, \prettyref{eq:joint_prob_input_output_app}
is more accurately represented as 
\begin{align*}
\rho[\boldsymbol{x},\boldsymbol{h}] & =\rho[\boldsymbol{x}(\circ+)|\boldsymbol{h}(\circ)]\,\rho[\boldsymbol{h}(\circ)|\boldsymbol{x}(\circ)]
\end{align*}
which is ordered in time, resulting in a spiraling structure.

Performing the disorder average $\langle\ldots\rangle_{\bJ}$ of \prettyref{eq:marginalization_x},
the only term affected is the last exponential factor in the second
line of \prettyref{eq:rho_h}, which yields
\begin{align}
 & \Big\langle\exp\big(-\bhh^{\T}\,\bJ\,\bx\big)\Big\rangle_{\bJ\stackrel{\text{i.i.d.}}{\sim}\mathcal{N}\left(\frac{\bar{g}}{N},\frac{g^{2}}{N}\right)}\label{eq:disorder_avgd_conn}\\
= & \exp\Big(-\frac{\bar{g}}{N}\,\sum_{i=1}^{N}\hat{h}_{i}^{\T}\,\sum_{j=1}^{N}x_{j}+\frac{g^{2}}{2N}\,\sum_{i,j=1}^{N}\big(\hat{h}_{i}^{\T}x_{j}\big)^{2}\Big).\nonumber 
\end{align}
Here, the scalar product in the last term in the exponent rewrites
explicitly as
\begin{align*}
\big(\hat{h}_{i}^{\T}x_{j}\big)^{2} & =\iint dt\,ds\,\hat{h}_{i}(t)\,\hat{h}_{i}(s)\,x_{j}(t)x_{j}(s).
\end{align*}
The terms suggest the introduction of the auxiliary fields $\mathcal{R}(t):=\frac{\bar{g}}{N}\sum_{j}x_{j}(t)$
and $\mathcal{Q}(t,s):=\frac{g^{2}}{N}\sum_{j}x_{j}(t)x_{j}(s)$ to
rewrite \prettyref{eq:disorder_avgd_conn} as
\begin{align}
\prod_{i}\,\exp\Big(-\hat{h}_{i}^{\T}\mathcal{R}+\frac{1}{2}\,\hat{h}_{i}^{\T}\mathcal{Q}\hat{h}_{i}\Big),\label{eq:factorization}
\end{align}
where the bi-linear form is to be read as 
\begin{align*}
\hat{h}_{i}^{\T}\mathcal{Q}\hat{h}_{i} & =\iint dt\,ds\,\hat{h}_{i}^{\T}(t)\mathcal{Q}(t,s)\hat{h}_{i}(s).
\end{align*}
The appearance of the product sign and the neuron-independent fields
$\mathcal{R}$ and $\mathcal{Q}$ signifies that the problem becomes
completely symmetric with regard to neurons. Enforcing the definitions
of the auxiliary fields by Dirac distributions, represented in Fourier
domain, analogous to \prettyref{eq:rho_h}, yields another pair of
fields $\hat{\mathcal{R}}$ and $\hat{\mathcal{Q}}$ and brings \prettyref{eq:marginalization_x}
into the form
\begin{align}
\big\langle\rho[\bh]\big\rangle_{\bJ} & \stackrel{(\ref{eq:marginalization_x}),(\ref{eq:rho_h})}{=}\int\D\bx\,\rho[\boldsymbol{x}|\boldsymbol{h}]\,\big\langle\delta[\bh-\bJ\bx]\big\rangle_{\bJ}\label{eq:rho_disorder_avgd}\\
\stackrel{(\ref{eq:disorder_avgd_conn}),(\ref{eq:factorization})}{=} & \int\D\{\mathcal{Q},\mathcal{R},\hat{\mathcal{Q}},\hat{\mathcal{R}}\}\exp\big(-\frac{N}{\bar{g}}\hat{\mathcal{R}}^{\T}\mathcal{R}-\frac{N}{g^{2}}\mathcal{\hat{Q}}^{\T}\mathcal{Q}\big)\nonumber \\
 & \times\prod_{i}\int\D\{x_{i},\hat{h}_{i}\}\,\rho[x_{i}|h_{i}]\nonumber \\
 & \times\exp\Big(\hat{h}_{i}^{\T}h_{i}-\hat{h}_{i}^{\T}\mathcal{R}+\frac{1}{2}\,\hat{h}_{i}^{\T}\mathcal{Q}\hat{h}_{i}+\hat{\mathcal{R}}^{\T}x_{i}+x_{i}^{\T}\hat{\mathcal{Q}}x_{i}\Big).\label{eq:rho_written_out}
\end{align}
Adding the normalization condition by integrating over $\b h$
\begin{align*}
1 & \equiv\int\D\bh\,\big\langle\rho[\bh]\big\rangle_{\bJ}
\end{align*}
we note that this integral affects only the last two lines in \prettyref{eq:rho_written_out}.
The exponent in the second line can be considered an action of a field
theory for the auxiliary fields $\{\mathcal{Q},\mathcal{R},\hat{\mathcal{Q}},\hat{\mathcal{R}}\}$.
The integral in the last two lines appears to the power of $N$, so
that one may rewrite the full expression as
\[
\int\D\{\mathcal{Q},\mathcal{R},\hat{\mathcal{Q}},\hat{\mathcal{R}}\}\exp\big(N\,\Omega[\mathcal{R},\mathcal{Q},\hat{\mathcal{R}},\hat{\mathcal{Q}}]\big)
\]
with
\begin{align*}
\Omega[\mathcal{R},\mathcal{Q},\hat{\mathcal{R}},\hat{\mathcal{Q}}] & :=-\frac{\mathcal{R}^{\T}\hat{\mathcal{R}}}{\bar{g}}-\frac{\mathcal{Q}^{\T}\mathcal{\hat{Q}}}{g^{2}}+\ln\,\int\D\{x,h,\hat{h}\}\,\rho[x|h]\\
 & \quad\times\exp\Big(\hat{h}^{\T}h-\hat{h}^{\T}\mathcal{R}+\frac{1}{2}\,\hat{h}^{\T}\mathcal{Q}\hat{h}+\hat{\mathcal{R}}^{\T}x+x^{\T}\hat{\mathcal{Q}}x\Big).
\end{align*}
We now compute the values of the auxiliary fields that provide the
dominant contribution to the probability mass. The appearance of $N$
in the exponent $N\,\Omega[\mathcal{R},\mathcal{Q}]$ suggests to
perform the integration over the fields $\{\mathcal{Q},\mathcal{R},\hat{\mathcal{Q}},\hat{\mathcal{R}}\}$
in saddle-point approximation, demanding $\frac{\delta\Omega}{\delta\{\mathcal{Q},\mathcal{R},\hat{\mathcal{Q}},\hat{\mathcal{R}}\}}\stackrel{!}{=}0$,
which yields four conditions for the saddle-point values $R,Q,\hat{R},\text{\ensuremath{\hat{Q}}}$
of the fields
\begin{align*}
R(t) & =\bar{g}\,\langle x(t)\rangle_{\Omega(R,Q)},\\
\hat{R}(t) & =\bar{g}\,\langle\hat{h}(t)\rangle_{\Omega(R,Q)}\equiv0,\\
Q(t,s) & =g^{2}\,\langle x(t)x(s)\rangle_{\Omega(R,Q)},\\
\hat{Q}(t,s) & =\frac{g^{2}}{2}\,\langle\hat{h}(t)\hat{h}(s)\rangle_{\Omega(R,Q)}\equiv0.
\end{align*}
Here the expectation value is $\langle\ldots\rangle_{\Omega(R,Q)}=\int\D x\,\ldots\,\int\D\{h,\hat{h}\}\,\rho[x|h]\,\exp\Big(\hat{h}^{\T}h-\hat{h}^{\T}R+\frac{1}{2}\,\hat{h}^{\T}Q\hat{h}\Big)$.
The denominator coming from the outer derivative of the logarithm
appearing in the expression for $\Omega$ does not contribute, because
the normalization condition of the latter distribution is unity, since
the exponential term is the moment-generating functional of a Gaussian
process $h\sim\N(R,Q)$ and $\rho[x|h]$ is normalized, allowing us
to rewrite
\begin{align}
\langle\ldots\rangle_{\Omega(R,Q)}= & \int\D x\,\ldots\,\langle\rho[x|h]\rangle_{h\sim\N(R,Q)}.\label{eq:def_av_Omega}
\end{align}
The auxiliary fields $\hat{h}$ and all their powers are zero on expectation,
which is a consequence of the normalization \citep[Section X]{Coolen00_arxiv_II,Helias20_970}.

The equations thus leads to the result in the main text, eqs. \prettyref{eq:saddle_point_R-1}
and \prettyref{eq:saddle_point_Q-1}.

\subsection{Derivation of the mean-field equation for binary networks\label{app:Deriv_MFeq_binary}}

Having obtained the saddle-point solution to the path integral developed
in the previous section, the first result is the time evolution of
the mean input activity $R(t)$. For binary networks, we need to insert
information specific to the neuron model in order to compute $\av{x(t)}{h\sim\N(R(t),Q(t,t))}$
in \prettyref{eq:saddle_point_R-1}. This means we need $\rho[x_{i}(t)|h_{i}]$
for a binary neuron. Note that only the probability distribution of
the activity at a single time point is needed, which is much simpler
to obtain than a distribution across all time points, which would
include not only the dependence on the input history, but also on
the neuron's own activity state. For the most compact presentation,
we will here use the bitlike $n\in\{0,1\}$ representation instead
of the Ising $x\in\{-1,1\}$ representation used in the main text.
The results between the two can be easily related by the mapping
\begin{align}
x & =2\,n-1.\label{eq:x_of_n}
\end{align}

The bit-like $n\in\{0,1\}$ representation has the advantage that
we only need to consider the active state in averages, since the inactive
$n=0$ state does not contribute. In this case, the probability of
finding a neuron active at some time $t$ 
\begin{equation}
p[n_{i}(t)=1|h_{i}]=\int_{-\infty}^{t}\,\frac{dt^{\prime}}{\tau}\,e^{-\frac{t-t^{\prime}}{\tau}}\,\mathrm{T}_{p}\left(h_{i}\left(t^{\prime}\right)\right).\label{eq:Time_evolution_p_appendix}
\end{equation}
is given by the probability $\mathrm{T}_{p}\left(h_{i}\left(t^{\prime}\right)\right)$
to be activated at any prior update time point $t^{\prime}$ and the
survivor function $e^{-\frac{t-t^{\prime}}{\tau}}$ \citep{Cox62},
the probability that no further update happened since. Therefore,
plugging into \prettyref{eq:saddle_point_R-1}
\begin{align*}
R(t)= & \bar{g}\av{x(t)}{\Omega(R,Q)}\\
\stackrel{(\ref{eq:x_of_n})}{=} & \bar{g}\,\Big(2\av{n(t)}{\Omega(R,Q)}-1\Big)\\
= & \bar{g}\,\int_{-\infty}^{t}\,\frac{dt^{\prime}}{\tau}\,e^{-\frac{t-t^{\prime}}{\tau}}\,\av{2\mathrm{T}_{p}\left(h\right)-1}{h\sim\N(R(t^{\prime}),Q(t^{\prime},t^{\prime}))}
\end{align*}
where we use $\int_{-\infty}^{t}\,\frac{dt^{\prime}}{\tau}e^{-\frac{t-t^{\prime}}{\tau}}=1$
from the second to the third line. Taking a time derivative and using
\prettyref{eq:relation_Tp_T}, we obtain the mean-field equation
\begin{equation}
\tau\frac{d}{dt}R(t)+R(t)=\bar{g}\left\langle \T\left(h\right)\right\rangle _{h\sim{\cal N}\left(R(t),\Qs(t,t)\right)}.\label{eq:ODE_mean-appendix}
\end{equation}
Note that here we need only the input variance $Q(t,t)$, which is
trivially given by the mean activity and, potentially, zero-time-lag
cross-correlations of the outputs. However, the mean-field equation
does not depend on the autocorrelation at nonzero time lag, which
is derived in the next section. Therefore, at least as far as cross-correlations
are negligible, \prettyref{eq:ODE_mean-appendix} is closed.

\subsection{Derivation of the ODE for autocorrelations in binary networks\label{app:Derivation-of-autocorr-in-binary-1}}

Here we derive the form \prettyref{eq:Eom_input_autocorr} for the
evolution of the autocorrelation.

The correlation functions in the Ising and bitlike representation
are, according to \prettyref{eq:x_of_n}, related as
\begin{align}
q_{I}(t,s): & =\langle x(t)x(s)\rangle\label{eq:rel_Ising_binary}\\
 & =\langle(2n(t)-1)\,(2n(s)-1)\rangle\nonumber \\
 & =4\langle n(t)n(s)\rangle-2\langle n(t)\rangle-2\langle n(s)\rangle+1,\nonumber 
\end{align}
where by $Q(t,s)=g^{2}q_{I}(t,s)$ we obtain the quantity considered
in the main text in \prettyref{eq:Eom_input_autocorr}. Defining $q(t,s):=\langle n(t)n(s)\rangle$
we have
\begin{align}
q\left(t,s\right) & =\int dh\,\rho\left(h\right)\sum_{n\left(t\right)=0}^{1}\sum_{n\left(s\right)=0}^{1}\,n\left(t\right)n\left(s\right)\rho[n\left(t\right),n\left(s\right)|h]\nonumber \\
 & =\langle\rho[n\left(t\right)=1,n\left(s\right)=1|h]\rangle_{h\sim\rho}.\label{eq:q_binary}
\end{align}
where we write $\langle\ldots\rangle_{h\sim\rho}$ as a short form
of $\int dh\,\rho(h)\ldots$. The latter joint probability is decomposed,
analogous to \prettyref{eq:decomp_input_output}, as
\begin{align}
\rho[n\left(t\right)=1,n\left(s\right)=1|h]= & \rho[x\left(t\right)=1|x\left(s\right)=1,h]\,\rho[x\left(s\right)=1|h].\label{eq:decomp_conditional}
\end{align}
We obtain the first conditional probability on the right by considering
the possibilities to reach the final state $x(t)=1$ given that $x(s)=1$
\begin{align}
\rho\left[x\left(t\right)=1|x\left(s\right)=1,h\right] & =P\left(\text{no updates in }[s,t]\right)\label{eq:cond_1}\\
 & +P\left(\text{last update in }[s,t]\text{ to up-state}\right)\nonumber \\
 & =e^{-\frac{t-s}{\tau}}\nonumber \\
 & +\int_{s}^{t}e^{-\frac{t-t^{\prime}}{\tau}}\,\Tbi\left(h\left(t^{\prime}\right)\right)\,\frac{dt^{\prime}}{\tau}.\nonumber 
\end{align}
Likewise we obtain the latter conditional probability on the right
of \prettyref{eq:decomp_conditional} as
\begin{align}
\rho[x\left(s\right)=1|h]= & \int_{-\infty}^{s}\,e^{-\frac{s-t^{\prime}}{\tau}}\,\Tbi\left(h\left(t^{\prime}\right)\right)\,\frac{dt^{\prime}}{\tau}.\label{eq:cond_2}
\end{align}
Combining \prettyref{eq:q_binary}, \prettyref{eq:cond_1} and \prettyref{eq:cond_2}
we get\begin{widetext}
\begin{align*}
q\left(t,s\right) & =\left\langle \int_{-\infty}^{s}\,e^{-\frac{s-t^{\prime}}{\tau}}\,\Tbi\left(h\left(t^{\prime}\right)\right)\,\frac{dt^{\prime}}{\tau}\,\left(e^{-\frac{t-s}{\tau}}+\int_{s}^{t}\,e^{-\frac{t-t^{\prime\prime}}{\tau}}\,\Tbi\left(h\left(t^{\prime\prime}\right)\right)\,\frac{dt^{\prime\prime}}{\tau}\right)\right\rangle _{h\sim\rho}\\
 & =\int_{-\infty}^{s}\,e^{-\frac{t-t^{\prime}}{\tau}}\,\langle\Tbi(h(t^{\prime}))\rangle_{h\sim\rho}\,\frac{dt^{\prime}}{\tau}+\int_{-\infty}^{s}e^{-\frac{s-t^{\prime}}{\tau}}\int_{s}^{t}\,e^{-\frac{t-t^{\prime\prime}}{\tau}}\,\langle\Tbi(h(t^{\prime}))\Tbi(h(t^{\prime\prime}))\rangle_{h\sim\rho}\,\frac{dt^{\prime\prime}}{\tau}\,\frac{dt^{\prime}}{\tau}.
\end{align*}
In the stationary state, the first integral in the last line reduces
to $\langle\Tbi(h)\rangle_{h\sim\rho}\,e^{-\frac{t-s}{\tau}}$. Also
$q(t,s)=:q(t-s)$ is then a function of the time lag $\Delta t:=t-s$
alone
\begin{align*}
q\left(\Delta t\right) & =\langle\Tbi(h)\rangle_{h\sim\rho}\,e^{-\frac{\Delta t}{\tau}}+\int_{-\infty}^{s}e^{-\frac{s-t^{\prime}}{\tau}}\,\int_{s}^{s+\Delta t}\,e^{-\frac{s+\Delta t-t^{\prime\prime}}{\tau}}\,\langle\Tbi(h(t^{\prime}))\Tbi(h(t^{\prime\prime}))\rangle_{h\sim\rho}\,\frac{dt^{\prime\prime}}{\tau}\,\frac{dt^{\prime}}{\tau}.
\end{align*}
Differentiating by $\Delta t$ we get
\begin{align}
\tau\,\frac{d}{d\Delta t}q(\Delta t) & =-q(\Delta t)+\int_{-\infty}^{s}e^{-\frac{s-t^{\prime}}{\tau}}\,\langle\Tbi(h(t^{\prime}))\Tbi(h(s+\Delta t))\rangle_{h\sim\rho}\frac{dt^{\prime}}{\tau}\label{eq:diffeq_q_final}\\
 & =-q(\Delta t)+\int_{0}^{\infty}e^{-\frac{t}{\tau}}\,\langle\Tbi(h(0))\Tbi(h(t+\Delta t))\rangle_{h\sim\rho}\frac{dt}{\tau},\nonumber 
\end{align}
where we substitute $s-t^{\prime}\to t$ in the last step and used
the stationarity to shift the time arguments of the $h$ by $t-s$.
Using \prettyref{eq:x_of_n}, \prettyref{eq:rel_Ising_binary}, and
\prettyref{eq:relation_Tp_T} and assuming stationarity we get the
result \prettyref{eq:IDE_Q_stationary} in the main text, where the
gain function $\T$ instead of $\Tbi$ appears. Shifting the integration
variable $t$ by $\Delta t$, we obtain
\[
\tau\frac{d}{d\Delta t}q(\Delta t)=-q(\Delta t)+\int_{\Delta t}^{\infty}e^{-\frac{t-\Delta t}{\tau}}\,\langle\Tbi(h(0))\Tbi(h(t))\rangle_{h\sim\rho}\frac{dt}{\tau}
\]
and performing another derivative $\tau\frac{d}{d\Delta t}$ yields
\begin{align}
\tau^{2}\frac{d^{2}}{d\Delta t^{2}}q(\Delta t)= & -\tau\frac{d}{d\Delta t}q(\Delta t)+\int_{\Delta t}^{\infty}e^{-\frac{t-\Delta t}{\tau}}\,\langle\Tbi(h(0))\Tbi(h(t))\rangle_{h\sim\rho}\frac{dt}{\tau}-\langle\Tbi(h(0))\Tbi(h(\Delta t))\rangle_{h\sim\rho}\nonumber \\
= & \,q(\Delta t)-\langle\Tbi(h(0))\Tbi(h(\Delta t))\rangle_{h\sim\rho},\label{eq:diffeq_pre_final}
\end{align}
\end{widetext}where for the second equality, we use the first-order
differential equation \eqref{eq:diffeq_q_final}. Closing the equation
in the mean-field approximation, amounts to setting the measure of
$h\sim\rho\equiv\N(R,Q)$ to the Gaussian process with mean $R$ and
variance $Q$, as determined by the saddle-point equations \prettyref{eq:saddle_point_R-1}
and \prettyref{eq:saddle_point_Q-1}. This approximation neglects
fluctuations of $\mathcal{R}$ and $\mathcal{Q}$, which is justified
if the system is not close to the critical point and the average connectivity
scales at most like $1/N$ \citep{Ginzburg94}, but even if the latter
condition is relaxed to a $1\sqrt{N}$-scaling, this merely leads
to an additional term in the input fluctuations taking into account
pairwise correlations \citep{Helias14}.

Moving to the $[-1,1]$ representation by using \prettyref{eq:x_of_n},
\prettyref{eq:rel_Ising_binary}, and \prettyref{eq:relation_Tp_T}
and multiplying \prettyref{eq:diffeq_pre_final} by $g^{2}$ changes
$\mathrm{T}_{p}\to\mathrm{T}$ and $g^{2}\,q\to Q$ so that we obtain
\prettyref{eq:Eom_input_autocorr}. Here, in addition, we introduce
$\mathcal{N}_{R,Q(0),Q}$ as the bivariate Gaussian with stationary
mean $R$ and covariance matrix $\left(\begin{array}{cc}
Q(0) & Q\\
Q & Q(0)
\end{array}\right)$. This Gaussian expectation value allows us to employ Price's theorem
\citep{PapoulisProb4th} which states that 
\begin{align*}
 & \frac{d}{dQ}\,\langle{\cal T}(h){\cal T}(h^{\prime})\rangle_{(h,h^{\prime})\sim\mathcal{N}_{R,Q(0),Q}}\\
 & \stackrel{\text{(Price's theorem)}}{\equiv}\langle\mathrm{T}(h)\mathrm{T}(h^{\prime})\rangle_{(h,h^{\prime})\sim\mathcal{N}_{R,Q(0),Q}},
\end{align*}
where ${\cal T}(x):=\int^{x}\mathrm{T}(x^{\prime})\,dx^{\prime}$
is the primitive of $\mathrm{T}$.

\subsection{Replica calculation for chaos\label{app:Replica-calculation-for-chaos-in-binary}}

\subsubsection*{Model-independent replica calculation}

To assess the transition to chaos, we perform a replica calculation
that considers a pair of networks with identical connectivity but
slightly different initial conditions for the neurons. We use superscripts
$(1)$ and $(2)$ to distinguish the two systems. The correlation
between the two replicas is a measure of the distance between their
respective states in terms of the squared Euclidean distance
\begin{align}
d^{(12)}(t) & :=||\bx^{(1)}(t)-\bx^{(2)}(t)||^{2}\label{eq:distance_replica}\\
 & =\sum_{\alpha=1}^{2}\sum_{i=1}^{N}x_{i}^{(\alpha)}x_{i}^{(\alpha)}-2\,\sum_{i=1}^{N}x_{i}^{(1)}x_{i}^{(2)}.\nonumber 
\end{align}
The first term, on expectation over realizations of the activity,
approaches the average autocorrelation in the two replicas and the
latter term the inter-replica correlation. For Ising spins the expression
simplifies to $2N-2\,\sum_{i=1}^{N}x_{i}^{(1)}x_{i}^{(2)}$.

The formal derivation of mean-field equations that approximate these
quantities proceeds analogous to \prettyref{app:Model-independent-mean-field-the_appendix}:
The analog expression to \prettyref{eq:marginalization_x} and \prettyref{eq:rho_h}
reads
\begin{align*}
\rho[\bh^{(1)},\bh^{(2)}]= & \int\D\{\bx^{(1)},\bx^{(2)}\}\,\rho[\bx^{(1)},\bx^{(2)}|\bh^{(1)},\bh^{(2)}]\,\\
 & \times\prod_{\alpha=1}^{2}\delta[\bh^{(\alpha)}-\bJ\,\bx^{(\alpha)}].
\end{align*}
Here the conditional density $\rho[\bx^{(1)},\bx^{(2)}|\bh^{(1)},\bh^{(2)}]$
is a joint distribution across the two replicas, because it must allow
the representation of update processes or stochastic activations of
corresponding neurons that have identical realizations between the
two replicas.

The important point is the identical matrix $\bJ$ appearing in the
product of the latter two Dirac distributions, which, after introducing
Fourier representations as in \prettyref{eq:rho_h} and taking the
disorder average over $\bJ$, analogous to \prettyref{eq:disorder_avgd_conn},
yields
\begin{align*}
 & \Big\langle\exp\big(-\bhh^{(1)\T}\bJ\bx^{(1)}-\hat{\bh}^{(2)\T}\bJ\bx^{(2)}\big)\Big\rangle_{\bJ\stackrel{\text{i.i.d.}}{\sim}\mathcal{N}\left(\frac{\bar{g}}{N},\frac{g^{2}}{N}\right)}\\
= & \prod_{i=1}^{N}\prod_{\alpha=1}^{2}\exp\Big(-\frac{\bar{g}}{N}\,\hat{h}_{i}^{(\alpha)\T}\,\sum_{j=1}^{N}x_{j}^{(\alpha)}+\frac{g^{2}}{2N}\,\sum_{j=1}^{N}\big(\hat{h}_{i}^{(\alpha)\T}x_{j}^{(\alpha)}\big)^{2}\Big)\\
 & \phantom{\prod_{i=1}^{N}}\times\exp\Big(\frac{g^{2}}{N}\,\sum_{j=1}^{N}\hat{h}_{i}^{(1)\T}x_{j}^{(1)}\,\hat{h}_{i}^{(2)\T}x_{j}^{(2)}\Big).
\end{align*}
The penultimate line is the same contribution for each replica as
in the single system; it is treated in the same manner by introducing
pairs of auxiliary fields $\{\mathcal{R}^{(\alpha)},\hat{\mathcal{R}}^{(\alpha)},\mathcal{Q}^{(\alpha\alpha)},\hat{\mathcal{Q}}^{(\alpha\alpha)}\}_{\alpha\in\{1,2\}}$.
The last line couples the two replicas and can be decoupled similarly
by defining
\begin{align*}
\mathcal{Q}^{(12)}(t,s) & :=\frac{g^{2}}{N}\,\sum_{j=1}^{N}x_{j}^{(1)}(t)\,x_{j}^{(2)}(s).
\end{align*}
This definition is enforced by inserting a $\delta$ constraint, represented
as a Fourier integral with the corresponding conjugate field $\hat{\mathcal{Q}}^{(12)}(s,t)$.
The integral over $\{\mathcal{R}^{(\alpha)},\hat{\mathcal{R}}^{(\alpha)},\mathcal{Q}^{(\alpha\beta)},\hat{\mathcal{Q}}^{(\alpha\beta)}\}_{\alpha,\beta\in\{1,2\}}$
is then taken in saddle-point approximation with the resulting nontrivial
saddle-point equations
\begin{align}
R^{(\alpha)}(t) & =\bar{g}\,\langle x^{(\alpha)}(t)\rangle_{\Omega(\{R^{(\alpha)},Q^{(\alpha\beta)}\}),}\label{eq:saddle_replica}\\
Q^{(\alpha\beta)}(t,s) & =g^{2}\,\langle x^{(\alpha)}(t)x^{(\beta)}(s)\rangle_{\Omega(\{R^{(\alpha)},Q^{(\alpha\beta)}\}).}\nonumber 
\end{align}
The remaining response fields vanish, $\hat{R}^{(\alpha)}=\hat{Q}^{(\alpha\beta)}\equiv0$.
The expectation value in \prettyref{eq:saddle_replica} is taken with
the measure
\begin{align}
 & \langle\ldots\rangle_{\Omega(\{R^{(\alpha)},Q^{(\alpha\beta)}\})}\label{eq:def_exp_replica}\\
= & \int\D\{x^{(1)},x^{(2)}\}\,\ldots\,\langle\rho[x^{(1)},x^{(2)}|h^{(1)},h^{(2)}]\rangle_{(h^{(1)},h^{(2)})},
\end{align}
where $(h^{(1)},h^{(2)})\sim\N(\{R^{(\alpha)},Q^{(\alpha\beta)}\})$
is a pair of Gaussian processes with cumulants
\begin{align*}
\llangle h^{(\alpha)}(t)\rrangle & =R^{(\alpha)}(t),\\
\llangle h^{(\alpha)}(t)h^{(\beta)}(s)\rrangle & =Q^{(\alpha\beta)}(t,s).
\end{align*}
The distance \prettyref{eq:distance_replica} between the replicas
in mean-field approximation can then be written as
\begin{align*}
d^{(12)}(t) & =N\,g^{-2}\,\big(\sum_{\alpha=1}^{2}Q^{(\alpha\alpha)}(t,t)-2\,Q^{(12)}(t,t)\big).
\end{align*}

\subsubsection*{Application to binary networks}

The zero-lag cross-replica correlation is then given with \prettyref{eq:saddle_replica}
and \prettyref{eq:def_exp_replica} as
\begin{align*}
Q^{(12)}\left(t,t\right) & \stackrel{(\ref{eq:saddle_replica},\ref{eq:def_exp_replica})}{=}\,g^{2}\sum_{x^{(1)}(t),x^{(2)}(t)=-1}^{1}x^{(1)}\cdot x^{(2)}\\
 & \times\big\langle\rho\big(x^{(1)},x^{(2)},t|h^{(1)},h^{(2)}\big)\big\rangle_{(h^{(1)},h^{(2)})\sim\N(\{R^{(\alpha)},Q^{(\alpha\beta)}\})}.
\end{align*}
To construct $\rho(x^{(1)},x^{(2)},t|h^{(1)},h^{(2)})$, first note
that both neurons are updated by the same stochastic realizations
of the update process. This process has two random components: The
drawing of the update time point $t^{\prime}$, which, for the Poisson
updates, has a distribution of $e^{-\frac{t-t^{\prime}}{\tau}}\,\frac{dt^{\prime}}{\tau}$
for the last event to have appeared in $[t^{\prime},t^{\prime}+dt]$,
and the stochastic activation depending on the gain function $\Tbi\in[0,1]$,
whose value for both replicas is compared to the same realization
of a uniformly distributed random number $r\in[0,1]$ .

The four possible outcomes of this update of states $(x^{(1)},x^{(2)})$
are $(-1,-1),$ $(1,1)$, both of which lead to $x^{(1)}\cdot x^{(2)}=+1$
and $(-1,1)$, $(1,-1)$, both of which lead to $x^{(1)}\cdot x^{(2)}=-1$.
One thus only needs to distinguish two outcomes: The event $x^{(1)}\cdot x^{(2)}=-1$
takes place if the random variable $r$ is in between the values of
the two gain functions, $\T_{p}(h^{(1)})<r<\T_{p}(h^{(2)})$, which
happens with probability $p_{\mathrm{diff}}=|\T_{p}(h^{(1)})-\T_{p}(h^{(2)})|$;
the other event $x^{(1)}\cdot x^{(2)}=+1$ with $1-p_{\mathrm{diff}}$.
So in total we get at the time $t^{\prime}$ of update
\begin{align}
\langle x^{(1)}(t^{\prime})x^{(2)}(t^{\prime})\rangle_{r} & =(-1)\cdot p_{\mathrm{diff}}+(+1)\cdot(1-p_{\mathrm{diff}})\label{eq:c_out_p_diff}\\
 & =1-2\,p_{\mathrm{diff}}\nonumber \\
 & =1-2\,|\T_{p}(h^{(1)}(t^{\prime}))-\T_{p}(h^{(2)}(t^{\prime}))|.
\end{align}
Taken together with the asynchronous update time point, we thus have\begin{widetext}
\begin{align*}
Q^{(12)}\left(t,t\right) & =g^{2}\,\int_{-\infty}^{t}\,\frac{dt^{\prime}}{\tau}\,e^{-\frac{t-t^{\prime}}{\tau}}\,\Big(1-2\,\big\langle|\T_{p}(h^{(1)}(t^{\prime}))-\T_{p}(h^{(2)}(t^{\prime}))|\big\rangle_{(h^{(1)},h^{(2)})\sim\N(\{R^{(\alpha)},Q^{(\alpha\beta)}\})}\Big).
\end{align*}
Taking a derivative with respect to $t$, we obtain an ODE governing
the time evolution of the cross-replica correlation 
\begin{align}
\tau\frac{d}{dt}Q^{(12)}\left(t,t\right) & =-Q^{(12)}\left(t,t\right)+g^{2}\,\left(1-2\big\langle|\T_{p}(h^{(1)}(t))-\T_{p}(h^{(2)}(t))|\big\rangle_{(h^{(1)},h^{(2)})\sim\N(\{R^{(\alpha)},Q^{(\alpha\beta)}\})}\right).\label{eq:ODE_Q12_appendix}
\end{align}
Note that $Q^{(12)}(t)$ also appears implicitly in the distribution
of $h^{(1)},h^{(2)}$, rendering the equation nonlinear. This implies
the result \prettyref{eq:full_ODE_crossreplicacorr_binary} in the
main text, which follows by replacing $2(\T_{p}(h^{(1)})-\T_{p}(h^{(2)}))=\T(h^{(1)})-\T(h^{(2)})$
due to \prettyref{eq:relation_Tp_T}. 

So far we have proceeded without approximation apart from the saddle-point
approximation. Perfect correlation of the replicas $Q^{(12)}(t,t)=Q_{0}=g^{2}$
is clearly a fixed point, since then $h^{(1)}(t)=h^{(2)}(t)$ and
the right-hand side vanishes. We now wish to assess the stability
of this solution, that is, whether a perturbation of one replica results
in recovery of perfect correlation (regular dynamics) or in a decorrelation
of the replicas (chaos). Making the ansatz $Q^{(12)}(t,t)=Q_{0}-\epsilon(t)$
and using the mean-field approximation of the input distribution,
the last term of the ODE \prettyref{eq:ODE_Q12_appendix} becomes,
by substituting $H\coloneqq(h^{(1)}+h^{(2)})/2,\,h\coloneqq(h^{(1)}-h^{(2)})/2$
and then expanding in $h$ and $\epsilon/(2Q_{0})$: 

\begin{align}
 & \left\langle \left|\T_{p}\left(h^{(1)}\right)-\T_{p}\left(h^{(2)}\right)\right|\right\rangle _{\begin{pmatrix}h^{(1)}\\
h^{(2)}
\end{pmatrix}\sim\mathcal{N}\left(\begin{pmatrix}R\\
R
\end{pmatrix},\begin{pmatrix}Q_{0} & Q_{0}-\epsilon\\
Q_{0}-\epsilon & Q_{0}
\end{pmatrix}\right)}\nonumber \\
= & \left\langle \left|\T_{p}\left(H+h\right)-\T_{p}\left(H-h\right)\right|\right\rangle _{\begin{pmatrix}H\\
h
\end{pmatrix}\sim\mathcal{N}\left(\begin{pmatrix}R\\
0
\end{pmatrix},\begin{pmatrix}Q_{0}-\frac{\epsilon}{2} & 0\\
0 & \frac{\epsilon}{2}
\end{pmatrix}\right)}\nonumber \\
= & \frac{1}{2\pi\sqrt{\frac{\epsilon}{2}\left(Q_{0}-\frac{\epsilon}{2}\right)}}\iint\,\exp\left(-\frac{H^{2}}{2\left(Q_{0}-\frac{\epsilon}{2}\right)}-\frac{h^{2}}{2\frac{\epsilon}{2}}\right)\left|\T_{p}\left(R+H+h\right)-\T_{p}\left(R+H-h\right)\right|\,dH\,dh\nonumber \\
= & \frac{1}{2\pi\sqrt{\frac{\epsilon}{2}\left(Q_{0}-\frac{\epsilon}{2}\right)}}\iint\,e^{-\frac{H^{2}}{2\left(Q_{0}-\frac{\epsilon}{2}\right)}}e^{-\frac{h^{2}}{\epsilon}}\left|2\T_{p}^{\prime}\left(R+H\right)h+\mathcal{O}\left(h^{3}\right)\right|\,dH\,dh\nonumber \\
= & 2\,\frac{1}{\sqrt{2\pi Q_{0}}}\int\,e^{-\frac{H^{2}}{2Q_{0}}}\left|\T_{p}^{\prime}\left(R+H\right)\right|\,dH\frac{1}{\sqrt{2\pi\frac{\epsilon}{2}}}\int\,\left|h\right|e^{-\frac{h^{2}}{\epsilon}}\,dh\left(1+\mathcal{O}\left(\epsilon\right)\right)\nonumber \\
= & 2\left\langle \T_{p}^{\prime}\left(H\right)\right\rangle _{H\sim\mathcal{N}\left(R,Q_{0}\right)}\,\sqrt{\frac{\epsilon}{\pi}}+\mathcal{O}\left(\epsilon^{\frac{3}{2}}\right).\label{eq:abs_calc_Ising-1-1}
\end{align}
Plugging this result and the ansatz $Q^{(12)}(t)=Q_{0}-\epsilon(t)$
back into \prettyref{eq:ODE_Q12_appendix} then yields \prettyref{eq:ODE_deviation_fully_correlated}.\end{widetext}

\subsection{Growth of perturbations in binary networks\label{app:Initial-growth-of-perturbations-in-binary}}

Starting from \prettyref{eq:ODE_deviation_fully_correlated}
\begin{align*}
\tau\frac{d}{dt}\epsilon(t) & =-\epsilon(t)+c\,\sqrt{\epsilon(t)},
\end{align*}
where $c=\frac{2}{\sqrt{\pi}}g^{2}\left\langle \T^{\prime}\left(h\right)\right\rangle _{h\sim\mathcal{N}\left(R,g^{2}\right)}$,
we integrate the differential equation

\begin{align*}
\tau^{-1}\int_{0}^{t}\,dt & =-\int_{\epsilon(0)}^{\epsilon(t)}\,\frac{d\epsilon}{\epsilon-c\,\sqrt{\epsilon}}\\
 & =-2\,\ln(c-\sqrt{\epsilon})\big|_{\epsilon(0)}^{\epsilon(t)},
\end{align*}
so the solution is

\begin{align*}
\epsilon(t) & =\big(c-\big(c-\sqrt{\epsilon(0)}\big)\,e^{-\frac{t}{2\tau}}\big)^{2}.
\end{align*}
Expressed in terms of the dimensionality $d(t)=\frac{N}{g^{2}}\epsilon(t)$
and $d_{\ast}=\frac{N}{g^{2}}c^{2}$ \prettyref{eq:d_star} 

\begin{align*}
d(t) & =\big(\sqrt{d_{\ast}}-\big(\sqrt{d_{\ast}}-\sqrt{d(0)}\big)\,e^{-\frac{t}{2\tau}}\big)^{2}.
\end{align*}
In the long-time limit the solution reaches the fixed point

\begin{align*}
d_{\ast} & =d(t\to\infty).
\end{align*}
The fastest increase happens for a Heaviside gain function $T=-1+2\theta\in\{-1,1\}$
for which $\left\langle \T^{\prime}\left(h\right)\right\rangle _{h\sim\N(0,g^{2})}=2/(\sqrt{2\pi}\,g)$
so $d^{\ast}=N\,g^{2}\,\frac{4}{\pi}\,\frac{4}{2\pi\,g^{2}}=N\,\frac{8}{\pi^{2}}$.
This shows that the binary network decorrelates only to a dimensionality
of $d^{\ast}/N=8/\pi^{2}\simeq0.81$. The distance as a function of
time is then
\begin{align*}
\frac{d(t)}{N} & =\frac{8}{\pi^{2}}\,\big(1-\big(1-\frac{\pi}{\sqrt{8}}\sqrt{\frac{d(0)}{N}}\big)\,e^{-\frac{t}{2\tau}}\big)^{2}.
\end{align*}
The maximal signal-to-noise ratio is obtained by using $d_{s}$ with
a nonzero initial $d_{s}(0)$ and a noise distance $d_{n}^{0}$ that
initially vanishes $d_{n}^{0}(0)=0$, leading to:
\begin{align*}
\frac{d_{s}(t)-d_{n}^{0}(t)}{d_{\ast}} & =\big(1-\big(1-\delta\big)\,e^{-\frac{t}{2\tau}}\big)^{2}-\big(1-e^{-\frac{t}{2\tau}}\big)^{2}.
\end{align*}
where we define $\delta:=\sqrt{\frac{d_{s}(0)}{d_{\ast}}}$. The
maximum of this function is at
\begin{align}
0 & =-\big(1-\delta\big)\,\big(1-\big(1-\delta\big)\,e^{-\frac{\hat{t}}{2\tau}}\big)+\big(1-e^{-\frac{\hat{t}}{2\tau}}\big)\nonumber \\
\Rightarrow & e^{-\frac{\hat{t}}{2\tau}}=\frac{\delta}{1-\big(1-\delta\big)^{2}}.\label{eq:t_hat_exact}
\end{align}
For small $\delta\ll1$ this expression yields
\begin{align*}
e^{-\frac{\hat{t}}{2\tau}} & \simeq\frac{\delta}{1-\big(1-2\delta\big)}=\frac{1}{2}
\end{align*}
so the time of the maximum becomes approximately independent of the
initial value $\delta$ and thus independent of the signal-to-noise
ratio 
\begin{align}
\hat{t} & \simeq2\tau\,\ln2\simeq1.39\,\tau.\label{eq:t_hat}
\end{align}
The maximum is with $e^{-\frac{\hat{t}}{2\tau}}\simeq\frac{1}{2}$
\begin{align*}
\frac{d_{s}(\hat{t})-d_{n}^{0}(\hat{t})}{d_{\ast}} & \simeq\big(1-\big(1-\delta\big)\,\frac{1}{2}\big)^{2}-\big(1-\frac{1}{2}\big)^{2}\\
 & =\frac{\delta}{2}+\frac{\delta^{2}}{4}.
\end{align*}
The latter expression shows that this maximum, relative to the initial
signal is

\begin{align}
\frac{d_{s}(\hat{t})-d_{n}^{0}(\hat{t})}{d_{s}(0)-d_{n}^{0}(0)} & =\frac{(d_{s}(\hat{t})-d_{n}^{0}(\hat{t}))/d_{\ast}}{d_{s}(0)/d_{\ast}}\label{eq:improvement_signal}\\
 & \simeq\frac{\frac{\delta}{2}+\frac{\delta^{2}}{4}}{\delta^{2}}=\frac{1}{2\,\delta}+\frac{1}{4}\nonumber \\
 & =\frac{1}{2}\sqrt{\frac{d_{\ast}}{d_{s}(0)}}+\frac{1}{4}.\nonumber 
\end{align}
For the Heaviside nonlinearity the result becomes with $d^{\ast}/N=8/\pi^{2}$
\begin{align*}
\frac{d_{s}(\hat{t})-d_{n}^{0}(\hat{t})}{d_{s}(0)-d_{n}^{0}(0)} & =\frac{1}{\pi}\sqrt{2\,\frac{N}{d_{s}(0)}}+\frac{1}{4}.
\end{align*}

\subsection{Growth of perturbations in rate networks\label{app:Initial-growth-of-perturbations-in-binary-1}}

On small timescales the distance evolves in proportion to the Lyapunov
exponent
\begin{align*}
d(t) & =d(0)\,\exp(\lambda_{\mathrm{max}}\,t).
\end{align*}
So $d_{s}(t)-d_{n}(t)=\big(d_{s}(0)-d_{n}(0)\big)\,\exp\big(\lambda_{\mathrm{max}}\,t\big)$.
In particular, there is no maximum expected on a timescale of the
neuronal dynamics. The typical timescale instead is determined by
the maximal Lyapunov exponent.

To derive an equation for the time evolution of the correlation between
replicas in the rate network, we use the pair of equations obtained
from the mean-field description
\begin{align}
\tau\,\partial_{t}x^{\alpha}(t) & =-x^{\alpha}(t)+h^{\alpha}(t)\label{eq:eom_mft_replica}\\
\langle h^{\alpha}(t)h^{\beta}(s)\rangle & =Q^{\alpha\beta}(t,s)\equiv g^{2}\,\langle T(x^{\alpha}(t))T(x^{\beta}(s))\rangle.\nonumber 
\end{align}
We may thus write
\begin{align*}
x^{\alpha}(t) & =\frac{1}{\tau}\,\int_{-\infty}^{t}\,e^{-\frac{t-t^{\prime}}{\tau}}\,h^{\alpha}(t^{\prime})\,dt^{\prime}.
\end{align*}
So the correlation function $c^{\alpha\beta}(t,s):=\langle x^{\alpha}(t)x^{\beta}(s)\rangle$
obeys

\begin{align*}
c^{\alpha\beta}(t,s) & =\frac{1}{\tau}\,\int_{-\infty}^{t}\,e^{-\frac{t-t^{\prime}}{\tau}}\,\frac{1}{\tau}\,\int_{-\infty}^{s}\,e^{-\frac{s-s^{\prime}}{\tau}}\,\langle h^{\alpha}(t^{\prime})h^{\beta}(s^{\prime})\rangle\,dt^{\prime}\,ds^{\prime},
\end{align*}
which becomes in differential form
\begin{align*}
\tau\partial_{t}c^{\alpha\beta}(t,s) & =-c^{\alpha\beta}(t,s)+\frac{1}{\tau}\,\int_{-\infty}^{s}\,e^{-\frac{s-s^{\prime}}{\tau}}\,\langle h^{\alpha}(t)h^{\beta}(s^{\prime})\rangle\,ds^{\prime}.
\end{align*}
This differential equation allows the integration along the $t$ direction
by one time-step $\delta$
\begin{align*}
c^{\alpha\beta}(t+\delta,s) & =(1-\frac{\delta}{\tau})\,c^{\alpha\beta}(t,s)+\frac{\delta}{\tau}\,\int_{-\infty}^{s}\,e^{-\frac{s-s^{\prime}}{\tau}}\,\langle h^{\alpha}(t)h^{\beta}(s^{\prime})\rangle\,ds^{\prime},
\end{align*}
which requires only $c^{\alpha\beta}$ and $\langle hh\rangle$ in
the $t-$past and in the $s-$past. The integration can be done for
$t\in[0,T]$, where $T$ is a desired final point.

As a result, one has $c^{\alpha\beta}(t,s)$ for $t\in[0,T]$. In
the next update step we move into the $s-$direction by

\begin{align*}
c^{\alpha\beta}(t,s+\delta) & =(1-\frac{\delta}{\tau})\,c^{\alpha\beta}(t,s)+\frac{\delta}{\tau}\,\int_{-\infty}^{t}\,e^{-\frac{t-t^{\prime}}{\tau}}\,\langle h^{\alpha}(t^{\prime})h^{\beta}(s)\rangle\,dt^{\prime},
\end{align*}
where the latter integral can be computed because it requires only
\begin{align}
\langle h^{\alpha}(t)h^{\beta}(s)\rangle & =g^{2}\langle T(x_{1})T(x_{2})\rangle_{(x_{1},x_{2})\sim\N(0,c^{\alpha\beta}(t,s))}\label{eq:input_output_corr}
\end{align}
 computed in the previous step.

It makes sense to introduce as an auxiliary variable 
\begin{align*}
q^{\alpha\beta}(t,s) & :=\frac{1}{\tau}\int_{-\infty}^{t}\,e^{-\frac{t-t^{\prime}}{\tau}}\,\langle h^{\alpha}(t^{\prime})h^{\beta}(s)\rangle\,dt^{\prime},
\end{align*}
for which we can assume the symmetry $q^{\alpha\beta}(t,s)=q^{\beta\alpha}(t,s)$
to write the updates

\begin{align}
c^{\alpha\beta}(t+\delta,s) & =(1-\frac{\delta}{\tau})\,c^{\alpha\beta}(t,s)+\frac{\delta}{\tau}\,q^{\alpha\beta}(s,t),\label{eq:it_1}\\
c^{\alpha\beta}(t,s+\delta) & =(1-\frac{\delta}{\tau})\,c^{\alpha\beta}(t,s)+\frac{\delta}{\tau}\,q^{\alpha\beta}(t,s).\label{eq:it_2}
\end{align}
The auxiliary variable $q^{\alpha\beta}(t,s)$ obeys the differential
equation $(\tau\partial_{t}+1)\,q^{\alpha\beta}(t,s)=\langle h^{\alpha}(t)h^{\beta}(s)\rangle$,
which yields the update equation
\begin{align}
q^{\alpha\beta}(t+\delta,s) & =(1-\frac{\delta}{\tau})\,q^{\alpha\beta}(t,s)+\frac{\delta}{\tau}\,\langle h^{\alpha}(t)h^{\beta}(s)\rangle.\label{eq:it_3}
\end{align}
So the required sequence of updates is:
\begin{enumerate}
\item Start at $t=0$.
\item Assume we have computed $c^{\alpha\beta}(t^{\prime},s^{\prime})$
and $q^{\alpha\beta}(t^{\prime},s^{\prime})$ until this point $t$
for all $(t^{\prime}<t,s^{\prime}<t)$.
\item Compute $c^{\alpha\beta}(t+\delta,\forall s^{\prime}\le t)$ using
\eqref{eq:it_1}.
\item Compute $c^{\alpha\beta}(\forall t^{\prime}\le t,t+\delta)$ using
\eqref{eq:it_2}.
\item Compute $\langle h^{\alpha}(t+\delta)h^{\beta}(s^{\prime})\rangle$
$\forall s^{\prime}\le t$ and $\langle h^{\alpha}(t^{\prime})h^{\beta}(t+\delta)\rangle$
$\forall t^{\prime}\le t$ using \eqref{eq:input_output_corr} and
the result from the previous step.
\item Compute $q^{\alpha\beta}(t^{\prime}\le t+\delta,t+\delta)$ by iterating
\eqref{eq:it_3} with zero initial condition and starting with $t^{\prime}$
sufficiently far back in the past.
\item Compute $c^{\alpha\beta}(t+\delta,t+\delta)$ by average of \eqref{eq:it_1}
and \eqref{eq:it_2} using new value $q^{\alpha\beta}(t+\delta,t)$
and $q^{\alpha\beta}(t,t+\delta)$, respectively.
\item Go to the next time slice $t\to t+\delta$, return to step 3.
\end{enumerate}
\textcolor{brown}{}

\textcolor{brown}{}

\textcolor{brown}{}

\textcolor{brown}{}

\subsection{Flux tubes in binary networks\label{app:Fluxtubes-in-binary-networks}}

It has been shown by \citet{Touzel19} that the borders of flux tubes
in spiking networks of inhibitory LIF neurons are related to changes
in the global order of spikes. In particular, if a perturbation creates
an additional spike or causes the omission of an expected one, the
mean firing rate will stay constant but the order of future spikes
is very likely to be irrevocably changed. The divergence rate of two
trajectories can be assessed by calculating the mean number of unexpected
spike order changes caused by a single such perturbation, resulting
in a branching process. In the context of binary networks, we can
ask the equivalent question: Given a flip of a single neuron's activity
variable, how many ``wrong'' update results will occur on average
in the following time $\tau$ ?

The flip of one neuron $x_{j}\to-x_{j}$ causes a change $\Delta h_{i}=-2J_{ij}x_{j}$
in the input of neurons it is connected to. Across different target
neurons $\Delta h$ is therefore distributed as
\[
\rho(\Delta h)=\mathcal{N}\left(\pm2\frac{\bar{g}}{N},4\frac{g^{2}}{N}\right)
\]
and the probability of a neuron to be updated into the wrong state
due to the perturbation in the input is
\begin{align*}
p\left(\text{flip}(x)|\Delta h\right) & =\av{\abs{\Tbi(h+\Delta h)-\T_{p}(h)}}{h\sim\mathcal{N}(R,Q_{0})}\\
 & \overset{\abs{\Delta h}\ll1,\,\Tbi^{\prime}\geq0}{\approx}\abs{\Delta h}\av{\T_{p}^{\prime}(h)}{h\sim\mathcal{N}(R,Q_{0})},
\end{align*}
where the absolute value enters because both directions of perturbation
cause a positive probability of ``wrong'' updating, and we assume
$T^{\prime}\geq0$ for simplicity. Now we ask the following question:
How many downstream flips $n_{\text{spawns}}$ will, on average, be
triggered in the network during one time constant, given a single
original flip? This quantity controls whether the decorrelating flips
will proliferate or not, because since every neuron is updated on
average once per time constant, if $n_{\text{spawns}}<1$ and the
neuron carrying the original flip is updated again, it is most likely
updated ``correctly'' again and the average number of flips in the
network has decreased. If $n_{\text{spawns}}>1$ on the other hand,
the average number of flips increases.

Being interested in the transition point, we can assume $n_{\text{spawns}}\approx1$
so that we do not need to take the interaction of several flips into
account. Then 
\begin{align}
\bar{n}_{\text{spawns}} & =N\av{p\left(\text{flip}(x)|\Delta h\right)}{\Delta h}\label{eq:n_spawns_def}
\end{align}
and while we take the mean input $R$ into account, we neglect the
perturbation of the mean input $\av{\Delta h}{}=\pm2\bar{g}/N=\text{\ensuremath{\mathcal{O}(N^{-1})}}\approx0$
as it is small compared to the standard deviation $\sigma_{\Delta h}=2g/\sqrt{N}=\mathcal{O}(N^{-\frac{1}{2}})$,
allowing the simple calculation
\begin{align*}
\bar{n}_{\text{spawns}} & =N\av{\T_{p}^{\prime}(h)}{h\sim\mathcal{N}(R,Q_{0})}\av{\abs{\Delta h}}{\Delta h\sim\mathcal{N}(0,\sigma_{\Delta h}^{2})}\\
 & =N\av{\T_{p}^{\prime}(h)}{h\sim\mathcal{N}(R,Q_{0})}\sqrt{\frac{2}{\pi}}\sigma_{\Delta h}\\
 & =2\sqrt{\frac{2N}{\pi}}g\av{\T_{p}^{\prime}(h)}{h\sim\mathcal{N}(R,Q_{0})}.
\end{align*}
Finally, accounting for $\T_{p}^{\prime}(h)=\T^{\prime}(h)/2$, given
by \prettyref{eq:relation_Tp_T}, the chaos transition is expected
at
\begin{align}
1\overset{!}{=}\bar{n}_{\text{spawns}} & =\frac{\sqrt{2N}}{\sqrt{\pi}}g\av{\mathrm{T}^{\prime}(h)}{h\sim\mathcal{N}(R,Q_{0})},\label{eq:chaos_crit_fluxtubes}
\end{align}
which is exactly the result \prettyref{eq:chaos_criterion_finite_size},
derived via the completely different route of the replica calculation.
While the derivation here is nicely and intuitively interpretable,
the derivation via field theory and replica calculation allows for
systematic generalizations. For example, it is not clear how to obtain
the residual correlation \prettyref{eq:residual_epsilon} in the ad
hoc approach.

\subsubsection*{Fluxtube size}

The flux-tube diameter is not a very informative measure for a binary
network, since the system trajectory in phase space is typically not
in the middle of a ``tube'' but close to some of its boundaries
(given by the thresholds). Therefore, the distance to a boundary strongly
depends on the direction of perturbation. As a relatively informative
measure, we consider smearing the trajectory in all directions with
some variance $\text{Var}(\Delta h)=\sigma_{\text{fl}}^{2}$, which
is chosen such that on average, one flux-tube boundary is crossed.
This procedure makes sense insofar, as it is similar to adding noise
onto the input. It is important to be aware that $\sigma_{\text{fl}}$
is not strictly the average distance to the closest boundary, although
the two quantities should covary.

The situation is analogous to the above calculation, because we again
need to consider the flips occurring during an update in (on average)
all $N$ neurons, which is given by \prettyref{eq:n_spawns_def} only
with $\sigma_{\Delta h}$ replaced by $\sigma_{\text{fl}}$. Demanding
$\bar{n}_{\text{spawns}}\overset{!}{=}1$ then yields 
\begin{align*}
1 & \overset{!}{=}N\av{p\left(\text{flip}(x)|\Delta h\right)}{\Delta h\sim\mathcal{N}(0,\sigma_{\text{fl}}^{2})}\\
\Rightarrow\quad\sigma_{\text{fl}} & =\frac{\sqrt{2\pi}}{N\av{\mathrm{T}'(h)}{h\sim\mathcal{N}(R,Q_{0})}}.
\end{align*}
Of course, the $1/N$ scaling needs to be taken with caution, since
our perturbation goes into all $N$ phase-space directions, resulting
in a total length scaling as $1/\sqrt{N}$ .

\subsection{Equivalence of dynamical mean-field theories of binary and rate networks\label{app:Equivalence-of-mean-field-rate-binary}}

The dynamics \prettyref{eq:rate_SCS_form} can equivalently be written
as
\begin{align}
\tau\partial_{t}\b x & =-\b x+\mathrm{T}\Bigl(\bh\Bigr),\label{eq:rate_natural_form}\\
\bh & =\b{Jx}+\sqrt{\tau}\b{\eta},\nonumber 
\end{align}
where the noise $\eta_{i}$ is an Ornstein-Uhlenbeck process \citep{Uhlenbeck30},
$(\tau\partial_{t}+1)\,\eta_{i}=\xi_{i}$. This form allows the
application of the model-independent field theory. The single-neuron,
single-time-slice probability functional is $\rho[x(t)|h]=\delta[x(t)-\int_{-\infty}^{t}\,e^{-\frac{t-t^{\prime}}{\tau}}\,\mathrm{T}\left(h_{i}\left(t^{\prime}\right)\right)\,\frac{dt^{\prime}}{\tau}]$,
and the noise term is taken into account in $\rho[\bh]=\delta[\b h-\b J\b x-\sqrt{\tau}\b{\eta}]$.
Plugging this expression into \prettyref{eq:saddle_point_R-1} we
obtain the same equation \prettyref{eq:ODE_mean} for the mean activity
as for the binary neuron, if we choose the strength of the noise $\sigma_{\xi}$
such that $Q(t,t)=g^{2}$ as well. The reason for the equivalence
is that the exponential function appearing in the convolution equation
is the Green's function of $\tau\partial_{t}+1$. In a stationary
state, the saddle-point solution for $Q(\Delta t)\coloneqq Q(t,t+\Delta t)=g^{2}\,\langle x(t)x(t+\tau)\rangle$,
moreover, follows the same Newtonian equation of motion \eqref{eq:Eom_input_autocorr}
as for the binary model \citep[eq. 7]{Sompolinsky88_259}. 

\subsubsection*{Matching initial conditions\label{app:App_Equivalence_binary_rate_matching_init}}

Knowing that the differential equations for the time-lagged autocorrelations
are the same, we have to adjust their respective initial conditions
to establish full equivalence. Here we use the subscripts $b$ and
$r$ to refer to the quantities of the binary and rate model, respectively.
Two initial conditions are needed for a unique solution. One is to
require that $\lim_{t\rightarrow\infty}\dot{Q}\left(t\right)=0$,
which is the same in both cases. So the autocorrelation for infinite
time-lags is described by a single value $Q_{\infty}$, which vanishes
for point-symmetric activation functions, but is in general nonzero
and self-consistently determined by the static variability across
neurons, caused by the disorder (compare \prettyref{fig:rate_binary_autocorr_match}).
In the binary case, the second condition is $Q_{b}(0)=g^{2}$ because
the zero-lag autocorrelation of a single spin is always one. In a
rate network, however, the input noise strength determines how quickly
the autocorrelation decays, resulting in the condition on the derivative
$\dot{Q}_{r}(0+)=-\sigma_{\xi}^{2}/2$ \citep{Schuecker18_041029}\footnote{(note that our notational convention differs by the factor $1/2$)}.

The idea is to choose the variance of the noise $\sigma_{\xi}^{2}$
in the rate network such that $Q_{r}(0)=Q_{b}(0)$, so that the time-lagged
solutions for the variance $Q(\Delta t)$ match.

To do so, using $\dot{Q}(\infty)=0$ and conservation of total ``energy''
$V_{Q_{0}}+\tau^{2}\dot{Q}^{2}/2$ implied by the Newtonian form of
\prettyref{eq:Eom_input_autocorr-Potential}, the condition $Q_{0}:=Q_{r}(0)\stackrel{!}{=}Q_{b}(0)=g^{2}$
can be expressed as a condition for the derivative and thus the noise
amplitude
\begin{align}
\frac{1}{2}\tau^{2}\dot{Q}_{0}^{2}+V_{Q_{0}}(Q_{0}) & =\left.V_{Q_{0}}(Q_{\infty})\right|_{Q_{0}=g^{2}}.\label{eq:Noise_condition_for_equivalence}
\end{align}
Plugging in $\dot{Q}_{0}=-\sigma_{\xi}^{2}/2$ and solving for $\sigma_{\xi}^{2}$
yields the condition \prettyref{eq:Initial_condition_Q_dot_binary}
in the main text. This proves that the binary and rate model with
appropriate noise have equivalent mean activities and time-lagged
autocorrelations in dynamical mean-field approximation.

\subsubsection*{Explanation of the result}

\begin{figure}
\centering{}\includegraphics[width=1\columnwidth]{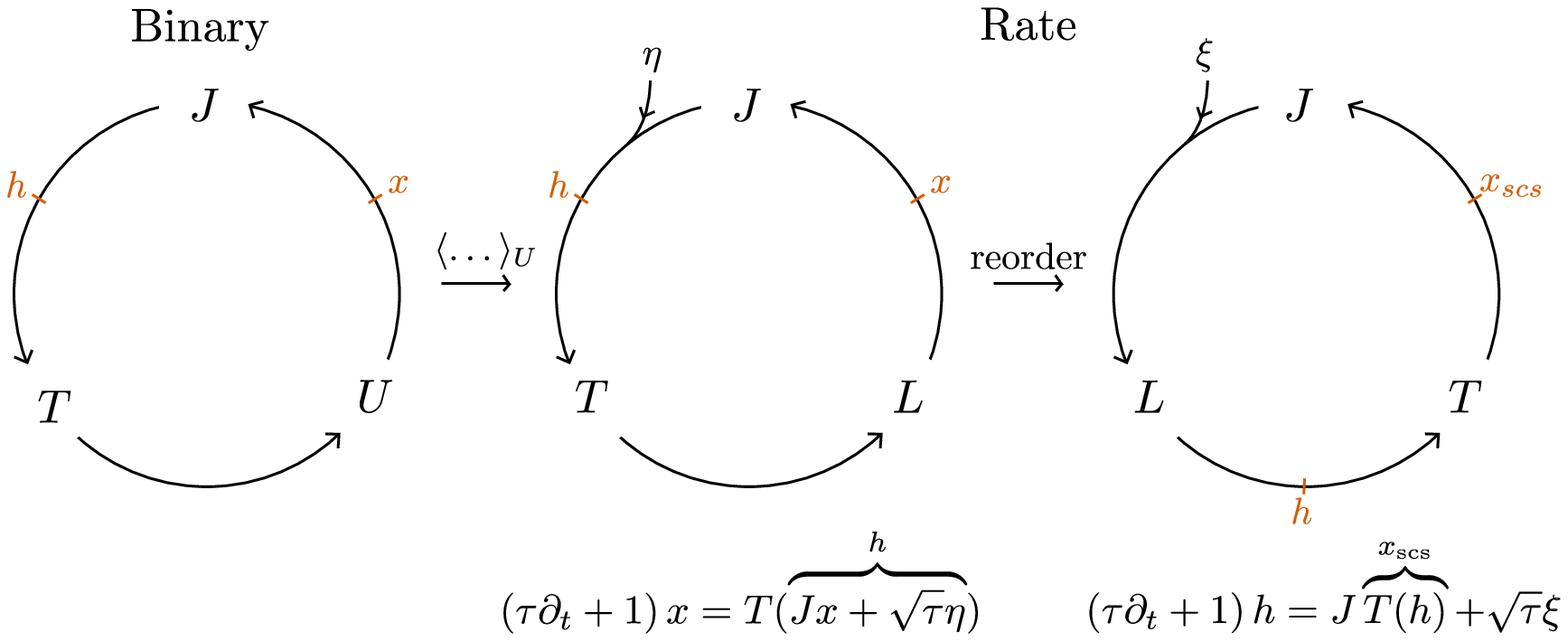}\caption{\textbf{Equivalence of binary (left) and rate models in the two different
forms \prettyref{eq:rate_natural_form} (middle) and \prettyref{eq:rate_SCS_form}
(right):} Mapping from output to input by identical matrix $J$; asynchronous
update process $U$ with rate $\tau^{-1}$ implies exponential convolution
kernel, leading to leaky-integration (cf. \prettyref{eq:ODE_mean}),
identical to operator $L=(\tau\partial_{t}+1)^{-1}$ present explicitly
in \prettyref{eq:rate_natural_form} and \prettyref{eq:rate_SCS_form}.
Transitions between discrete binary states effect red noise $\eta$
in input $h$ (middle), which corresponds to white noise $\xi$ that
is low-pass filtered by $L$ (right). Rate models differ in the order
of application of this kernel and the connectivity, which yields equivalent
dynamics because the two operators commute. \label{fig:rate_binary_match_conceptual}
}
\end{figure}

Taking a step back, what is the intuition behind this result? When
the binary neurons are averaged over realizations of the update time
disorder, the Poisson update process with rate $\tau^{-1}$ becomes
an exponential kernel corresponding to that of the rate network. The
discrete jumps of the binary neurons around their mean become red
noise \citep{Lindner2009,Frey17_046601}, corresponding to the low-pass-filtered
noise $\eta$ of the rate network \prettyref{eq:rate_natural_form}.
By nice conspiracy, this noise corresponds to simple white noise $\xi$
in \prettyref{eq:rate_SCS_form}, which is also the version treated
in most works on rate networks with noise, such as \citep{rajan10_011903,Aljadeff15_088101,Kadmon15_041030,Marti18_062314,Schuecker18_041029,Crisanti18_062120}.
This tight relation between the binary and rate models is summarized
conceptually in \prettyref{fig:rate_binary_match_conceptual}.

\subsection{Slope of correlation transmission in binary and rate neurons\label{app:Diverging-slope-of-binary}}

Here we show that the difference between discrete signaling and continuous
signaling leads to a qualitative difference in the slope of the correlation-transmission
curve and thus the transition to chaos.

Assume, as an approximation, that two neurons receive inputs that
are jointly Gaussian distributed as
\begin{align*}
(h_{1},h_{2}) & \sim\N(0,K),
\end{align*}
where the covariance matrix is given by 
\begin{align*}
K(c_{\mathrm{in}}) & :=q\,\left(\begin{array}{cc}
1 & c_{\mathrm{in}}\\
c_{\mathrm{in}} & 1
\end{array}\right).
\end{align*}
Here $c_{\mathrm{in}}\in[-1,1]$ controls the correlation between
the inputs.

\subsubsection*{Continuous signaling}

A neuron with continuous signaling has the output
\begin{align*}
y_{i} & =\T(h_{i}),
\end{align*}
where $\T\in[-1,1]$ is an activation function. The mean output is
thus
\begin{align}
\langle y_{i}\rangle & =\langle\T(h)\rangle_{h\sim\N(0,q)}.\label{eq:mean_output}
\end{align}
For a point-symmetric gain function that we assume in the following
the mean vanishes so that the variance of the outputs is
\begin{align*}
a & =\langle\T^{2}(h)\rangle_{h\sim\N(0,q)}.
\end{align*}
The correlation coefficient between the outputs of a pair of neurons
is
\begin{align}
c_{\mathrm{out}}^{\text{cont.}}(c_{\mathrm{in}}) & :=a^{-1}\,\langle y_{1}y_{2}\rangle\label{eq:cout_cont}\\
 & =a^{-1}\,\langle\T(h_{1})\T(h_{2})\rangle_{(h_{1},h_{2})\sim\N(0,K(c_{\mathrm{in}}))}\nonumber 
\end{align}
which has the slope
\begin{align*}
\frac{dc_{\mathrm{out}}^{\mathrm{cont.}}(c_{\mathrm{in}})}{dc_{\mathrm{in}}} & =a^{-1}\,\langle\T^{\prime}(h_{1})\T^{\prime}(h_{2})\rangle_{(h_{1},h_{2})\sim\N(0,K(c_{\mathrm{in}}))},
\end{align*}
by Price's theorem \citep{PapoulisProb4th}. Evaluated at $c_{\mathrm{in}}=1$
this is
\begin{align}
\frac{dc_{\mathrm{out}}^{\mathrm{cont.}}(1)}{dc_{\mathrm{in}}} & =a^{-1}\,\langle(\T^{\prime}(h))^{2}\rangle_{h\sim\N(0,q)}\stackrel{\T^{\prime}<\infty}{<}\infty.\label{eq:finite_slope}
\end{align}
For activation functions $\T$ with finite slope $\T^{\prime}<\infty$
this slope is thus finite. For the signum function $\T(x)=2H(x)-1$
we get $a=1$ and
\begin{align}
\frac{dc_{\mathrm{out}}^{\mathrm{cont.}}(c_{\mathrm{in}})}{dc_{\mathrm{in}}} & =2\,\langle\delta(h_{1})\delta(h_{2})\rangle_{(h_{1},h_{2})\sim\N(0,K(c_{\mathrm{in}}))}\nonumber \\
 & =\frac{1}{\pi\,\sqrt{\det(K(c_{\text{in}}))}}=(q\pi)^{-1}\,(1-c_{\mathrm{in}}^{2})^{-\frac{1}{2}},\label{eq:inifinite_slope_gain_rate}
\end{align}
where the latter line comes from the normalization condition of the
two-dimensional Gaussian distribution. Thus, the slope diverges if
and only if the output of the neuron becomes discrete.

\subsubsection*{Discrete signaling}

Now consider a neuron with discrete output, but smooth activation
function $T\in[-1,1]$; a smooth function here corresponds to a probabilistic
activation
\begin{align*}
y_{i} & =\begin{cases}
1 & \text{with prob. }\big(\T(h)+1\big)/2\\
-1 & \text{with prob. }1-\big(\T(h)+1\big)/2
\end{cases}.
\end{align*}
The mean output is thus
\begin{align*}
\langle y_{i}\rangle & =\langle1\cdot\big(\T(h)+1\big)/2-1\cdot(1-\big(\T(h)+1\big)/2)\rangle_{h\sim\N(0,q)}\\
 & =\langle\T(h)\rangle_{h\sim\N(0,q)},
\end{align*}
the same as for the continuous signaling \prettyref{eq:mean_output}.
For a point-symmetric gain function that we assume in the following
the mean vanishes so that the variance of the outputs is $a=1$. The
correlation coefficient of the output is then identical to the second
moment between the outputs of a pair of neurons
\begin{align}
c_{\mathrm{out}}^{\text{disc}}(c_{\mathrm{in}}) & :=\langle y_{1}y_{2}\rangle\label{eq:cout_disc}\\
 & =1\cdot(1-p_{\mathrm{diff}})-1\cdot p_{\mathrm{diff}}=1-2\,p_{\mathrm{diff}}\nonumber \\
 & =1-\langle|\T(h_{1})-\T(h_{2})|\rangle_{(h_{1},h_{2})\sim\N(0,K(c_{\mathrm{in}}))}.\nonumber 
\end{align}
The latter expression is related to the probability $p_{\mathrm{diff}}=\frac{1}{2}\langle|\T(h_{1})-\T(h_{2})|\rangle$
that the two neurons are in different states. This expression is of
course the same as found in \prettyref{eq:abs_calc_Ising-1-1}. In
the limit of $c_{\mathrm{in}}\to1$ we thus have
\begin{align*}
c_{\mathrm{out}}^{\text{disc}}(c_{\mathrm{in}}) & \stackrel{c_{\mathrm{in}}=1-\hat{\epsilon}}{\simeq}1-2\,\left\langle \T^{\prime}\left(h\right)\right\rangle _{h\sim\mathcal{N}\left(0,q\right)}\,\sqrt{\frac{\hat{\epsilon}}{\pi}}+\mathcal{O}\left(\hat{\epsilon}^{3/2}\right)
\end{align*}
where $\hat{\epsilon}=\frac{\epsilon}{g^{2}}$. So the slope diverges
for $c_{\mathrm{in}}\to1$ as
\begin{align*}
 & \frac{dc_{\mathrm{out}}^{\mathrm{disc}}(c_{\mathrm{in}})}{dc_{\mathrm{in}}}\big|_{c_{\mathrm{in}}=1-\hat{\epsilon}}\\
 & \stackrel{(\ref{eq:abs_calc_Ising-1-1})}{=}\frac{d}{d(-\hat{\epsilon})}\Big[1-2\,\left\langle \T^{\prime}\left(h\right)\right\rangle _{h\sim\mathcal{N}\left(0,q\right)}\,\sqrt{\frac{\hat{\epsilon}}{\pi}}+\mathcal{O}\left(\hat{\epsilon}^{3/2}\right)\Big]\\
 & =\left\langle \T^{\prime}\left(h\right)\right\rangle _{h\sim\mathcal{N}\left(0,q\right)}\,(\pi\hat{\epsilon})^{-\frac{1}{2}}+\mathcal{O}\left(\hat{\epsilon}^{1/2}\right)\,\propto(1-c_{\mathrm{in}})^{-\frac{1}{2}}.
\end{align*}
This divergence is present even if the gain function has a finite
slope $\T^{\prime}<\infty$. This is in qualitative contrast to the
finite slope found for the continuous signaling in \prettyref{eq:finite_slope}.

The infinite slope for continuous signaling in the limit of a sharp
activation function, \prettyref{eq:inifinite_slope_gain_rate}, can
be shown to have the same form of divergence for $c_{\mathrm{in}}\to1$
when expanded for $c_{\text{in}}=1-\frac{1}{q}\epsilon$ in the limit
of small $\epsilon\ll1$.

\subsection{Noisy binary pattern classification task\label{app:Noisy-binary-pattern-task}}

We implement a classification task by training one linear readout
\begin{align}
S_{\alpha^{\prime}}(t) & =w_{\alpha^{\prime}}(t)^{\T}(x_{\alpha}(t)+\xi_{\text{pre}})+\xi_{\text{post}}\label{eq:def_signal}
\end{align}
of the network state $x_{\alpha}(t)$ at time $t$ for each of the
$\alpha^{\prime}=1,\ldots,P=50$ patterns to be detected. Here $\xi_{\text{pre}}$
and $\xi_{\text{post}}$ are additional Gaussian readout noises of
standard deviation $\sigma_{\xi,\text{pre}}$ and $\sigma_{\xi,\text{post}}$,
respectively. $\xi_{\text{pre}}$ controls how precisely a single
neuron's state can be read out. $\xi_{\text{post}}$ represents a
noise component of the classification mechanism. Training of the readout
$w_{\alpha^{\prime}}(t)$ is performed for each time point $t$ by
linear regression (see \prettyref{app:Linear-regression}), minimizing
the quadratic error of detecting the stimulus identity, i.e. minimizing
$(S_{\alpha^{\prime}}-\delta_{\alpha\alpha^{\prime}})^{2}$.

The patterns are presented to the network by initializing the first
$L=10$ of the $N=500$ neurons to the stimulus. All other neurons
are in an initial state corresponding to the stationary statistics.
Each stimulus $\alpha$ is a random binary pattern of length $L$
with $\{-1,1\}$ appearing equally likely, superimposed with Gaussian
noise of standard deviation $\sigma$. Note that because of the noise
added to the binary values, these initial states are not strictly
$\in\{-1,1\}$. This freedom in the initial states is just a way to
introduce the noise; after their first update the neurons' states
are strictly $\{-1,1\}$ again. The resulting evolution of the network
state given this initial condition is termed $x_{\alpha}(t)$.

\subsubsection{Linear regression\label{app:Linear-regression}}

Minimizing the quadratic error over all patterns amounts to linear
regression; we consider a single scalar readout target value $y_{\alpha}\in\bR$
for each pattern $\alpha$; in the example above $y_{\alpha}\in\{0,1\}$.
Then $w,x\in\bR^{N}$ and the quadratic error is
\begin{align}
\epsilon & :=\min_{w}\sum_{\alpha=1}^{P}\big(w^{\T}x_{\alpha}-y_{\alpha}\big)^{2}.\label{eq:error-1}
\end{align}
Demanding stationarity with regard to $w$ by differentiating by $\partial_{w_{i}}$
we get $N$ equations 
\begin{align*}
0 & =\sum_{\alpha=1}^{P}2\,\big(w^{\T}x_{\alpha}-y_{\alpha}\big)\,x_{\alpha i}\quad\forall i\\
w^{\T}\sum_{\alpha=1}^{P}x_{\alpha}x_{\alpha}^{\T} & =\sum_{\alpha=1}^{P}y_{\alpha}x_{\alpha}^{\T}.
\end{align*}
The value $w^{\ast}$ to achieve stationarity is
\begin{align}
w^{\ast} & =C^{-1}\,\sum_{\alpha=1}^{P}y_{\alpha}x_{\alpha}^{\T},\label{eq:w_star-2}\\
\text{with }C & :=\sum_{\alpha=1}^{P}x_{\alpha}x_{\alpha}^{\T},\nonumber 
\end{align}
where we use the symmetry of $C$. Inserted into \prettyref{eq:error-1}
\begin{align}
\epsilon & =w^{\ast\T}\,C\,w^{\ast}-2\,w^{\ast\T}\sum_{\alpha}y_{\alpha}x_{\alpha}+\sum_{\alpha}y_{\alpha}^{2}\nonumber \\
 & =\sum_{\alpha=1}^{P}y_{\alpha}^{2}-S,\nonumber \\
S & =\big(\sum_{\alpha=1}^{P}y_{\alpha}x_{\alpha}^{\T}\big)\,C^{-1}\,\big(\sum_{\alpha=1}^{P}y_{\alpha}x_{\alpha}\big).\label{eq:def_signal-1}
\end{align}
In the case of classification, the latter expression simplifies even
further: The first term is a constant $\sum_{\alpha=1}^{P}y_{\alpha}^{2}=1$
for labels $y_{\alpha}\in\{0,1\}$, where $y_{\alpha}=1$ if the presented
pattern $\alpha$ is the pattern $\alpha^{\prime}$ to be detected
and $y_{\alpha}=0$ else, $y_{\alpha}=\delta_{\alpha\alpha^{\prime}}$.
The second term is then identical to the definition \prettyref{eq:def_signal}
for $\xi=0$, obtained by inserting $w^{\ast}$ from \prettyref{eq:w_star-2}.
The expression shows that the signal amplitude $S_{\alpha^{\prime}}$
actually depends on the signal-to-noise ratio, the length of the selected
vector $x_{\alpha^{\prime}}$ measured with regard to the variability
$C$ across all patterns
\begin{align}
S_{\alpha^{\prime}} & =x_{\alpha^{\prime}}^{\T}\,C^{-1}\,x_{\alpha^{\prime}}.\label{eq:signal_classification}
\end{align}
The generalization to stochastic realizations of $x_{\alpha}$, for
example due to the presentation of noisy patterns is straightforward.
We need to replace $\sum_{\alpha=1}^{P}\ldots$ by $\sum_{\alpha=1}^{P}\langle\ldots\rangle$
in the measure for the error \prettyref{eq:error-1} and thus throughout
this calculation, where $\langle\ldots\rangle$ is the expectation
over the noise realizations.

\subsubsection{Approximation of orthogonal patterns and uniform noise\label{app:Approximation-orthogonal-patterns}}

If the patterns $x$ are sufficiently orthogonal in the signal subspace,
we can think of the entries $k$ of any state vector $x_{\alpha}$
to be drawn independently. So the $x_{\alpha,k}\in\{-1,1\}$ appear
with equal probability for those entries that lie in the subspace
of dimension $d_{\mathrm{s}}$. All remaining entries are assumed
to be constant across patterns. We may thus restrict the space to
the $d_{\mathrm{s}}$ informative components. The $kl$-th element
of the covariance matrix for independently drawn entries is
\begin{align}
C_{kl} & =\sum_{\alpha=1}^{P}x_{\alpha k}x_{\alpha l}\nonumber \\
 & \simeq P\delta_{kl}.\label{eq:uncorrelated_patterns-1}
\end{align}
The signal of the readout $\alpha^{\prime}$, following from \prettyref{eq:signal_classification},
then takes the simple form
\begin{align}
S_{\alpha^{\prime}} & \simeq P^{-1}\,||x_{\alpha^{\prime}}||_{d_{s}}^{2}.\label{eq:signal_approx-1}
\end{align}
If the signal is perfectly reliable, that is, if for all noise realizations
$i$ the response $x_{\alpha i}$ is equal to the stereotypical response
$x_{\alpha i}=\bar{x}_{\alpha}$, and if the dimension of the informative
subspace is $d_{s}$, so $\bar{x}\in\bR^{d_{s}}$, we get with $||\bar{x}||_{d_{s}}^{2}=d_{\mathrm{s}}$
\begin{align*}
S_{\alpha^{\prime},\mathrm{max}} & \apprle\frac{d_{\mathrm{s}}}{P}.
\end{align*}
If noisy realizations of patterns cause flips in random entries of
$x_{\alpha^{\prime}}$, which is an approximation since spins are
expected to differ in their susceptibility, the responses are not
perfectly reliable, so we need to replace $x_{\alpha^{\prime}}$ by
$\langle x_{\alpha^{\prime}}\rangle$ in \prettyref{eq:signal_classification}
and thus
\begin{align*}
S_{\alpha^{\prime}} & \simeq P^{-1}\,||\langle x_{\alpha^{\prime}}\rangle||_{d_{s}}^{2}.
\end{align*}
The above expressions thereby link the readout signal to the dimensionality
of the responses, as discussed in the main text in \prettyref{sec:Computation-by-transient-dim-expansion}.

\subsubsection{Nonzero plateau of the signal.\label{app:Nonzero-plateau-of-signal}}

In the simulations, the noise distance somewhat unintuitively saturates
slightly below the signal distance. This is explainable by taking
into account that not all the initial noise realizations actually
cause a crossing of the flux-tube boundary. Instead, those realizations
simply follow the unperturbed pattern trajectory, so that $d_{n,i}=0$
in those cases. Then it is clear that the average noise distance is
smaller than the one predicted based on the assumption of diverging
trajectories:
\begin{align*}
\left\langle d_{n,i}\right\rangle _{i}= & \left(1-p_{\text{no flip}}\right)d_{n}^{\text{residual}}.
\end{align*}
We can estimate the probability that no flip occurred due to the noise
by using the results from \prettyref{app:Fluxtubes-in-binary-networks},
where we calculate the average number of flips in the network after
one time constant given an additional (noise) variance in the input
of the neurons $\sigma_{\Delta h}^{2}$. In our present case the
noise is given by adding $\xi_{i}\sim\mathcal{N}(0,\sigma^{2})$ on
the output activities of the $L$ original neurons of the pattern,
so that the corresponding input variance felt by all neurons in the
network is 
\begin{align*}
\tilde{\sigma}_{\Delta h}^{2}= & \frac{g^{2}}{N}\sigma^{2}L.
\end{align*}
Now we need only to consider that actually the variance is not constant
for a complete round of $N$ updates, but linearly diminishes every
time one of the $L$ source neurons is updated until none is left.
Since the flip probability depends on the square-root of the variance,
there is a corrective factor $c_{\text{dim}}=\sqrt{1-\frac{k}{N}}$
in each term of the product: 
\begin{align}
p_{\text{no flip}} & =\prod_{k=0}^{N-1}\left(1-p_{\text{single flip}}(k)\right)\nonumber \\
 & =\prod_{k=0}^{N-1}\left(1-\frac{1}{N}\bar{n}_{\mathrm{spawns}}\left(\tilde{\sigma}_{\Delta h}\right)c_{\text{dim}}(k)\right)\nonumber \\
 & =\prod_{k=0}^{N-1}\left(1-g\sigma\sqrt{\frac{2L}{\pi N}}\av{\T_{p}^{\prime}(h)}{h\sim\mathcal{N}(R,Q_{0})}\sqrt{1-\frac{k}{N}}\right).\label{eq:p_noflip_correction}
\end{align}
This result fits well with the simulations, yielding the predicted
offset of the asymptotic average signal- and noise distances shown
in \prettyref{fig:transient_dim_expansion_binary}\textbf{b} and the
asymptotic plateau of the approximated average signal in \prettyref{fig:transient_dim_expansion_binary}\textbf{c}.

\subsection{Description of simulations\label{app:Description-of-simulations}}

Simulations for \prettyref{fig:rate_binary_autocorr_match} and \prettyref{fig:simulation-theory_chaos_comparison}\textbf{b},\textbf{d
}were implemented using NEST \citep{Nest2160}. NEST treats binary
neurons in the bitlike $\{0,1\}$ representation. To let every neuron
\textquotedblleft see\textquotedblright{} inputs from $\{-1,1\}$
(Ising spins) we add to each neuron $i$ a bias $\sum_{j}J_{ij}$
and then connect the neurons by the connections $2J_{ij}$ instead
of $J_{ij}$; thereby effectively simulating an Ising system. To obtain
the autocorrelations for the Ising case, \prettyref{eq:rel_Ising_binary}
is used, leading to the result shown in \prettyref{fig:rate_binary_autocorr_match}.
Furthermore, we use a non-point-symmetric activation function $\T(h)=\text{tanh}(h-\Theta)$
by choosing a $\Theta$ to be nonzero for this plot. The reason is,
first, that $\Theta=0$ leads to the theoretical prediction of maximal
output variance $\llangle x^{2}\rrangle=1$ because the mean output
activity $\av x{}$ is $0$. However, due to disorder, the time-averaged
activity is actually a fluctuating quantity across the population
and therefore, the population-averaged variance is always below $1$
for finite systems. This systematic underestimation of the peak of
the autocorrelation at zero time lag can be avoided only by choosing
an activation function that is not point symmetric. Second, this choice
is also natural because mean activity $0$ would imply that neurons
are active half of the time on average \citep[Supp. Mat. II B]{Kuehn17},
which is considerably more than indicated by the low firing rates
measured in cortex \citep{Roxin11_16217}. For \prettyref{fig:rate_binary_autocorr_match},
we have therefore shift the working point by numerically inverting
$\av x{}=\av{\text{tanh}(h-\Theta)}{h\sim\mathcal{N}(R,Q)}$ to obtain
the value for $\Theta$, which in mean-field approximation corresponds
to $\left\langle x\right\rangle =-0.5$. In the simulations it turns
out that $\left\langle x\right\rangle =-0.501$, which agrees well
with mean-field theory.

For \prettyref{fig:simulation-theory_chaos_comparison}\textbf{b}
and \textbf{d}, for each point of the grid one simulation of two identical
networks was performed. After $1000\,\text{ms}$, in one replica,
the first two neurons are set to the active state and the third and
fourth neurons are set to the inactive state, after which the simulation
continues for $2500\,\text{ms}$. This method of perturbation entails
the small probability that these four neurons are already in exactly
this state, so that nothing is changed; this is the explanation for
the scattered single green dots in \prettyref{fig:simulation-theory_chaos_comparison}\textbf{b},\textbf{d}.
The advantage of the method is, however, that it guarantees the same
state of the random number generators across both replicas.

The simulations for \prettyref{fig:transient_dim_expansion_binary},
\prettyref{fig:rate_chaos_A} and \prettyref{fig:rate_chaos_B} are
performed using a custom FORTRAN kernel.

The simulations of the LIF network for \prettyref{fig:LIF-SNR-amplification}
are implemented in NEST, using the ``iaf\_psc\_delta'' neuron with
its default parameter settings $V_{\mathrm{rest}}=V_{\mathrm{reset}}=-70\,\mathrm{mV,}\,V_{th}=-55\,\mathrm{mV,}\,\tau_{m}=10\,\mathrm{ms},\,t_{\mathrm{refr}}=2\,\mathrm{ms},\,C_{m}=250\,\mathrm{pF}$.
Asynchronous-irregular firing in the inhibitory network is evoked
by supplying external excitatory Poisson input with rate of $3000\,\mathrm{Hz}$
and unit weight, leading to an average network activity of $19\,\mathrm{Hz}$.
After simulating for $10\,\tau_{m}$ to obtain a state with stationary
statistics, the input pattern is applied by causing a spike in the
corresponding neurons. Noise is added to the pattern by additional
external input spikes perturbing the membrane potentials of the $L$
pattern neurons. Further parameters as mentioned in figure caption:
$N=500,\,J_{ii}=-1\mathrm{mV},\,K=125,\,\tau_{m}=10\,\mathrm{ms}$.

The LSTM network for \prettyref{fig:LSTM-SNR-amplification} is simulated
using the vanilla PYTORCH implementation with a hidden layer of size
$N=200$. To obtain chaotic fluctuations over a range of timescales,
the hidden weights $W_{hi},\,W_{hf},\,W_{hg}$ are initialized as
$\sim\mathcal{N}(0,5.8/N)$. This excludes the $W_{ho}$ weights,
which would cause very rapid, erratic dynamics. Instead, these and
all remaining weights and biases used the default uniformly distributed
$\sim\mathcal{U}(-1/\sqrt{N},1/\sqrt{N})$ initialization. After simulating
for $50$ time steps to obtain stationary statistics, the input patterns
are supplied using the standard input function.

Analysis of simulation data and numerical solutions are implemented
in PYTHON. The code to generate all figures \textcolor{black}{is available
as a Zenodo archive at \href{http://doi.org/10.5281/zenodo.4705262}{doi.org/10.5281/zenodo.4705262}.}

\end{document}